\documentclass[twocolumn,traditabstract,longauth]{aa}
\usepackage{graphicx,amsmath,amsfonts,amssymb}
\usepackage{epsf}
\usepackage{url}
\usepackage[breaklinks, colorlinks, citecolor=blue]{hyperref}

\usepackage{float}
\usepackage{hhline}
\usepackage{placeins}
\usepackage{array}
\newcolumntype{L}[1]{>{\raggedright\let\newline\\\arraybackslash\hspace{0pt}}m{#1}}
\usepackage{numprint}
\npdecimalsign{.}
\nprounddigits{2}

\usepackage{tikz}
\usepackage[absolute]{textpos}
\usepackage{natbib}
\bibpunct{(}{)}{;}{a}{}{,} 

\def\setsymbol#1#2{\expandafter\def\csname #1\endcsname{#2}}
\def\getsymbol#1{\csname #1\endcsname}

\def\Planck{\textit{Planck}}

\def\alltwentyfifteenresultspapers{\nocite{planck2014-a01, planck2014-a03, planck2014-a04, planck2014-a05, planck2014-a06, planck2014-a07, planck2014-a08, planck2014-a09, planck2014-a11, planck2014-a12, planck2014-a13, planck2014-a14, planck2014-a15, planck2014-a16, planck2014-a17, planck2014-a18, planck2014-a19, planck2014-a20, planck2014-a22, planck2014-a24, planck2014-a26, planck2014-a28, planck2014-a29, planck2014-a30, planck2014-a31, planck2014-a35, planck2014-a36, planck2014-a37, planck2015-ES}}

\newbox\tablebox    \newdimen\tablewidth
\def\leaderfil{\leaders\hbox to 5pt{\hss.\hss}\hfil}
\def\endPlancktable{\tablewidth=\columnwidth 
    $$\hss\copy\tablebox\hss$$
    \vskip-\lastskip\vskip -2pt}
\def\endPlancktablewide{\tablewidth=\textwidth 
    $$\hss\copy\tablebox\hss$$
    \vskip-\lastskip\vskip -2pt}
\def\tablenote#1 #2\par{\begingroup \parindent=0.8em
    \abovedisplayshortskip=0pt\belowdisplayshortskip=0pt
    \noindent
    $$\hss\vbox{\hsize\tablewidth \hangindent=\parindent \hangafter=1 \noindent
    \hbox to \parindent{$^#1$\hss}\strut#2\strut\par}\hss$$
    \endgroup}
\def\doubleline{\vskip 3pt\hrule \vskip 1.5pt \hrule \vskip 5pt}

\def\L2{\ifmmode L_2\else $L_2$\fi}

\def\DeltaT{\ifmmode \Delta T\else $\Delta T$\fi}
\def\deltat{\ifmmode \Delta t\else $\Delta t$\fi}
\def\fknee{\ifmmode f_{\rm knee}\else $f_{\rm knee}$\fi}
\def\Fmax{\ifmmode F_{\rm max}\else $F_{\rm max}$\fi}
\def\solar{\ifmmode{\rm M}_{\mathord\odot}\else${\rm M}_{\mathord\odot}$\fi}
\def\Msolar{\ifmmode{\rm M}_{\mathord\odot}\else${\rm M}_{\mathord\odot}$\fi}
\def\Lsolar{\ifmmode{\rm L}_{\mathord\odot}\else${\rm L}_{\mathord\odot}$\fi}
\def\inv{\ifmmode^{-1}\else$^{-1}$\fi}
\def\mo{\ifmmode^{-1}\else$^{-1}$\fi}
\def\sup#1{\ifmmode ^{\rm #1}\else $^{\rm #1}$\fi}
\def\expo#1{\ifmmode \times 10^{#1}\else $\times 10^{#1}$\fi}
\def\,{\thinspace}
\def\lsim{\mathrel{\raise .4ex\hbox{\rlap{$<$}\lower 1.2ex\hbox{$\sim$}}}}
\def\gsim{\mathrel{\raise .4ex\hbox{\rlap{$>$}\lower 1.2ex\hbox{$\sim$}}}}

\def\simprop{\mathrel{\raise .4ex\hbox{\rlap{$\propto$}\lower 1.2ex\hbox{$\sim$}}}}
\def\deg{\ifmmode^\circ\else$^\circ$\fi}
\def\pdeg{\ifmmode $\setbox0=\hbox{$^{\circ}$}\rlap{\hskip.11\wd0 .}$^{\circ}
          \else \setbox0=\hbox{$^{\circ}$}\rlap{\hskip.11\wd0 .}$^{\circ}$\fi}
\def\arcs{\ifmmode {^{\scriptstyle\prime\prime}}
          \else $^{\scriptstyle\prime\prime}$\fi}
\def\arcm{\ifmmode {^{\scriptstyle\prime}}
          \else $^{\scriptstyle\prime}$\fi}
\newdimen\sa  \newdimen\sb
\def\parcs{\sa=.07em \sb=.03em
     \ifmmode \hbox{\rlap{.}}^{\scriptstyle\prime\kern -\sb\prime}\hbox{\kern -\sa}
     \else \rlap{.}$^{\scriptstyle\prime\kern -\sb\prime}$\kern -\sa\fi}
\def\parcm{\sa=.08em \sb=.03em
     \ifmmode \hbox{\rlap{.}\kern\sa}^{\scriptstyle\prime}\hbox{\kern-\sb}
     \else \rlap{.}\kern\sa$^{\scriptstyle\prime}$\kern-\sb\fi}
\def\ra[#1 #2 #3.#4]{#1\sup{h}#2\sup{m}#3\sup{s}\llap.#4}
\def\dec[#1 #2 #3.#4]{#1\deg#2\arcm#3\arcs\llap.#4}
\def\deco[#1 #2 #3]{#1\deg#2\arcm#3\arcs}
\def\rra[#1 #2]{#1\sup{h}#2\sup{m}}

\def\dots{\relax\ifmmode \ldots\else $\ldots$\fi}
\def\WHzsr{\ifmmode $W\,Hz\mo\,sr\mo$\else W\,Hz\mo\,sr\mo\fi}
\def\mHz{\ifmmode $\,mHz$\else \,mHz\fi}
\def\GHz{\ifmmode $\,GHz$\else \,GHz\fi}
\def\mKs{\ifmmode $\,mK\,s$^{1/2}\else \,mK\,s$^{1/2}$\fi}
\def\muKs{\ifmmode \,\mu$K\,s$^{1/2}\else \,$\mu$K\,s$^{1/2}$\fi}
\def\muKRJs{\ifmmode \,\mu$K$_{\rm RJ}$\,s$^{1/2}\else \,$\mu$K$_{\rm RJ}$\,s$^{1/2}$\fi}
\def\muKHz{\ifmmode \,\mu$K\,Hz$^{-1/2}\else \,$\mu$K\,Hz$^{-1/2}$\fi}
\def\MJysr{\ifmmode \,$MJy\,sr\mo$\else \,MJy\,sr\mo\fi}
\def\MJysrmK{\ifmmode \,$MJy\,sr\mo$\,mK$_{\rm CMB}\mo\else \,MJy\,sr\mo\,mK$_{\rm CMB}\mo$\fi}
\def\microns{\ifmmode \,\mu$m$\else \,$\mu$m\fi}

\def\muK{\ifmmode \,\mu$K$\else \,$\mu$\hbox{K}\fi}
\def\microK{\ifmmode \,\mu$K$\else \,$\mu$\hbox{K}\fi}
\def\muW{\ifmmode \,\mu$W$\else \,$\mu$\hbox{W}\fi}
\def\kms{\ifmmode $\,km\,s$^{-1}\else \,km\,s$^{-1}$\fi}
\def\kmsMpc{\ifmmode $\,\kms\,Mpc\mo$\else \,\kms\,Mpc\mo\fi}
\providecommand{\sorthelp}[1]{}

\usepackage{ifthen}

\newcommand{\redmapper}{{\tt redMaPPer}}
\newcommand{\thetafive}{\theta_{500}}
\newcommand{\sigl}{\sigma_{\ln\lambda}}
\newcommand{\Dln}{\Delta_{\rm ln}}
\newcommand{\poserr}{\theta_{\rm err}}

\newcommand{\yfive}{$Y_{500}$}
\newcommand{\yfiver}{$Y_{5\text{R}500}$}
\newcommand{\pszone}{PSZ1}
\newcommand{\psztwo}{PSZ2}
\newcommand{\mmfone}{\texttt{MMF1}}
\newcommand{\mmfthree}{\texttt{MMF3}}
\newcommand{\pws}{\texttt{PwS}}
\newcommand{\snr}{S/N}

\begin{document}
\author{\small
Planck Collaboration: P.~A.~R.~Ade\inst{99}
\and
N.~Aghanim\inst{68}
\and
M.~Arnaud\inst{83}
\and
M.~Ashdown\inst{79, 6}
\and
J.~Aumont\inst{68}
\and
C.~Baccigalupi\inst{97}
\and
A.~J.~Banday\inst{113, 10}
\and
R.~B.~Barreiro\inst{74}
\and
R.~Barrena\inst{72, 43}
\and
J.~G.~Bartlett\inst{1, 77}
% ...
}
\institute{\small
APC, AstroParticule et Cosmologie, Universit\'{e} Paris Diderot, CNRS/IN2P3, CEA/lrfu, Observatoire de Paris, Sorbonne Paris Cit\'{e}, 10, rue Alice Domon et L\'{e}onie Duquet, 75205 Paris Cedex 13, France\goodbreak
\and
Academy of Sciences of Tatarstan, Bauman Str., 20, Kazan, 420111, Republic of Tatarstan, Russia\goodbreak
\and
African Institute for Mathematical Sciences, 6-8 Melrose Road, Muizenberg, Cape Town, South Africa\goodbreak
\and
Agenzia Spaziale Italiana Science Data Center, Via del Politecnico snc, 00133, Roma, Italy\goodbreak
\and
Aix Marseille Universit\'{e}, CNRS, LAM (Laboratoire d'Astrophysique de Marseille) UMR 7326, 13388, Marseille, France\goodbreak
\and
Astrophysics Group, Cavendish Laboratory, University of Cambridge, J J Thomson Avenue, Cambridge CB3 0HE, U.K.\goodbreak
\and
Astrophysics \& Cosmology Research Unit, School of Mathematics, Statistics \& Computer Science, University of KwaZulu-Natal, Westville Campus, Private Bag X54001, Durban 4000, South Africa\goodbreak
\and
Atacama Large Millimeter/submillimeter Array, ALMA Santiago Central Offices, Alonso de Cordova 3107, Vitacura, Casilla 763 0355, Santiago, Chile\goodbreak
\and
CITA, University of Toronto, 60 St. George St., Toronto, ON M5S 3H8, Canada\goodbreak
\and
CNRS, IRAP, 9 Av. colonel Roche, BP 44346, F-31028 Toulouse cedex 4, France\goodbreak
\and
CRANN, Trinity College, Dublin, Ireland\goodbreak
\and
California Institute of Technology, Pasadena, California, U.S.A.\goodbreak
\and
Centre for Theoretical Cosmology, DAMTP, University of Cambridge, Wilberforce Road, Cambridge CB3 0WA, U.K.\goodbreak
\and
Centro de Estudios de F\'{i}sica del Cosmos de Arag\'{o}n (CEFCA), Plaza San Juan, 1, planta 2, E-44001, Teruel, Spain\goodbreak
\and
Computational Cosmology Center, Lawrence Berkeley National Laboratory, Berkeley, California, U.S.A.\goodbreak
\and
Consejo Superior de Investigaciones Cient\'{\i}ficas (CSIC), Madrid, Spain\goodbreak
\and
DSM/Irfu/SPP, CEA-Saclay, F-91191 Gif-sur-Yvette Cedex, France\goodbreak
\and
DTU Space, National Space Institute, Technical University of Denmark, Elektrovej 327, DK-2800 Kgs. Lyngby, Denmark\goodbreak
\and
D\'{e}partement de Physique Th\'{e}orique, Universit\'{e} de Gen\`{e}ve, 24, Quai E. Ansermet,1211 Gen\`{e}ve 4, Switzerland\goodbreak
\and
Departamento de Astrof\'{\i}sica, Universidad de La Laguna, E-38206 La Laguna, Tenerife, Spain\goodbreak
\and
Departamento de F\'{\i}sica, Universidad de Oviedo, Avda. Calvo Sotelo s/n, Oviedo, Spain\goodbreak
\and
Department of Astronomy and Astrophysics, University of Toronto, 50 Saint George Street, Toronto, Ontario, Canada\goodbreak
\and
Department of Astronomy and Geodesy, Kazan Federal University,  Kremlevskaya Str., 18, Kazan, 420008, Russia\goodbreak
\and
Department of Astrophysics/IMAPP, Radboud University Nijmegen, P.O. Box 9010, 6500 GL Nijmegen, The Netherlands\goodbreak
\and
Department of Physics \& Astronomy, University of British Columbia, 6224 Agricultural Road, Vancouver, British Columbia, Canada\goodbreak
\and
Department of Physics and Astronomy, Dana and David Dornsife College of Letter, Arts and Sciences, University of Southern California, Los Angeles, CA 90089, U.S.A.\goodbreak
\and
Department of Physics and Astronomy, University College London, London WC1E 6BT, U.K.\goodbreak
\and
Department of Physics, Florida State University, Keen Physics Building, 77 Chieftan Way, Tallahassee, Florida, U.S.A.\goodbreak
\and
Department of Physics, Gustaf H\"{a}llstr\"{o}min katu 2a, University of Helsinki, Helsinki, Finland\goodbreak
\and
Department of Physics, Princeton University, Princeton, New Jersey, U.S.A.\goodbreak
\and
Department of Physics, University of Arizona, 1118 E 4th St, Tucson, AZ, 85721, USA\goodbreak
\and
Department of Physics, University of California, One Shields Avenue, Davis, California, U.S.A.\goodbreak
\and
Department of Physics, University of California, Santa Barbara, California, U.S.A.\goodbreak
\and
Department of Physics, University of Illinois at Urbana-Champaign, 1110 West Green Street, Urbana, Illinois, U.S.A.\goodbreak
\and
Dipartimento di Fisica e Astronomia G. Galilei, Universit\`{a} degli Studi di Padova, via Marzolo 8, 35131 Padova, Italy\goodbreak
\and
Dipartimento di Fisica e Scienze della Terra, Universit\`{a} di Ferrara, Via Saragat 1, 44122 Ferrara, Italy\goodbreak
\and
Dipartimento di Fisica, Universit\`{a} La Sapienza, P. le A. Moro 2, Roma, Italy\goodbreak
\and
Dipartimento di Fisica, Universit\`{a} degli Studi di Milano, Via Celoria, 16, Milano, Italy\goodbreak
\and
Dipartimento di Fisica, Universit\`{a} degli Studi di Trieste, via A. Valerio 2, Trieste, Italy\goodbreak
\and
Dipartimento di Fisica, Universit\`{a} di Roma Tor Vergata, Via della Ricerca Scientifica, 1, Roma, Italy\goodbreak
\and
Dipartimento di Matematica, Universit\`{a} di Roma Tor Vergata, Via della Ricerca Scientifica, 1, Roma, Italy\goodbreak
\and
Discovery Center, Niels Bohr Institute, Blegdamsvej 17, Copenhagen, Denmark\goodbreak
\and
Dpto. Astrof\'{i}sica, Universidad de La Laguna (ULL), E-38206 La Laguna, Tenerife, Spain\goodbreak
\and
European Southern Observatory, ESO Vitacura, Alonso de Cordova 3107, Vitacura, Casilla 19001, Santiago, Chile\goodbreak
\and
European Space Agency, ESAC, Planck Science Office, Camino bajo del Castillo, s/n, Urbanizaci\'{o}n Villafranca del Castillo, Villanueva de la Ca\~{n}ada, Madrid, Spain\goodbreak
\and
European Space Agency, ESTEC, Keplerlaan 1, 2201 AZ Noordwijk, The Netherlands\goodbreak
\and
Facolt\`{a} di Ingegneria, Universit\`{a} degli Studi e-Campus, Via Isimbardi 10, Novedrate (CO), 22060, Italy\goodbreak
\and
GEPI, Observatoire de Paris, Section de Meudon, 5 Place J. Janssen, 92195 Meudon Cedex, France\goodbreak
\and
Gran Sasso Science Institute, INFN, viale F. Crispi 7, 67100 L'Aquila, Italy\goodbreak
\and
HGSFP and University of Heidelberg, Theoretical Physics Department, Philosophenweg 16, 69120, Heidelberg, Germany\goodbreak
\and
Helsinki Institute of Physics, Gustaf H\"{a}llstr\"{o}min katu 2, University of Helsinki, Helsinki, Finland\goodbreak
\and
INAF - Osservatorio Astrofisico di Catania, Via S. Sofia 78, Catania, Italy\goodbreak
\and
INAF - Osservatorio Astronomico di Padova, Vicolo dell'Osservatorio 5, Padova, Italy\goodbreak
\and
INAF - Osservatorio Astronomico di Roma, via di Frascati 33, Monte Porzio Catone, Italy\goodbreak
\and
INAF - Osservatorio Astronomico di Trieste, Via G.B. Tiepolo 11, Trieste, Italy\goodbreak
\and
INAF/IASF Bologna, Via Gobetti 101, Bologna, Italy\goodbreak
\and
INAF/IASF Milano, Via E. Bassini 15, Milano, Italy\goodbreak
\and
INFN, Sezione di Bologna, Via Irnerio 46, I-40126, Bologna, Italy\goodbreak
\and
INFN, Sezione di Roma 1, Universit\`{a} di Roma Sapienza, Piazzale Aldo Moro 2, 00185, Roma, Italy\goodbreak
\and
INFN, Sezione di Roma 2, Universit\`{a} di Roma Tor Vergata, Via della Ricerca Scientifica, 1, Roma, Italy\goodbreak
\and
INFN/National Institute for Nuclear Physics, Via Valerio 2, I-34127 Trieste, Italy\goodbreak
\and
IPAG: Institut de Plan\'{e}tologie et d'Astrophysique de Grenoble, Universit\'{e} Grenoble Alpes, IPAG, F-38000 Grenoble, France, CNRS, IPAG, F-38000 Grenoble, France\goodbreak
\and
IUCAA, Post Bag 4, Ganeshkhind, Pune University Campus, Pune 411 007, India\goodbreak
\and
Imperial College London, Astrophysics group, Blackett Laboratory, Prince Consort Road, London, SW7 2AZ, U.K.\goodbreak
\and
Infrared Processing and Analysis Center, California Institute of Technology, Pasadena, CA 91125, U.S.A.\goodbreak
\and
Institut N\'{e}el, CNRS, Universit\'{e} Joseph Fourier Grenoble I, 25 rue des Martyrs, Grenoble, France\goodbreak
\and
Institut Universitaire de France, 103, bd Saint-Michel, 75005, Paris, France\goodbreak
\and
Institut d'Astrophysique Spatiale, CNRS (UMR8617) Universit\'{e} Paris-Sud 11, B\^{a}timent 121, Orsay, France\goodbreak
\and
Institut d'Astrophysique de Paris, CNRS (UMR7095), 98 bis Boulevard Arago, F-75014, Paris, France\goodbreak
\and
Institute of Astronomy, University of Cambridge, Madingley Road, Cambridge CB3 0HA, U.K.\goodbreak
\and
Institute of Theoretical Astrophysics, University of Oslo, Blindern, Oslo, Norway\goodbreak
\and
Instituto de Astrof\'{\i}sica de Canarias, C/V\'{\i}a L\'{a}ctea s/n, La Laguna, Tenerife, Spain\goodbreak
\and
Instituto de Astrof\'{\i}sica de Canarias, E-38200 La Laguna, Tenerife, Spain\goodbreak
\and
Instituto de F\'{\i}sica de Cantabria (CSIC-Universidad de Cantabria), Avda. de los Castros s/n, Santander, Spain\goodbreak
\and
Istituto Nazionale di Astrofisica - Osservatorio Astronomico di Roma, Via Frascati 33, 00040, Monte Porzio Catone (RM), Italy\goodbreak
\and
Istituto Nazionale di Fisica Nucleare, Sezione di Padova, via Marzolo 8, I-35131 Padova, Italy\goodbreak
\and
Jet Propulsion Laboratory, California Institute of Technology, 4800 Oak Grove Drive, Pasadena, California, U.S.A.\goodbreak
\and
Jodrell Bank Centre for Astrophysics, Alan Turing Building, School of Physics and Astronomy, The University of Manchester, Oxford Road, Manchester, M13 9PL, U.K.\goodbreak
\and
Kavli Institute for Cosmology Cambridge, Madingley Road, Cambridge, CB3 0HA, U.K.\goodbreak
\and
LAL, Universit\'{e} Paris-Sud, CNRS/IN2P3, Orsay, France\goodbreak
\and
LAPTh, Univ. de Savoie, CNRS, B.P.110, Annecy-le-Vieux F-74941, France\goodbreak
\and
LERMA, CNRS, Observatoire de Paris, 61 Avenue de l'Observatoire, Paris, France\goodbreak
\and
Laboratoire AIM, IRFU/Service d'Astrophysique - CEA/DSM - CNRS - Universit\'{e} Paris Diderot, B\^{a}t. 709, CEA-Saclay, F-91191 Gif-sur-Yvette Cedex, France\goodbreak
\and
Laboratoire Traitement et Communication de l'Information, CNRS (UMR 5141) and T\'{e}l\'{e}com ParisTech, 46 rue Barrault F-75634 Paris Cedex 13, France\goodbreak
\and
Laboratoire de Physique Subatomique et Cosmologie, Universit\'{e} Grenoble-Alpes, CNRS/IN2P3, 53, rue des Martyrs, 38026 Grenoble Cedex, France\goodbreak
\and
Laboratoire de Physique Th\'{e}orique, Universit\'{e} Paris-Sud 11 \& CNRS, B\^{a}timent 210, 91405 Orsay, France\goodbreak
\and
Lawrence Berkeley National Laboratory, Berkeley, California, U.S.A.\goodbreak
\and
Lebedev Physical Institute of the Russian Academy of Sciences, Astro Space Centre, 84/32 Profsoyuznaya st., Moscow, GSP-7, 117997, Russia\goodbreak
\and
Max-Planck-Institut f\"{u}r Astrophysik, Karl-Schwarzschild-Str. 1, 85741 Garching, Germany\goodbreak
\and
Max-Planck-Institut f\"{u}r Extraterrestrische Physik, Giessenbachstra{\ss}e, 85748 Garching, Germany\goodbreak
\and
McGill Physics, Ernest Rutherford Physics Building, McGill University, 3600 rue University, Montr\'{e}al, QC, H3A 2T8, Canada\goodbreak
\and
Moscow Institute of Physics and Technology, Dolgoprudny, Institutsky per., 9, 141700, Russia\goodbreak
\and
National University of Ireland, Department of Experimental Physics, Maynooth, Co. Kildare, Ireland\goodbreak
\and
Niels Bohr Institute, Blegdamsvej 17, Copenhagen, Denmark\goodbreak
\and
Optical Science Laboratory, University College London, Gower Street, London, U.K.\goodbreak
\and
SB-ITP-LPPC, EPFL, CH-1015, Lausanne, Switzerland\goodbreak
\and
SISSA, Astrophysics Sector, via Bonomea 265, 34136, Trieste, Italy\goodbreak
\and
SLAC National Accelerator Laboratory, Menlo Park, CA 94025, USA\goodbreak
\and
School of Physics and Astronomy, Cardiff University, Queens Buildings, The Parade, Cardiff, CF24 3AA, U.K.\goodbreak
\and
School of Physics and Astronomy, University of Nottingham, Nottingham NG7 2RD, U.K.\goodbreak
\and
Sorbonne Universit\'{e}-UPMC, UMR7095, Institut d'Astrophysique de Paris, 98 bis Boulevard Arago, F-75014, Paris, France\goodbreak
\and
Space Research Institute (IKI), Russian Academy of Sciences, Profsoyuznaya Str, 84/32, Moscow, 117997, Russia\goodbreak
\and
Space Sciences Laboratory, University of California, Berkeley, California, U.S.A.\goodbreak
\and
Special Astrophysical Observatory, Russian Academy of Sciences, Nizhnij Arkhyz, Zelenchukskiy region, Karachai-Cherkessian Republic, 369167, Russia\goodbreak
\and
Sterrewacht Leiden, P.O. Box 9513, NL-2300 RA Leiden, The Netherlands\goodbreak
\and
Sub-Department of Astrophysics, University of Oxford, Keble Road, Oxford OX1 3RH, U.K.\goodbreak
\and
T\"{U}B\.{I}TAK National Observatory, Akdeniz University Campus, 07058, Antalya, Turkey\goodbreak
\and
Theory Division, PH-TH, CERN, CH-1211, Geneva 23, Switzerland\goodbreak
\and
UCLA Astronomy, PO Box 951547, Los Angeles CA 90095-1547, USA\goodbreak
\and
UPMC Univ Paris 06, UMR7095, 98 bis Boulevard Arago, F-75014, Paris, France\goodbreak
\and
Universidad AndrŽs Bello, Dpto. de Ciencias Fisicas, Facultad de Ciencias Exactas, 8370134 Santiago de Chile, Chile\goodbreak
\and
Universit\'{e} Denis Diderot (Paris 7), 75205 Paris Cedex 13, France\goodbreak
\and
Universit\'{e} de Toulouse, UPS-OMP, IRAP, F-31028 Toulouse cedex 4, France\goodbreak
% ...
}

\title{\Planck\ 2015 results. XXVII. The Second \Planck\ Catalogue of Sunyaev-Zeldovich Sources}
\authorrunning{\Planck\ Collaboration}
\titlerunning{\Planck\ Legacy SZ}

\abstract{We present the all-sky \Planck~catalogue of Sunyaev-Zeldovich (SZ) sources detected from the 29 month full-mission data.  The catalogue (\psztwo) is the largest SZ-selected sample of galaxy clusters yet produced and the deepest all-sky catalogue of galaxy clusters.  It contains 1653 detections, of which 1203 are confirmed clusters with identified counterparts in external data-sets, and is the first SZ-selected cluster survey containing $>10^3$ confirmed clusters.  
We present a detailed analysis of the survey selection function in terms of its completeness and statistical reliability, placing a lower limit of 83\% on the purity.  Using simulations, we find that the \yfiver~estimates are robust to pressure-profile variation and beam systematics, but accurate conversion to \yfive~requires. the use of prior information on the cluster extent.
We describe the multi-wavelength search for counterparts in ancillary data, which makes use of radio, microwave, infra-red, optical and X-ray data-sets, and which places emphasis on the robustness of the counterpart match.  We discuss the physical properties of the new sample and identify a population of low-redshift X-ray under-luminous clusters revealed by SZ selection. These objects appear in optical and SZ surveys with consistent properties for their mass, but are almost absent from \emph{ROSAT} X-ray selected samples.

\begin{textblock}{87}(14,270)
\noindent \tikz\draw(0,0)--(0.4\textwidth,0);
\vskip2pt
 \hspace{3pt} $^{*}$ Corresponding~author:~D.~Sutton, \protect\url{sutton@ast.cam.ac.uk}
\end{textblock}
}


\maketitle


\alltwentyfifteenresultspapers

\section{Introduction}
\label{sec:intro}

This paper is one of a set associated with the 2015
\Planck\footnote{\Planck\ (\url{http://www.esa.int/Planck}) is a project of the European Space Agency  (ESA) with instruments provided by two scientific consortia funded by ESA member states and led by Principal Investigators from France and Italy, telescope reflectors provided through a collaboration between ESA and a scientific consortium led and funded by Denmark, and additional contributions from NASA (USA).} full mission data release and
describes the production and properties of the legacy catalogue of
Sunyaev Zeldovich sources (\psztwo).

In the framework of hierarchical structure formation, peaks in the
cosmological density field collapse and merge to form gravitationally
bound haloes of increasing mass \citep{pee80}.  The galaxy clusters
are the most massive of these bound structures and act as signposts
for the extrema of the cosmological density field on the relevant
scales.  The evolution of galaxy cluster abundance with mass and
redshift is thus a sensitive cosmological probe of the late-time
universe, providing unique constraints on the normalisation of the
matter density fluctuations, $\sigma_8$, the mean matter density,
$\Omega_{\text{m}}$, the density and equation of state of the dark
energy field, $\Omega_{\text{DE}}$ and $w$, as well as constraining
some extensions of the minimal cosmological model, such as massive
neutrinos, and non-standard scenarios such as modified gravity (see
eg: \citealt{bor09,all11}).  In recent years, cluster data from the
microwave through to the X-ray have been used to constrain
cosmological parameters
\citep{vik09c,rozo10,has13,ben13,planck2013-p15,zu14}.

Galaxy clusters are multi-component objects composed of dark matter,
which dominates the mass, stars, cold gas and dust in galaxies, and a
hot ionised intra-cluster medium (ICM).  These different components
make clusters true multi-wavelength objects.  The galaxies emit in the
optical and infrared.  The ICM, which is the majority of the baryonic
material by mass, emits in the X-rays via thermal bremsstrahlung and
line emission, and energy-boosts cosmic microwave background photons
via inverse Compton scattering.

This last effect, the thermal Sunyaev Zeldovich (SZ) effect
\citep{sun70,sun80}, imprints a redshift-independent spectral
distortion on the cosmic-microwave background (CMB) photons
reaching us along the line of sight to the cluster.  This results
in an increase in intensity at frequencies above 220\ GHz, and a
decrease in intensity at lower frequencies.  The
\Planck\ High-Frequency Instrument (HFI) is unique in providing
high-precision data for both the increment and the decrement across
the whole sky.

The utility of a cluster survey for cosmological work depends on our
ability to determine accurately its selection function and to obtain
unbiased measurements of cluster mass and redshift.  The first cluster
surveys consisted of galaxy overdensities identified by eye from
photographic plates \citep{abe58}.  The construction of large optical
catalogues improved significantly with the data from the SDSS
\citep{koe07}, whose five photometric bands have allowed robust
photometric classification of red-sequence cluster galaxies and
accurate photometric redshifts to $z<0.6$ across 1/4 of the sky
\citep{hao10,sza11,wen09,rykoff2014}.  These catalogues now typically
contain $10^4-10^5$ clusters and provide cluster richness as an
observable that correlates with mass with an intrinsic scatter
$\sigma_{\text{int}}$ of about $25$\% \citep{rozo14}.

Construction of X-ray cluster surveys is now a mature activity, with
several catalogues now available based on all-sky data from the
\emph{ROSAT} satellite, alongside additional catalogues of
serendipitous detections from pointed observations
\citep{ebe98,boe04,rei02,ebe10,pif11,bur07,meh12}.  The most basic
X-ray survey observable, the X-ray luminosity $L_{500}$ measured within $r_{500}$
\footnote{$r_{500}$ is the cluster-centric distance within which the mean density is 500
 times the critical density of the Universe at the cluster redshift.}, has been
shown correlate with mass with intrinsic scatter of about $40$\%
\citep{pra09}.  Observables with lower intrinsic scatter against mass
can be defined when pointed X-ray follow-up information is available,
including the core-excised X-ray luminosity \citep{mau07,pra09} and
$Y_{\text{x}}$, the product of the gas mass and the core-excised
spectroscopic temperature \citep{kra06, vik09, mahd13}.  While X-ray
surveys are unique in their purity, they do suffer from selection
biases that favour low-redshift systems, due to flux limitations, and
dynamically relaxed clusters with an X-ray bright cooling core
\citep{eck11,sch03,vik09c,che07}.

SZ surveys offer a different window on the cluster population: their
selection function flattens towards higher redshifts, providing a
nearly mass-limited census of the cluster population at high redshift,
where abundance is strongly sensitive to cosmological parameters
\citep{car02,planck2013-p05a}.  The SZ survey observable is the
spherically integrated Comptonisation parameter, $Y_{\text{sz}}$,
which is related to the integrated electron pressure and hence the
total thermal energy of the cluster gas.  It is also expected to
correlate with mass with a low intrinsic scatter and little dependence
on the dynamical state of the cluster (eg:
\citealt{das04,kay12,hoe12,planck2012-III,sif13}).
%
%

The spherically-integrated pressure profiles of X-ray and SZ clusters
have been observed to follow a near universal profile with little
dispersion \citep{arn10,planck2012-V}, permitting the detection of
clusters with a matched multi-frequency filter based on some assumed
pressure profile \citep{herranz2002,mel06}.  Samples constructed this
way have well understood selection functions, though discrepancies due
to profile mismatch or contaminating infra-red emission may still be present to some level.  Large SZ
surveys have only appeared recently, with catalogues of order $\sim
10^2$ clusters released by the Atacama Cosmology Telescope
\citep{has13}, the South Pole Telescope \citep{rei13} and
\Planck\ satellite collaborations.

This is the third all-sky catalogue produced from \Planck\ SZ data.
The early Sunyaev-Zeldovich (ESZ) catalogue presented 189 clusters
detected from 10 months of survey data \citep{planck2011-1.10sup},
while the \pszone, the full-sky catalogue assembled from the nominal
mission data, presented 1227 cluster candidates detected from 15.5
months of data \citep{planck2013-p05a}.  This paper presents 1653
candidates detected from the full mission survey of 29 months.  1203
of these have been confirmed in ancillary data, and 1094 have redshift
estimates.  The \psztwo\ expands the scope and sensitivity of the SZ view
of galaxy clusters by substantially increasing the number of lower
mass clusters available for study. It is also expected to contain many
new, as yet unconfirmed, high-redshift clusters.  We report on the
construction and characterisation of the catalogue, presenting the
survey selection functions and a compilation of multi-wavelength
ancillary information including redshifts. We also briefly discuss the
physical properties of the sample.

This paper is organised as follows. In Sect.~\ref{sec:extraction_algo}
we summarise the three extraction algorithms used to build the
catalogue, focussing on the changes in the algorithms since they were
used to construct the \pszone.  In Sect.~\ref{sec:construction} we
describe the construction of the catalogue.  In
Sect.~\ref{sec:selection_function} we present the survey selection
functions (completeness and statistical reliability) and the
complementary approaches used to estimate them, while in
Sect.~\ref{sec:param_est} we discuss and validate the estimation of
the $Y_{\text{sz}}$ parameters, both blindly and when using prior
information, and we compare the consistency of the new catalogue with
the \pszone~in Sect.~\ref{sec:consistency}.  In
Sect.~\ref{sec:ancillary_info} we report on the search for
multi-wavelength counterparts in ancillary catalogues and follow-up
observations.  Finally we present the physical properties of the
sample in Sect.~\ref{sec:sample_prop} and conclude in
Sect.~\ref{sec:conclusions}.  A full description of the available data
products is given in Appendix~\ref{appendix:products}

\section{Extraction Algorithms}
\label{sec:extraction_algo}

The SZ detection and parameter estimation algorithms used to construct
the PSZ2 extend and refine those used to construct the PSZ1.  In this
section we recall the principles of the three algorithms. The
refinements of each algorithm since the PSZ1 release are detailed in
Appendix~\ref{appendix:extraction_refine}.  Two of the algorithms
(\mmfone~and \mmfthree) are based on the same technique (Matched
Multi-filters) but have been implemented independently\footnote{The
  MMF numbers were given after the comparison of twelve algorithms in
  an earlier phase of the
  \Planck\ mission~\citep{melin2012}. \mmfone~and \mmfthree~were
  respectively the first and third algorithm based on Matched Multi
  Filters to enter the comparison}. The third one (\pws~for
PowellSnakes) relies on Bayesian inference.

\subsection{Matched Multi-filters:  \mmfone~and \mmfthree}
\label{sec:mmf}

The matched filtering technique was first proposed for SZ studies
by~\cite{hae1996}. It was subsequently developed by~\cite{herranz2002}
and~\cite{melin2006} for SZ cluster extraction in multifrequency data
sets such as \Planck. The method was later adopted by the SPT and ACT
collaborations~\citep{sta09,mar11}.

We model the vector of map emission at each frequency $\vec{m}(\vec{x})$, at a given position on the sky $\vec{x}$ as
\begin{equation}
\label{eq:datamodel}
   \vec{m}(\vec{x}) =  y_{\rm o} \vec{t_{\theta_s}}(\vec{x}) + \vec{n}(\vec{x})
\end{equation}
\noindent where $\vec{t_{\theta_s}}(\vec{x})$ is the signal vector describing the 
spatial distribution at each frequency of the SZ emission from a cluster with angular size $\theta_{s}$, 
$\vec{n}(\vec{x})$ is the total astrophysical and instrumental
noise. 
The $i^{th}$ frequency component of the signal vector is the normalized
cluster profile $\tau_{\theta_s}(\vec{x})$~\citep{arn10} convolved by
the \Planck\ beams $b_i(\vec{x})$ and scaled with the characteristic frequency
dependance $j_\nu(\nu_i)$ of the thermal SZ effect: $\vec{t_{\theta_s}}(\vec{x})_i =
j_\nu(\nu_i) [b_i\ast \tau_{\theta_s}](\vec{x})$. $\theta_s$ is the
cluster scale radius, which is related to $\theta_{500}$ through the
 concentration parameter $c_{500}$ by $\theta_{500}=c_{500} \times \theta_s$.
The Matched Multi-filter $\vec{\Psi_{\theta_s}}$ allows us to recover an
unbiased estimate ${\hat y_{\rm o}}$ of the central Comptonization
parameters $y_{\rm o}$ with minimal variance $\sigma_{\theta_s}^2$:
\begin{equation}
{\hat y_{\rm o}} = \int d^2x \; \vec{\Psi_{\theta_s}}^T(\vec{x}) \cdot \vec{m}(\vec{x}),
\end{equation}
where
\begin{equation}
\vec{\Psi_{\theta_s}}(\vec{k}) = \sigma_{\theta_s}^2 \vec{P}^{-1}(\vec{k}) \cdot \vec{t_{\theta_s}}(\vec{k}),
\end{equation}
with
\begin{eqnarray}
\label{eq:sigmat}
\sigma_{\theta_s}          & \equiv & \left[\int d^2k\; 
     \vec{t_{\theta_s}}^T(\vec{k}) \cdot \vec{P}^{-1} \cdot
     \vec{t_{\theta_s}}(\vec{k}) \right]^{-1/2},
\end{eqnarray}
$\vec{P}(\vec{k})$ being the cross-channel power spectrum matrix of the maps.  It
is effectively the noise matrix for the MMF, because the tSZ is small compared to other 
astrophysical signals, and is estimated directly from the maps.

The MMF algorithms first divide each \Planck\ all-sky map in 640/504
tangential maps (14.66/10 degrees on a side) for
\mmfone~/\mmfthree~respectively.  Each set of tangential maps is
filtered by $\Psi_{\theta_s}$ with the assumed cluster size varying from $\theta_s$=0.8 to
32 arcmin. We then locate peaks in the filtered maps above a \snr\
threshold of four. The locations of the peaks give the positions
of our cluster candidates. These are then merged into a single all-sky
catalogue by merging candidates separated by less than 10 arcmin. 
For \mmfthree, we performed a second step by creating sets of smaller rectangular 
frequency maps centered on the cluster candidates identified in the first step. 
We re-apply the MMF on these centred tangential maps which
allows a better estimation of the background. During the second step,
the sizes and fluxes are estimated more precisely.  This second step is only performed for \mmfthree~because the overlap
of the tangential maps in the first step is small with compared to \mmfone and \pws.

We define the blind cluster size as the filter scale that maximizes the \snr\ at the location of
the cluster candidate and the blind flux is defined as the corresponding ${\hat
  y_{\rm o}}$ parameter.  We then define the integrated blind flux as:
\begin{equation}
Y_{5R500}={\hat y_{\rm o}} \int_{\theta<5\times \theta_{500}}dr  \tau_{\theta_s}(r)
\end{equation}

Each of the algorithms produces probability distributions in the
($\theta_s$, $Y_{5\text{R}500}$) plane for each detection, marginalising over
 the parameters for the centre of the cluster, which possess a Gaussian likelihood.
   The algorithms also return an estimate of the radial position uncertainty, $\theta_\text{err}$ from the
   position likelihood.

Although the two implementations of the MMF are quite close, they
produce noticeably different catalogues because the extraction is very
sensitive to the estimation of the background (Eq. \ref{eq:sigmat}). Both the size adopted
for the tangential maps and the details of the estimation of the
matrix $\vec{P}(k)$ impact the \snr\ and hence which peaks are detected. 

\subsection{PowellSnakes (\pws)}
\label{sec:PowellSnakes}

PowellSnakes (\pws) is a fast, fully Bayesian, multifrequency
detection algorithm designed to identify and characterize compact
objects buried in a diffuse background as described in
\cite{car09,car12}.  \pws~operates using about 2800 square patches
of 14.66 degree on a side, in
order to ensure highly redundant sky coverage. \pws~detects candidate
clusters and at the same time computes the evidence ratio and samples
from the posterior distributions of the cluster parameters. Then, it
merges the sub-catalogues from each patch map and
applies criteria for acceptance or rejection of the
detection, as described in \cite{car12}. Priors may be provided for the position, integrated flux
and radius of the clusters.  For cluster detection, we apply flat priors on the position and 
non-informative priors in the radius and integrated flux, as determined using Jeffrey's method.
\pws ~uses a calibration of
the cross-power spectrum that uses an iterative scheme 
to reduce the contamination of the background by the SZ signal itself. This makes \pws\ particularly
robust to small changes in the background.  

\section{Catalogue Construction}
\label{sec:construction}

The main catalogue is constructed by combining the detections made by the
three methods into a union catalogue, while merging the detections made by
multiple more than one method.  Half of the detections in this union are also in the
intersection catalogue, defined as those detections made by all three codes simultaneously.
  This section describes the technical details of the construction of these catalogues.


\subsection{Pipeline}

\begin{table}
\caption{Effective frequencies and Gaussian beam widths assumed for
  extraction per channel.  The beam widths are the mean
  full-width-at-half-maximum of the FEBeCoP Gaussian beam fits across
  the sky, in arcmin.  The effective frequencies $\nu_{\text{eff}}$,
  shown in GHz, encapsulate band-pass effects in each channel.}
\begin{center}
\begin{tabular}{ccc}
\hline\hline
Channel & FWHM & $\nu_{\text{eff}}$ \\
\hline
100 & 9.659 & 103.416 \\
143 & 7.220 & 144.903 \\
217 & 4.900 & 222.598 \\
353 & 4.916 & 355.218 \\
545 & 4.675 & 528.400 \\
857 & 4.216 & 776.582 \\
\hline
\end{tabular}
\end{center}
\label{tab:fwhm_efffreq}
\end{table}%

\begin{figure}
\begin{center}
\includegraphics[angle=0,width=0.5\textwidth]{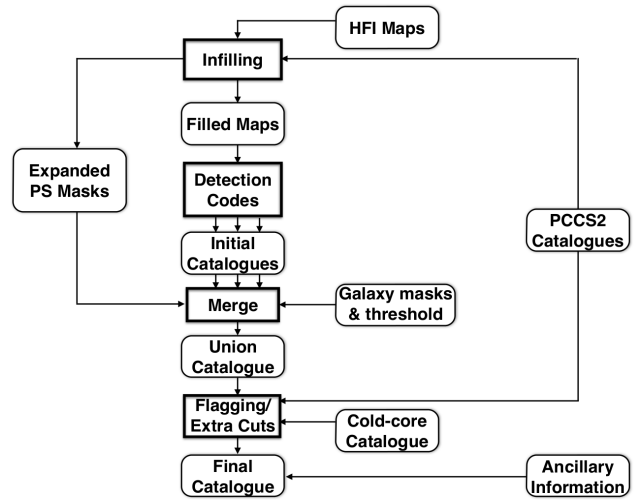}
\caption{Pipeline for catalogue construction.}
\label{fig:pipeline}
\end{center}
\end{figure}

The SZ catalogue construction pipeline is shown in schematic form in
Fig.~\ref{fig:pipeline} and largely follows the process used to build
the \pszone.  The \Planck\ data required for the construction of the
catalogue comprises the HFI maps, point source catalogues for each of
the HFI channels, effective frequencies and beam widths per HFI
channel as shown in Table~\ref{tab:fwhm_efffreq}, survey masks based
on dust emission as seen in the highest \Planck\ channels, and the
catalogue of extended galactic cold-clump detections.

The HFI maps are pre-processed to fill areas of missing data
(typically a few pixels), or areas with unusable data, specifically
bright point sources.  Point sources with \snr~$ > 10$ in any channel
are masked out to a radius of $3\sigma_{\text{beam}}$, using a harmonic infilling
algorithm.  This prevents spurious detections caused by Fourier
ringing in the filtered maps used by the detection algorithms.  As a
further guard against such spurious detections, we reject any
detections within $5\sigma_{\text{beam}}$ of a filled point source.
We have verified that this treatment reduces spurious detections due
to bright point sources to negligible levels in simulations, while
reducing the effective survey area by just 1.4\% of the sky.  Together
with the 15\% galactic dust and Magellanic cloud mask, this defines a
survey area of 83.6\% of the sky.

After infilling, the three detection codes produce individual
candidate catalogues down to a threshold \snr~$>4.5$.  The catalogues
are then merged to form a union catalogue, using the dust and extended
point source masks discussed above to define the survey area.  The
merging procedure identifies the highest \snr~detection as the
reference position during the merge: any detections by other codes
within 5 arcmin are identified with the reference position.  The
reference position and \snr~are reported in the union catalogue.

\pws\ can produce a small number of high-significance spurious
detections associated with galactic dusty emission.  We apply an extra
cut of \pws-only detections at \snr~$>10$ where the spectrum has a poor
goodness-of-fit to the SZ effect, $\chi^{2} > 16$.

We also remove five PSZ2 detections that match \pszone~detections
confirmed to be spurious by the \pszone~follow-up program (these were
the ones that we re-detected: there were many more confirmed spurious
detections from the program).

Finally the sample is flagged to identify the various sub-samples
discussed in Sect. \ref{sec:subsamples}.  The most important of these
flags is discussed in the next section.

\subsection{Infra-red spurious detections}
\label{sec:ir_spurious}

Cold compact infra-red emission, particularly that due to galactic
cold-clumps, can lead to high-significance spurious detections.  We
identify these detections by searching for 7 arcmin matches with the
\Planck\ cold-clump catalogue (C3PO), or with PCCS2 detections at both
545GHz and 857GHz.  This matching radius was chosen because it is the
typical size of a \Planck\ detected cold-clump \citep{planck2014-a37}.

\begin{figure}
\begin{center}
\includegraphics[angle=270,width=0.5\textwidth]{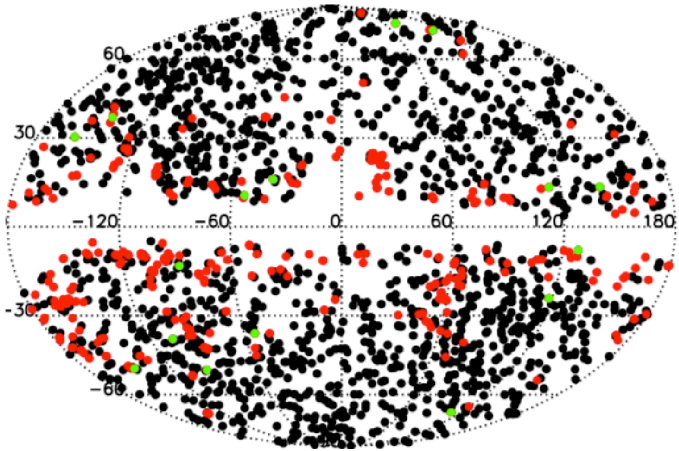}
\caption{The distribution of raw SZ detections, with deleted infra-red
  flagged candidates in red and retained infra-red flagged detections
  in green.}
\label{fig:cc_spu_distr}
\end{center}
\end{figure}

318 raw union detections match these criteria. They tightly follow the
distribution of galactic emission (see Fig. \ref{fig:cc_spu_distr}),
such that if the 65\% galactic dust mask (used for cluster cosmology )
is used to define the sample instead of the 85\% dust mask, the number
of IR-matched raw detections drops to 40. For the high-purity sample
formed from the intersection of all three codes, the numbers are 82
and 13 for the 85\% and 65\% dust masks respectively. Some high
latitude spurious candidates remain.  To minimise the effect of these
probably spurious detections on the catalogue, we delete them.

We have retained in the sample all 15 confirmed clusters that match
these criteria.  These IR contaminated clusters represent about
$1.5$\% of the total confirmed clusters in the \psztwo.  In the
catalogue, we define a flag, IR\_FLAG, to denote the retained clusters
that match these criteria. They can be expected to have heavily
contaminated SZ signal.

A small fraction of the unconfirmed detections deleted due to
IR-contaminations may have been real clusters. Assuming that optical
and X-ray confirmation is unbiased with regard to the presence of IR
emission, we estimate these deletions to bias our completeness
estimates by less than 1\%.

\subsection{Catalogue sub-samples}
\label{sec:subsamples}

The union catalogue can be decomposed into separate sub-samples,
defined as the primary catalogues of the three individual detection
codes (\pws,\mmfone,\mmfthree), as well as into unions and
intersections thereof.  The intersection subsample of candidates
detected by all three algorithms can be used as a high-reliability
catalogue with less than $2\%$ spurious contamination outside of the
galactic plane (see Sect.~\ref{sec:reliability}).

 \subsubsection{The Cosmology Catalogues}
\label{ssubec:cosmocats}
We constructed two cluster catalogues for cosmology studies from the
\mmfthree~and the intersection sub-samples respectively. For these
catalogues, our goal was to increase as much as possible the number of
detections while keeping contamination negligible. A good compromise
is to set the \snr~threshold to 6 and apply a 65\% galactic and point
source mask as in our 2013 cosmological
analysis~\citep{planck2013-p15}. In this earlier paper, our baseline
\mmfthree~cosmological sample was constructed using a threshold of 7
on the 15.5 month maps, which is equivalent to 8.5 on the full mission
maps. Estimations from the QA (Fig.~\ref{fig:reliability1}) suggest
that our 2014 intersection sample should be $>99$\% pure for a
threshold of 6.

The \mmfthree~cosmological sample contains 439 detections with 433
confirmed redshifts.  The intersection cosmological sample contains
493 detections with 479 confirmed redshifts.  Assuming that all
detections having VALIDATION flag greater than zero are clusters, the
empirical purity of our samples are $>99.8$\% for \mmfthree~ and
$>99.6$\% for the intersection. Note that the intersection sample
contains more clusters than the \mmfthree~sample for the same
\snr~threshold. This is expected since the definition of the \snr~for
the intersection sample is to use the highest value from the three
detection methods.

The completeness is also a crucial piece of information. It is
computed more easily with the single method catalogue for which the
analytical error-function (ERF) approximation can be used (as defined
in \citealt{planck2013-p05a}). In Sect.~\ref{sec:ext_comp} and
in~\cite{planck2014-a30}, we show that this analytical model is still
valid for the considered threshold. For the intersection sample, we
rely on the Monte-Carlo estimation of the completeness described in
Sect.~\ref{sec:completeness}.

These two samples are used in the cosmology analysis of
~\cite{planck2014-a30}.  Detections that are included in either of the
cosmology samples are noted in the main catalogue (see Appendix
\ref{appendix:products}).

\subsection{Consistency between codes}
\label{sec:code_consistency}

\begin{figure*}
\begin{center}
\includegraphics[angle=0,width= 0.32\textwidth]{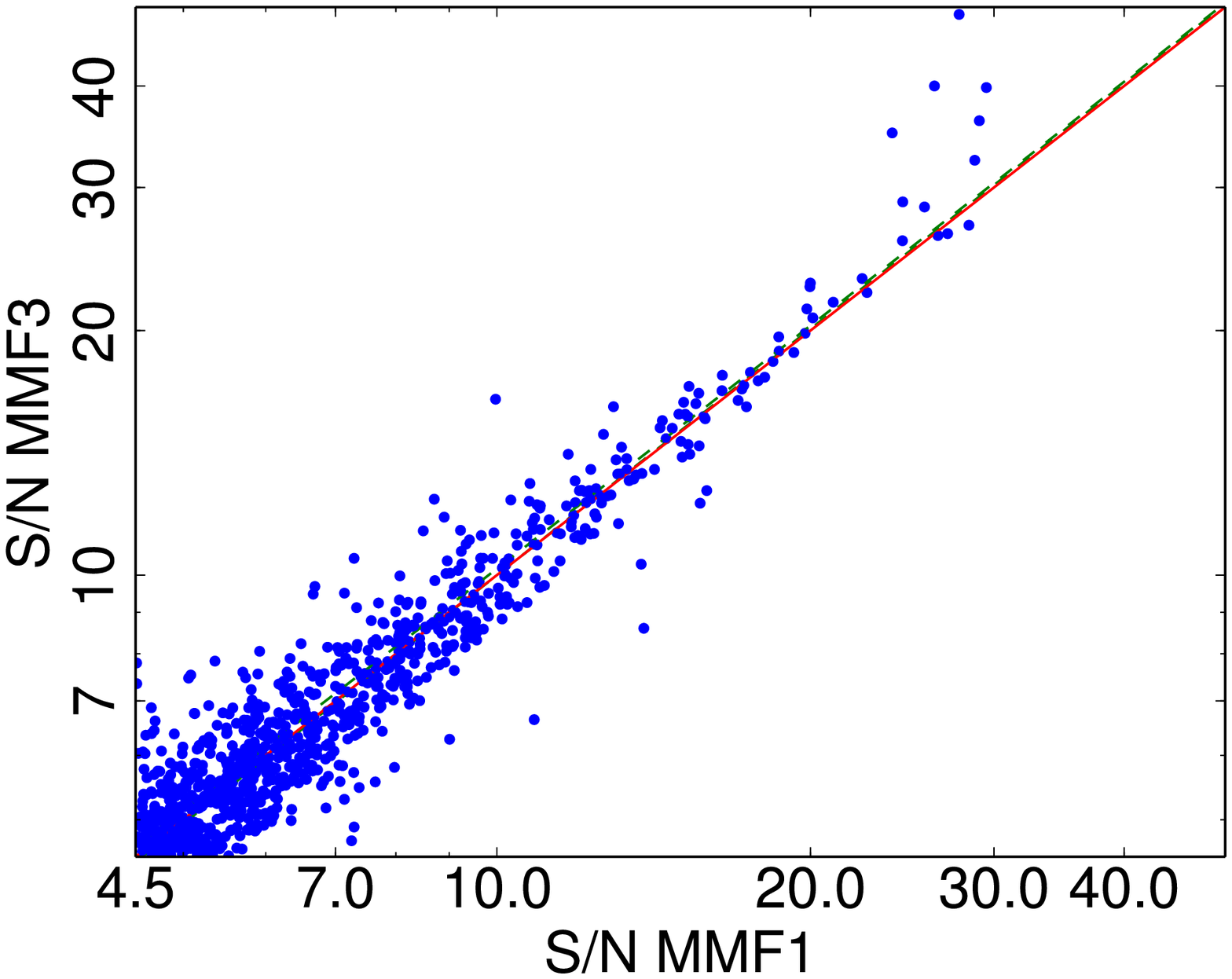}
\includegraphics[angle=0,width= 0.32\textwidth]{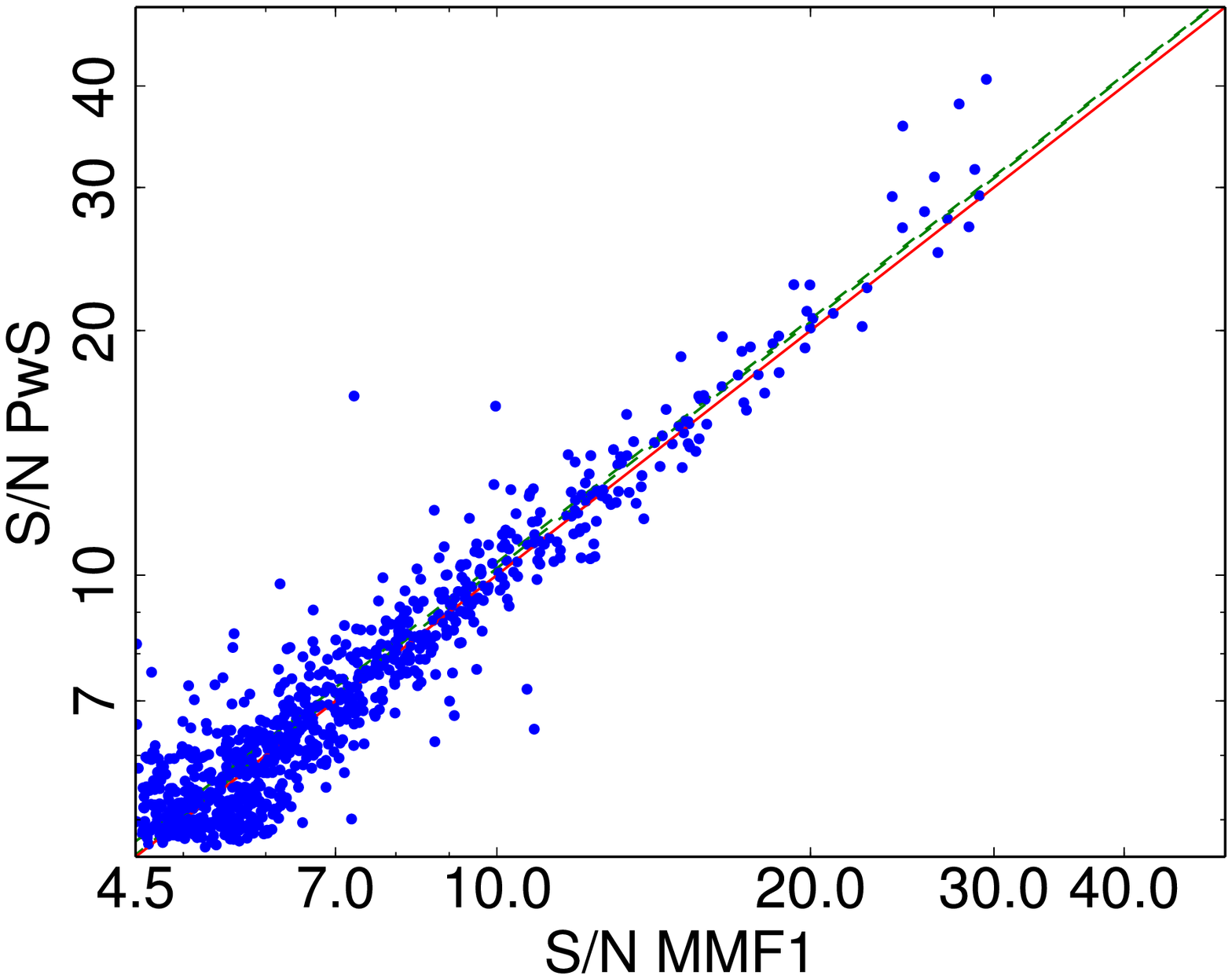}
\includegraphics[angle=0,width= 0.32\textwidth]{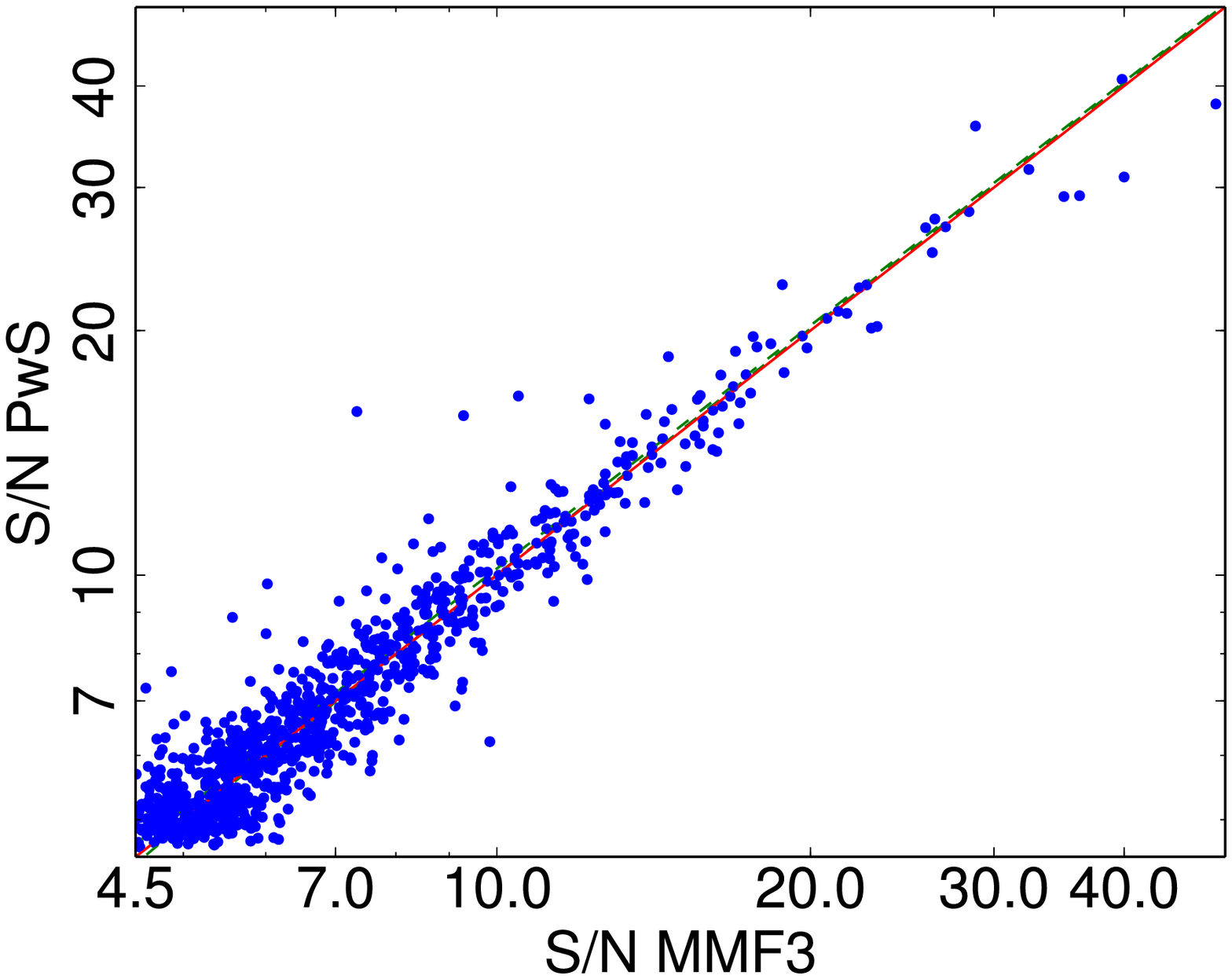}
\caption{Comparison of the \snr~estimates from the three detection
  codes.  The dashed green curves show the best fit relation for 0.8
  correlation and the red line is the line of equality.}
\label{fig:code_snr_comparison}
\end{center}
\end{figure*}

\begin{table*}
\caption{Results of fits between \snr~from the three detection codes,
  using the fitting function in equation \ref{equ:snr_fitfunc}. The
  assumed correlation of the uncertainties of $s_{1}$ and $s_{2}$ was
  0.8.}
\begin{center}
\begin{tabular}{ccrrrrr}
\hline\hline
 $s_{1}$ & $s_{2}$& \multicolumn{1}{c}{N} & \multicolumn{1}{c}{A} & \multicolumn{1}{c}{$\alpha$} & \multicolumn{1}{c}{$\sigma_\text{int}$} \\ 
\hline
\mmfone & \mmfthree & 1032 &$-0.01 \pm 0.01$ & $1.01 \pm 0.01$ &$0.033 \pm 0.001$\\
\mmfone & \pws & 985 & $-0.02\pm 0.01$ & $1.03 \pm 0.02$ & $0.030\pm 0.001$& \\ 
\mmfthree & \pws & 1045 & $0.0\pm 0.01$ & $1.01\pm 0.01$ & $0.031 \pm 0.001$&\\ 
\hline
\end{tabular}
\end{center}
\label{tab:consistency_snr_codes}
\end{table*}%

We construct the union sample using the code with the most significant
detection to supply the reference position and \snr.  This contrasts
with the \pszone, which used a pre-defined code ordering to select the
reference position and \snr.  In this section, we demonstrate the
consistency of the detection characteristics of the codes for common
detections, motivating this change in catalogue construction.

We fit the \snr~relation between codes using the Bayesian approach
described by \cite{hogg2010} for linear fits with covariant errors in
both variables.  We consider the catalogue \snr~values to be estimates
of a true underlying variable, $s$, with Gaussian uncertainties with 
standard deviation $\sigma = 1$.

We relate the $s$ values for two different catalogues using a simple
linear model
\begin{equation}
s_{2}= \alpha s_{1} + A
\label{equ:snr_fitfunc},
\end{equation}
where we assume flat priors for the intercept $A$, a flat prior on the
arc-tangent of the gradient $\alpha$, such that $p(\alpha) \propto 1 /
(1+\alpha^2)$.  We also allow for a Gaussian intrinsic scatter between
the $s$ values that includes any variation beyond the measurement
uncertainty on $s$. This is parameterised by $\sigma_{\text{int}}$
with an uninformative prior $p(\sigma_{\text{int}}) \propto 1 /
\sigma_{\text{int}}$.

We assume a fiducial correlation of $\rho_{\text{corr}}=0.8$ between
the \snr~estimates of each code pair, which is typical of the
correlation between the matched-multifrequency-filtered patches of
each code.  The fit results are shown in Table
\ref{tab:consistency_snr_codes}.

The \snr~estimates from the three codes are compared in
Fig. \ref{fig:code_snr_comparison}, which also shows the best fit
relation.  \mmfone~produces noticeably weaker
detection than the other two codes for the 14 very strong detections
at \snr~$>20$.  Excluding these exceptional cases from the comparison,
the best-fit relations between the \snr s from each code show no
significant deviations from equality between any of the codes.

There are a small number of highly significant outliers in the
relation between \pws~and the MMF codes.  These are clusters imbedded
in dusty regions where the different recipes for the filtered patch
cross-power spectrum vary significantly and the likelihood assumptions
common to all codes break down.  \pws~shows outlier behaviour relative
to the other codes as its recipe is most different from the other
codes.

\begin{figure}
\begin{center}
\includegraphics[angle=0,width= 0.48\textwidth]{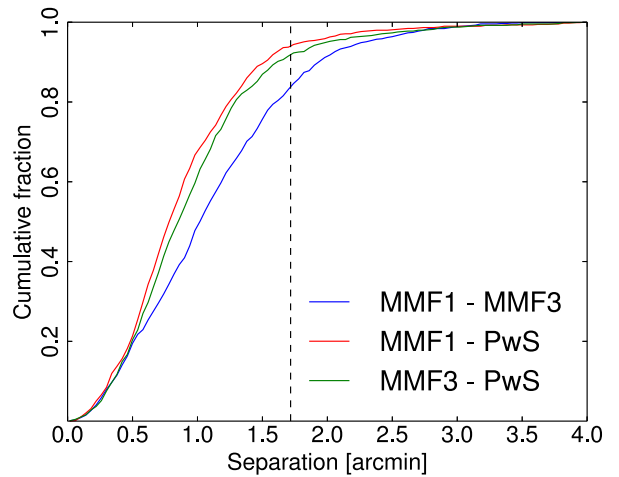}
\caption{Cumulative distribution of angular separation between matched
  detections for each possible code pair.  The vertical dashed line
  indicates the width of a Healpix pixel at the \Planck\ resolution.}
\label{fig:code_pos_comparison}
\end{center}
\end{figure}

Fig. \ref{fig:code_pos_comparison} shows the consistency of the
position estimates between the codes.  The positions of \mmfone~and
\mmfthree~are more inconsistent with one another than any other code
combination.  The 67\% bound on the \mmfone - \mmfthree~separation is
1.34 arcmin, while for \mmfone - \pws~it is $0.98$ arcmin and for
\mmfthree - \pws~it is $1.1$ arcmin.  This is consistent with the
observation from the quality assessment that the \pws~positions are
the most robust (Sect.~\ref{sec:qa_position}).

\section{Selection Function}
\label{sec:selection_function}

A necessary element of any cluster sample is the selection function
that relates the detected sample to the underlying population of
objects.  The selection function comprises two complementary
functions: the completeness, which defines the probability that a
given real object will be detected; and the statistical reliability,
also known as purity, which defines the probability that a given
detection corresponds to a real object.  As a function of underlying
object attributes, the completeness is a function of underlying SZ
observables, $\theta_{500}$ and \yfive.  The reliability is a
statistical function of detection attributes and is presented as a
function of detection \snr.

\subsection{Monte-Carlo Injection}
\label{sec:MC_injection}

The selection function is determined by the Monte-Carlo injection of
simulated clusters into both real and simulated \Planck\ maps.  A
common segment is the injection of cluster SZ signal.  The cluster
signal is assumed to be spherically symmetric and to follow a pressure
profile similar to the generalised Navarro-Frenk-White (GNFW) profile
assumed in the catalogue extraction.

\begin{figure}
\begin{center}
\includegraphics[angle=0,width=0.48\textwidth]{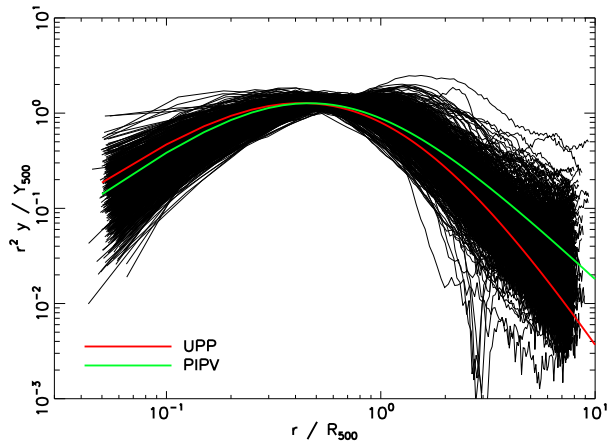}
\caption{The 910 simulated pressure profiles from the cosmo-OWLS
  simulations used for cluster injection.  Also shown are the assumed
  extraction profile (UPP) and the best-fit profile from a sample of
  62 pressure profiles fitted using \Planck\ and x-ray data (PIPV,
  \citealt{planck2012-V}).}
\label{fig:cosmoOWLS_profiles}
\end{center}
\end{figure}

To include the effects of system-on-system variation in the pressure
distribution, we draw the spherically-averaged individual pressure
profiles from a set of 910 pressure profiles from simulated clusters
from the cosmo-OWLS simulations (\citealt{lebrun14,mccarth14}), an
extension of the OverWhelmingly Large Simulations project
\citep{sch10}.  These pressure profiles are empirical in the sense
that they have not been fitted using a GNFW profile: the mean pressure
is used within concentric radial shells (after the subtraction of
obvious sub-structures) and the injected profiles are interpolated
across these shells.  The simulated clusters were selected for this
sample by requiring that their mass be above the approximate limiting
mass for \Planck\ at that redshift.  The ensemble of simulated
profiles are shown in Fig. \ref{fig:cosmoOWLS_profiles}.  Each profile
is normalised such that the spherically integrated Y parameter matches
the fiducial injected (\yfive,$\theta_{500}$) parameters for the halo.
The injected (\yfive,$\theta_{500}$) are different for completeness
and reliability simulations and each is discussed below.

Effective beam variation is an important consideration for the
unresolved clusters at the intermediate and high redshifts of
cosmological interest. The injected clusters are convolved with
effective beams in each pixel including asymmetry computed following
\citet{mitra2010}


\subsection{Completeness}
\label{sec:completeness}

The completeness is defined as the probability that a cluster with a
given set of true values for the observables (\yfive,$\theta_{500}$)
will be detected, given a set of selection criteria.  A good
approximation to the completeness can be defined using the assumption
of Gaussianity in the detection noise.  In this case, the completeness
for a particular detection code follows the error function (ERF),
parametrised by a selection threshold $q$ and the local detection
noise at the clusters radial size, $\sigma_{\text{Y}}(\phi,\psi)$ (see
the discussion in \citealt{planck2013-p05a}).  This approach is not suited
to the union and intersection catalogues from \Planck\, due to the
difficulty in modelling correlations between detection codes.  We
determine the completeness by brute force: injecting and detecting
simulated clusters into the \Planck\ sky maps.  This approach has the
advantage that all algorithmic effects are encoded into the
completeness, and the effects of systematic errors such as beam and
pressure profile variation can be characterised.  This approach also
fully accounts for the non-Gaussianity of the detection noise due to
foreground emission.

The injected (\yfive,$\theta_{500}$) parameters are drawn from a
uniform distribution in the logarithm of each variable, ensuring that
our logarithmically spaced completeness bins have approximately equal
numbers of injected sources.

As the completeness is estimated from injection into real data,
injected sources can contribute to the detection noise.  We therefore
use an injection mode, as was the case for the \pszone~completeness,
where injected clusters are removed from the maps used to estimate the
noise statistics.  We also avoid superimposing injected clusters on
top of one another, or on top of real data detections.  Together,
these ensure that the noise statistics for injected clusters are the
same as for the real detections in the map.

\begin{figure*}
\begin{center}
\includegraphics[angle=0,width=0.32\textwidth]{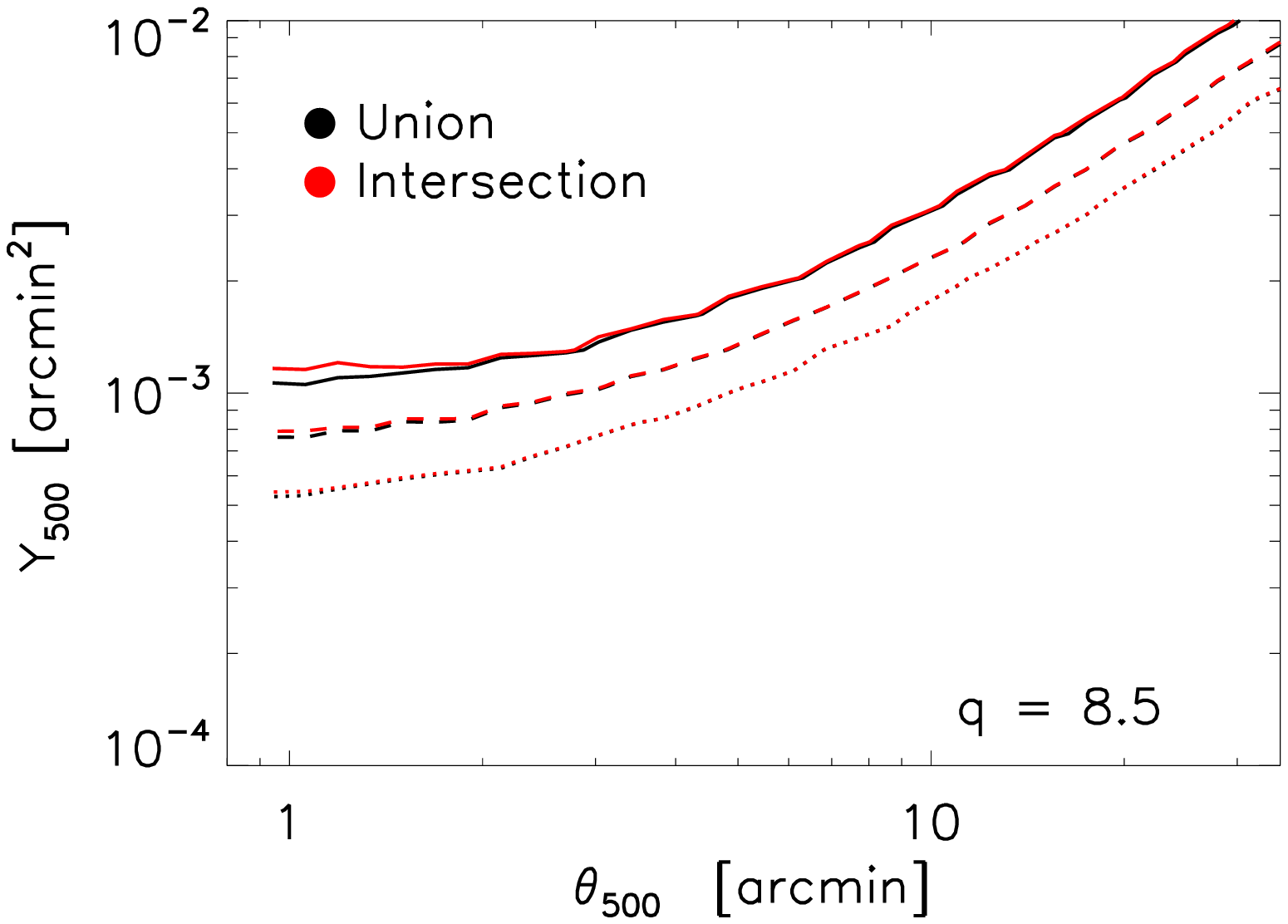}
\includegraphics[angle=0,width=0.32\textwidth]{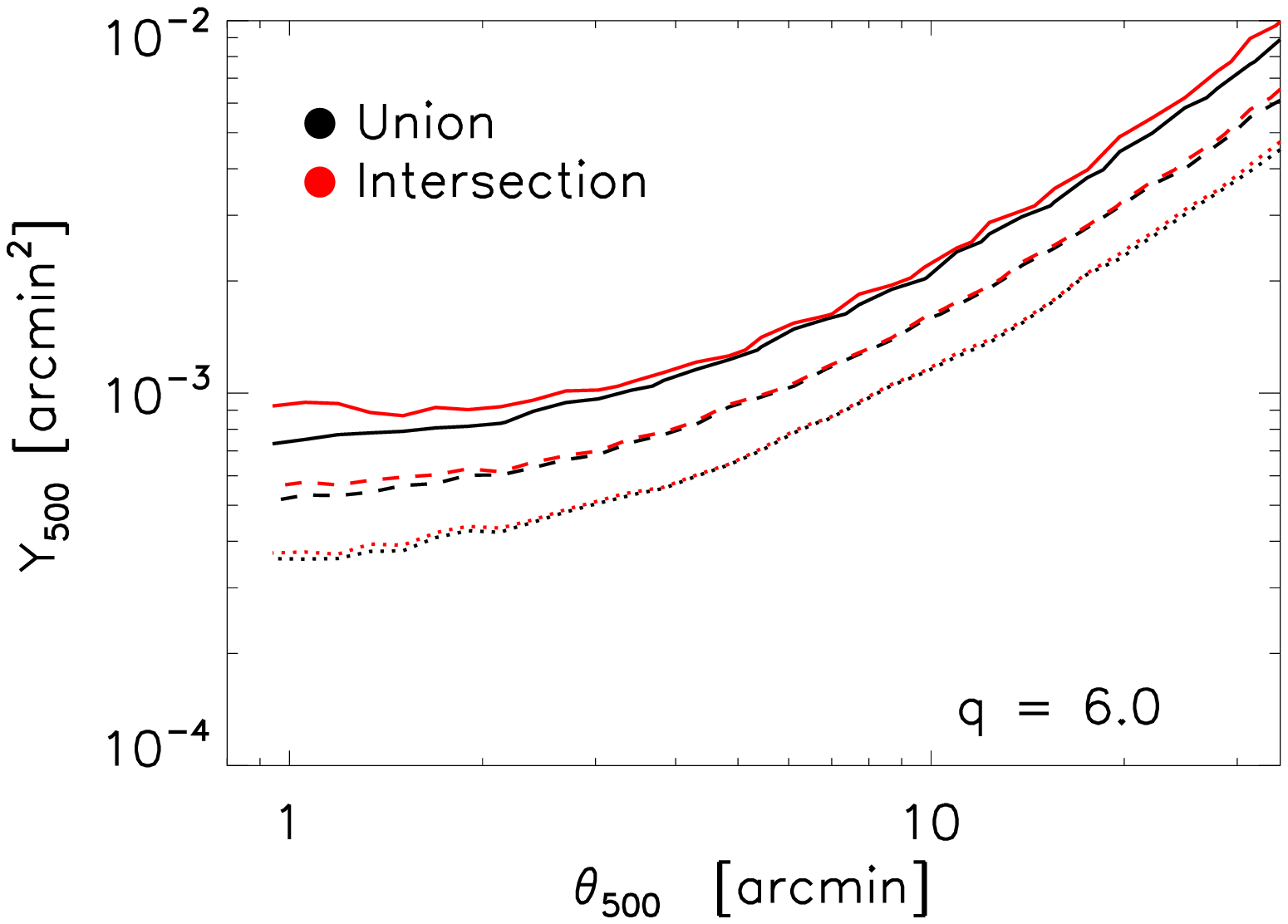}
\includegraphics[angle=0,width=0.32\textwidth]{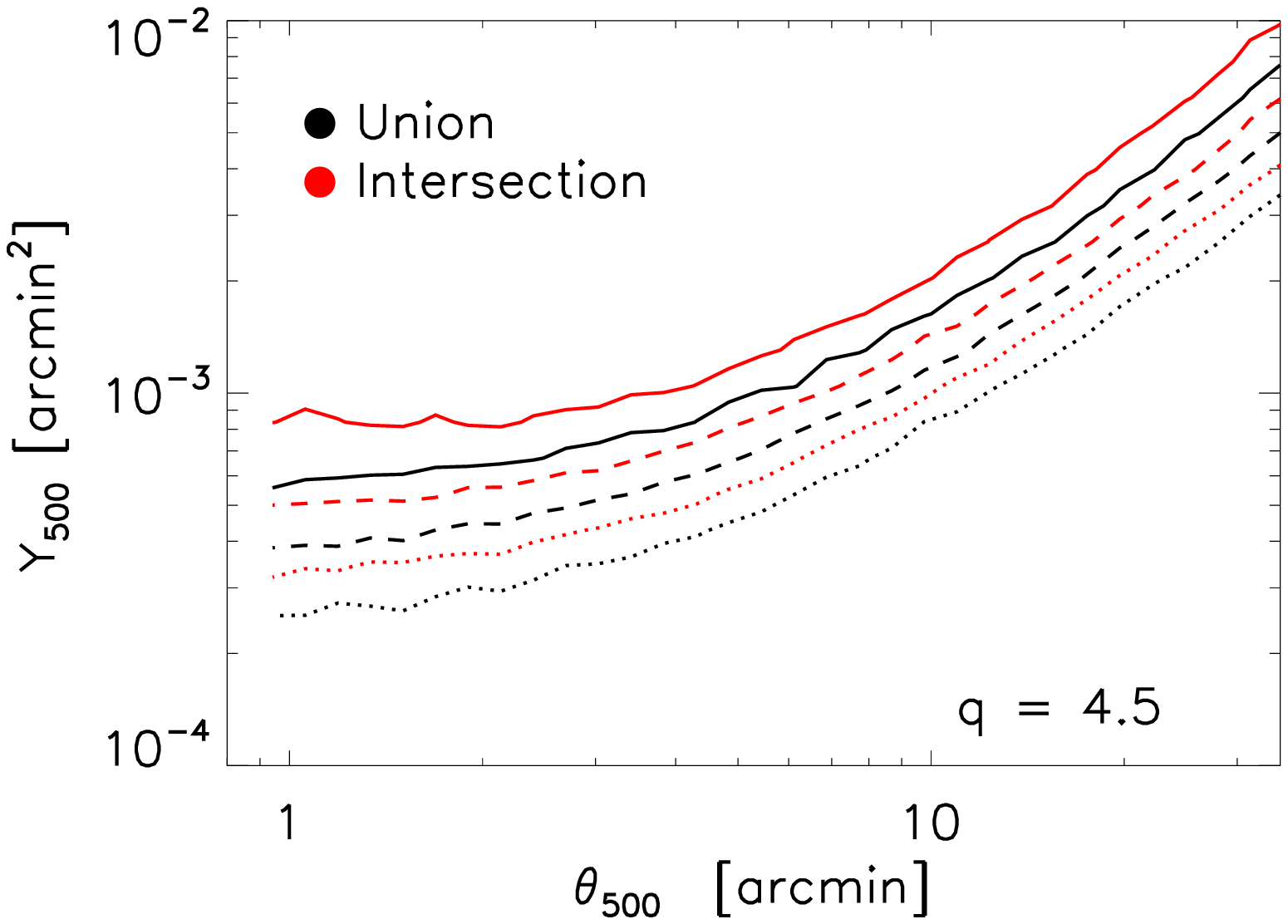}\\
\includegraphics[angle=0,width=0.32\textwidth]{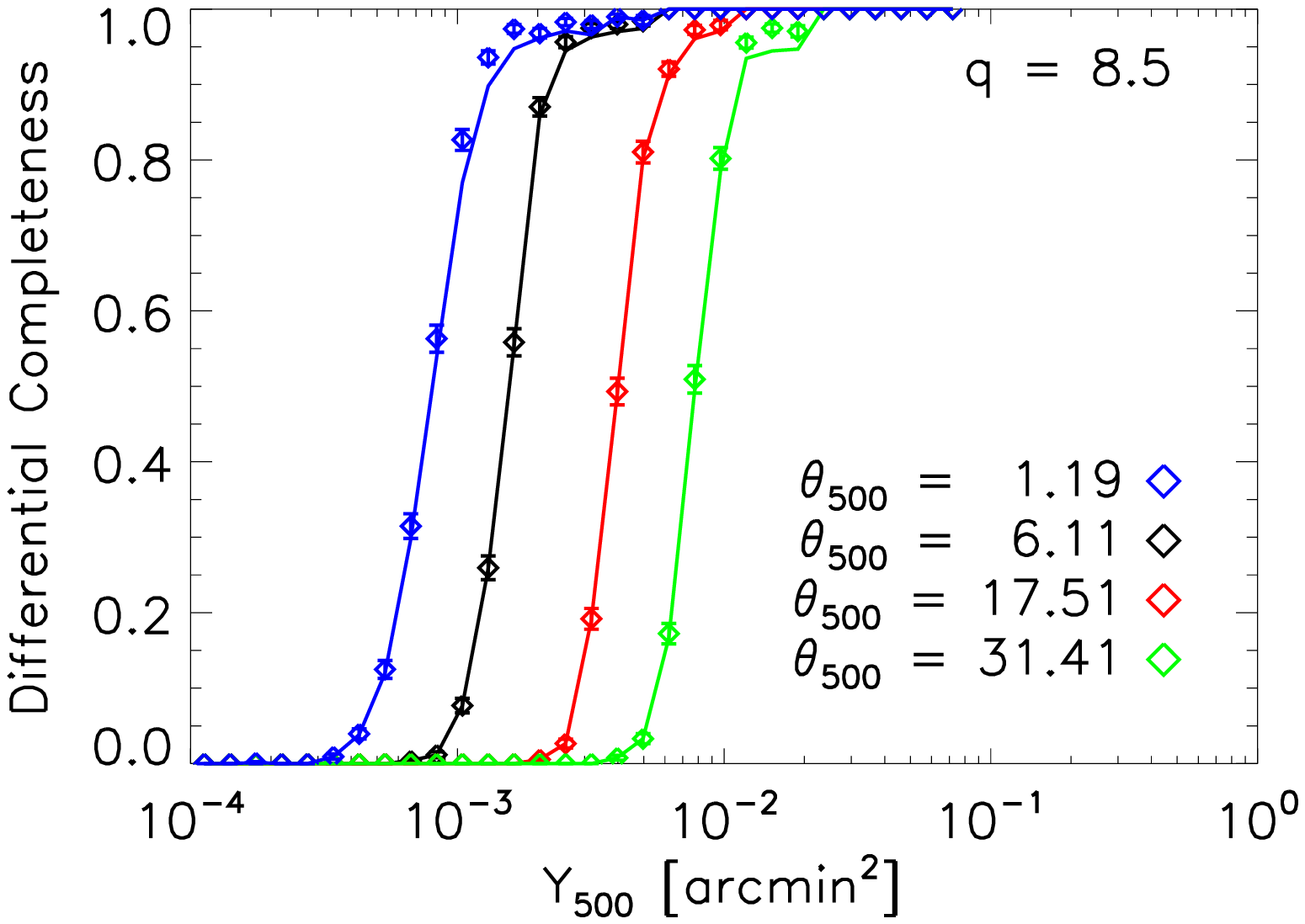}
\includegraphics[angle=0,width=0.32\textwidth]{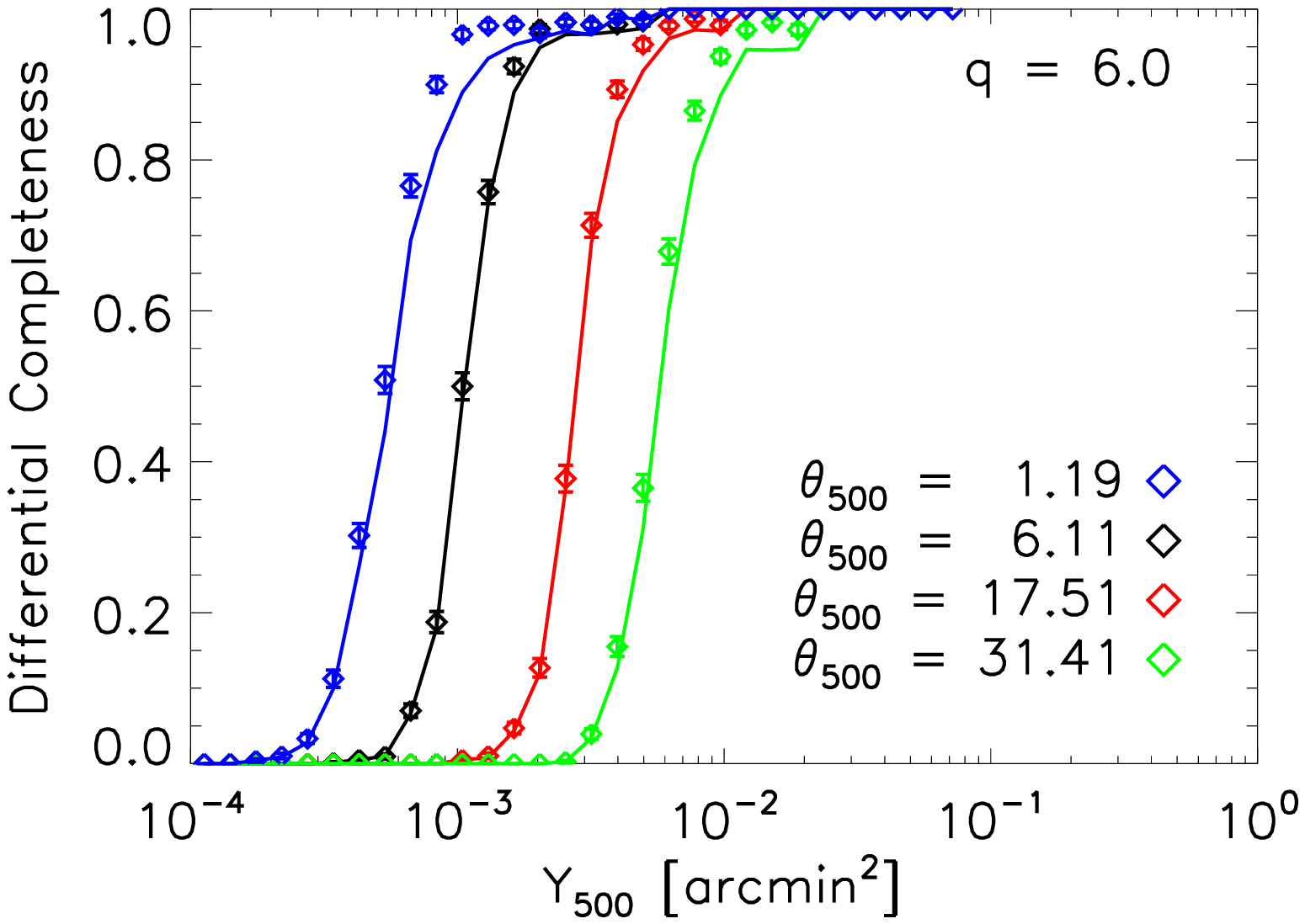}
\includegraphics[angle=0,width=0.32\textwidth]{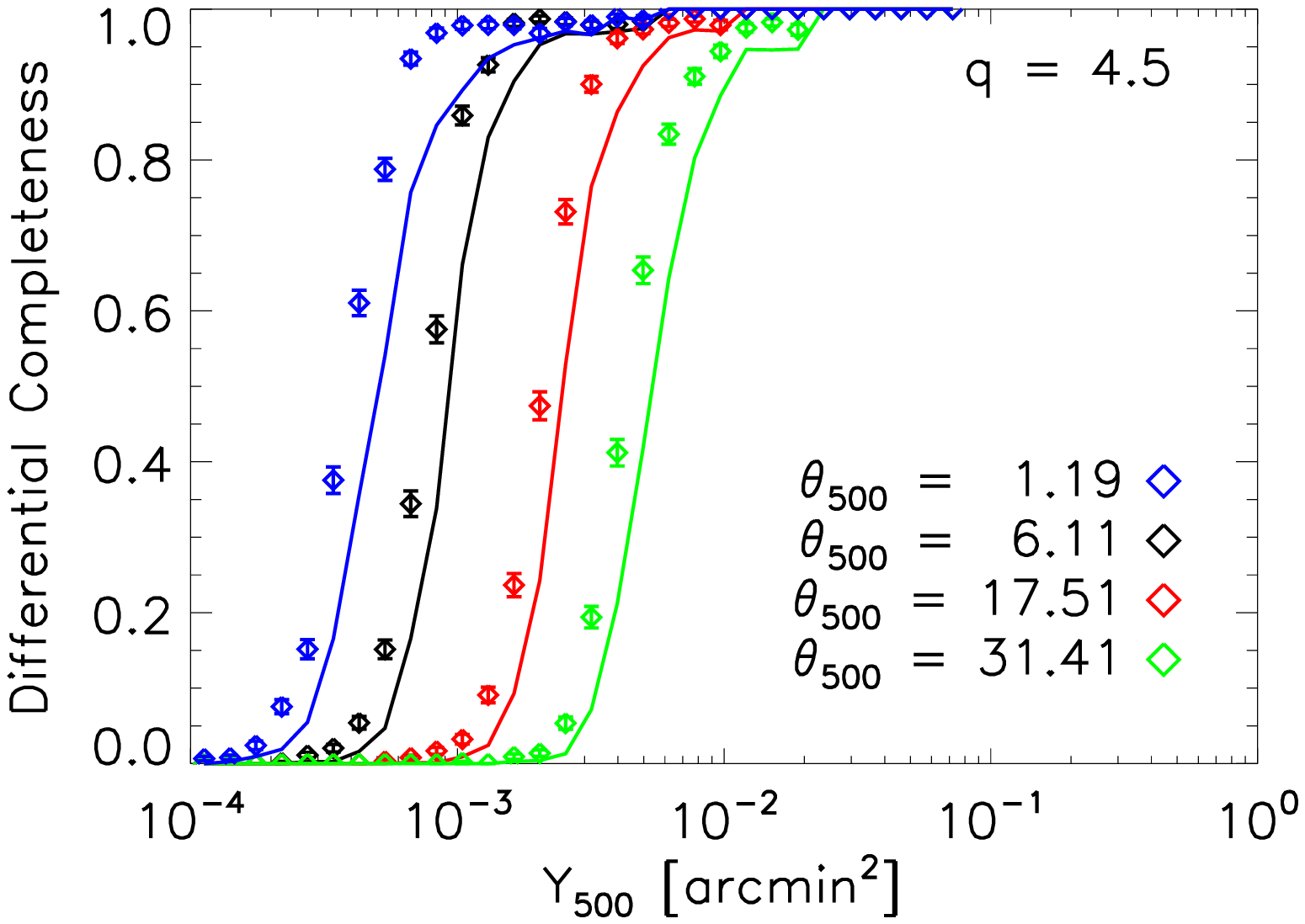}
\caption{Completeness of the union and intersection samples at
  progressively lower \snr~thresholds.  From left to right, the
  thresholds are 8.5, 6.0 and 4.5 (the survey threshold).  In the top
  panels, the dotted lines denote 15\% completeness, the dashed lines
  50\% and the solid lines 85\% completeness.  In the bottom panels,
  the union is denoted by the diamonds with Monte-Carlo uncertainties
  based on binomial statistics, and the intersection is denoted by the
  solid lines.}
\label{fig:completeness1}
\end{center}
\end{figure*}

We release as a product the Monte-Carlo completeness of the catalogues
at thresholds stepped by 0.5 in \snr~over the range $4.5 \leq
$\snr$\leq 10$.  Fig.~\ref{fig:completeness1} shows the completeness
of the union and intersection catalogues as functions of input
(\yfive,$\theta_{500}$) and at representative values of
$\theta_{500}$, for three detection thresholds.  The union and
intersection catalogues are most similar at high \snr, where they
match well except at small scales. Here the intersection catalogue
follows the lower completeness of \mmfone. This is due to an extra
selection step in that code which removes spurious detections caused
by point sources.  The union and the intersection catalogues mark the
upper and lower limits of the completeness values for the
sub-catalogues based on the individual codes.

\begin{figure*}
\begin{center}
\includegraphics[angle=0,width=0.48\textwidth]{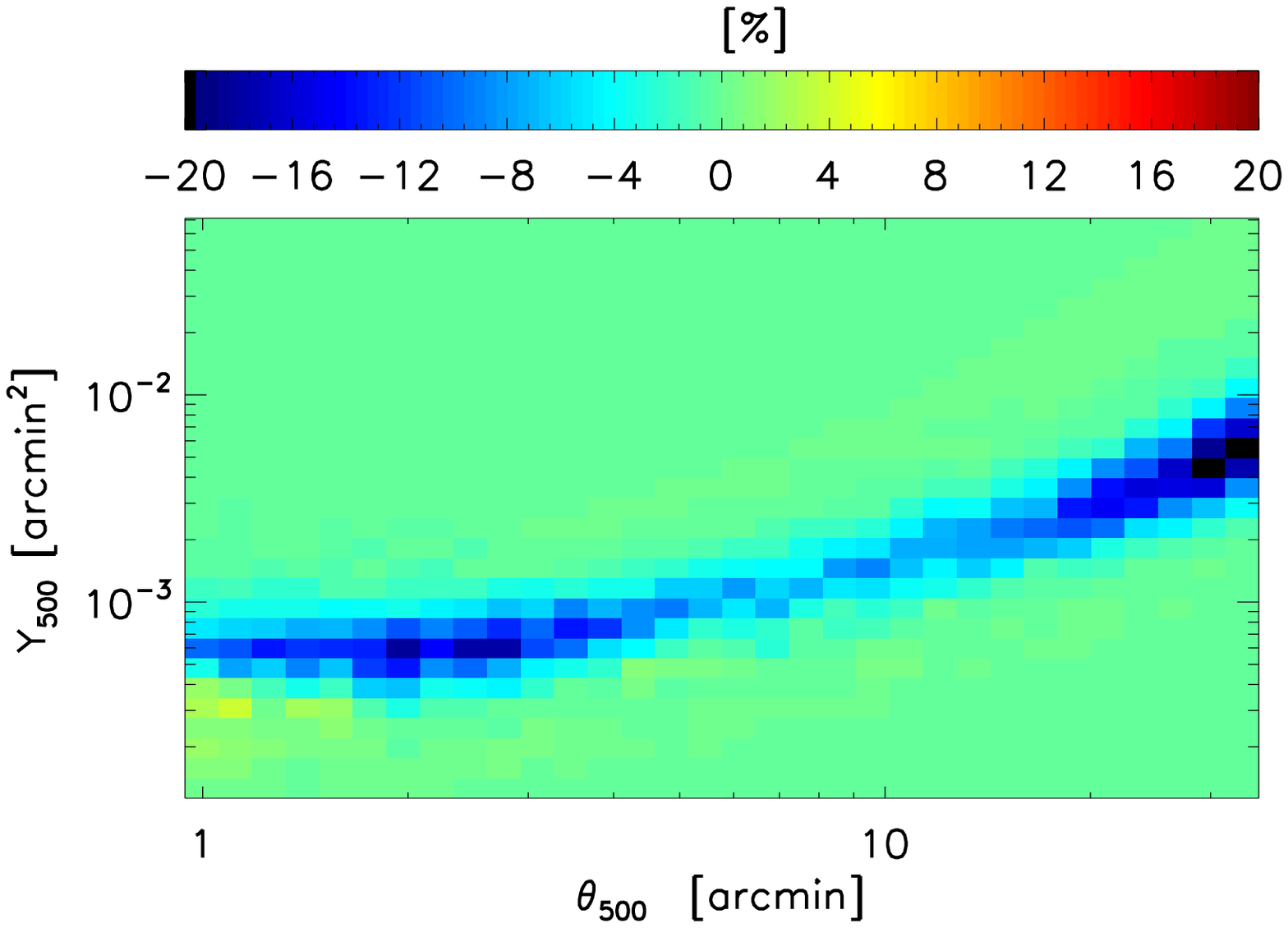}
\includegraphics[angle=0,width=0.48\textwidth]{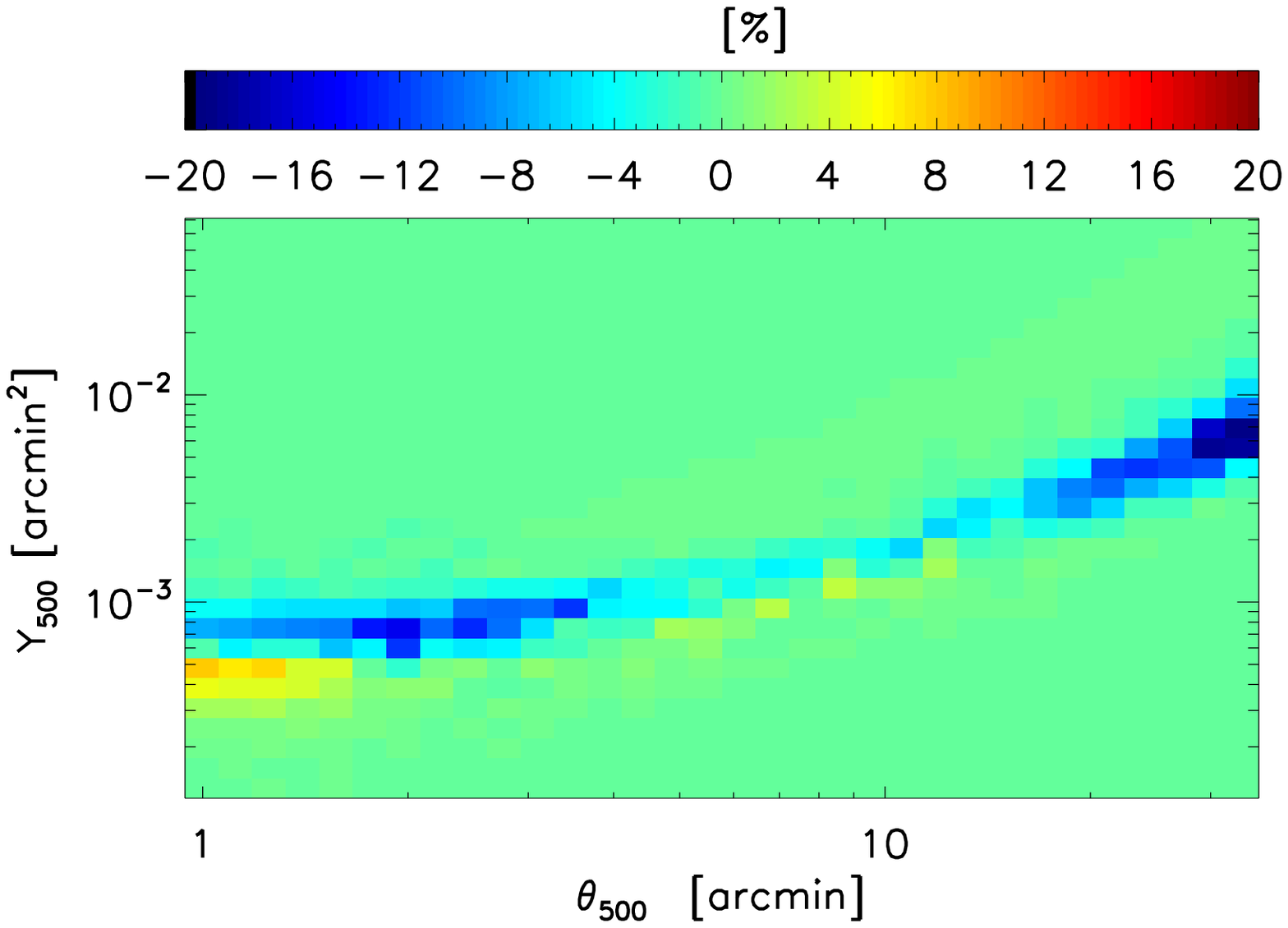} \\
\includegraphics[angle=0,width=0.48\textwidth]{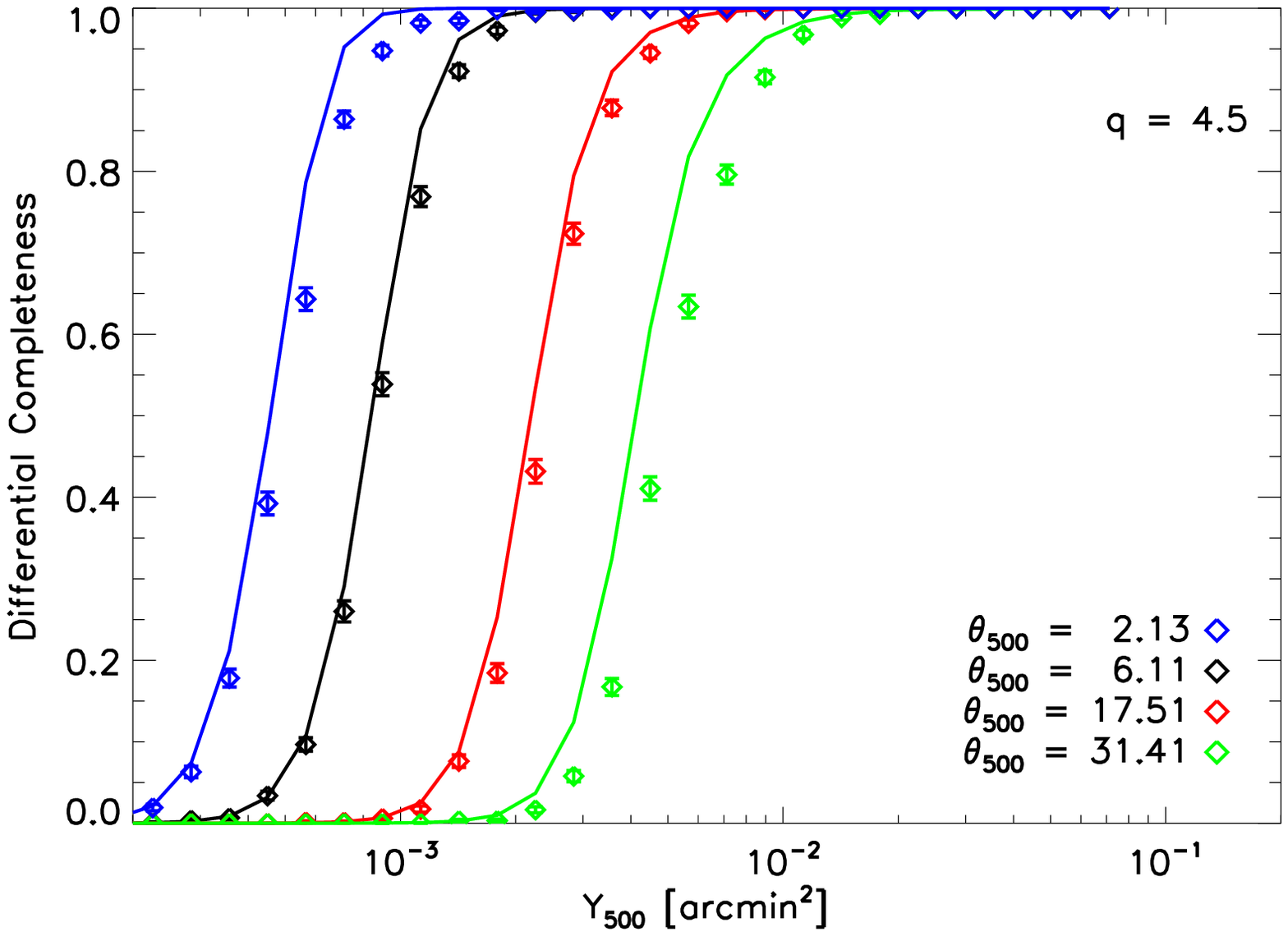}
\includegraphics[angle=0,width=0.48\textwidth]{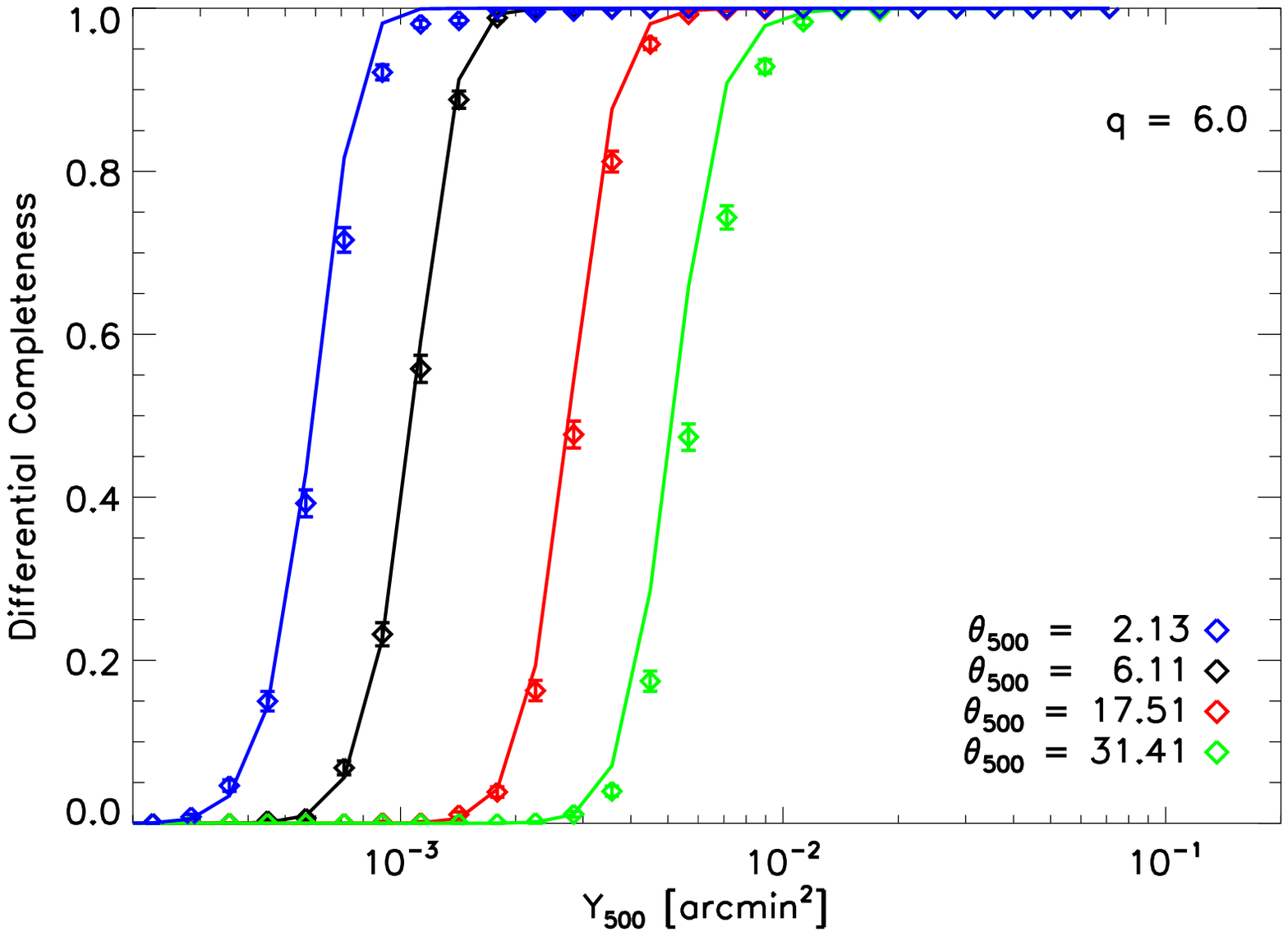}
\caption{Differences between the semi-analytic and Monte-Carlo
  completenesses for \mmfthree.  The left panels show the difference
  for the full survey over 85\% of the sky with a $q=4.5$ threshold.
  The right panels show the difference for the \mmfthree~cosmology
  sample, covering 65\% of the sky to a threshold of $q=6.0$.  The top
  panels show the difference $MC - ERF$ in percent.  The bottom panels
  compare the completenesses at particular $\theta_{500}$: the
  Monte-Carlo completeness is denoted by diamonds and the ERF
  completeness by solid lines.}
\label{fig:compl_mmf3}
\end{center}
\end{figure*}

The completeness of the \Planck\ cluster catalogue is robust with
respect to deviations of the real SZ profiles of galaxy clusters from
the one assumed by the algorithms for filter construction.  To
demonstrate this, we compare $C_\text{MC}$, the Monte-Carlo
completeness for the \mmfthree~sample, using the cosmoOWLs profile
variation prescription and effective beam variation, to
$C_\text{erf}$, the semi-analytic ERF completeness.  This comparison
is shown in Fig.~\ref{fig:compl_mmf3}, where we show the difference
between the two estimates as a function of \yfive\ and $\theta_{500}$
as well the individual completeness values as functions of \yfive for
representative values of $\theta_{500}$ slices through the
2D completeness and show the difference.

The error function is a good approximation to the MC completeness for
the cosmology sample, which uses a higher \snr~cut and a larger
Galactic mask than the full survey.  The MC estimate corrects this
analytic completeness by up to 20\% for large resolved clusters, where
$C_\text{MC}$ is systematically less complete than the ERF
expectation, primarily due to variation in the cluster pressure
profiles.  For unresolved clusters, the drop-off in $C_\text{MC}$ is
slightly wider than the ERF expectation, reflecting variation both of
pressure profiles and of effective beams.

The impact of these changes in completeness on expected number
  counts and inferred cosmological parameters for the cosmology sample is analysed in
  \cite{planck2014-a30}.  The difference between the Monte-Carlo and
  ERF completeness results in a change in modelled number counts of
  typically $\sim 2.5$\% (with a maximum of 9\%) in each redshift bin.
  This translates into a $0.26\sigma$ shift of the posterior peak for
  the implied linear fluctuation amplitude,$\sigma_8$.

The MC completeness is systematically lower than the analytic
approximation for the full survey.  One of the causes of this is
galactic dust contamination, which is stronger in the extra 20\% of
the sky included in the full survey area relative to the cosmology sample
area. This tends to reduce the \snr~of clusters on affected lines
of sight.

We note that this approach ignores other potential astrophysical
effects that could affect the completeness.  Radio emission is known
to be correlated with cluster positions, potentially `filling in' the
SZ decrement, though recent estimates suggest that this effect is
typically small in \Planck\ data \citep{rod15}.  Departures of the
pressure distribution from spherical symmetry may also affect the
completeness, though this effect is only likely to be significat for
nearby and dynamically disturbed clusters which may be large compared
to the \Planck\ beams.  We test for some of these effects through 
external validation of the completeness in the next section, and
explicitly through simulation in
Sect.~\ref{sec:astrophysics_completeness}.

Another source of bias is the presence of correlated IR
  emission from cluster member galaxies.  \cite{planck2014-a29} show
  that IR point sources are more numerous in the direction of galaxy
  clusters, especially at higher redshift, and contribute
  significantly to the cluster SED at the \Planck\ frequencies.
  Initial tests, injecting clusters signal with the combined IR+tSZ
  spectrum of $z>0.22$ clusters observed by \cite{planck2014-a29},
  suggest that this reduces the completeness for unresolved clusters.
  Future work is warranted to characterise the evolution and scatter
  of this IR emission and to propagate the effect on completeness
  through to cosmological parameters.

\subsection{External validation of the completeness}
\label{sec:ext_comp}

\begin{figure*}
\begin{center}
\includegraphics[angle=0,width=0.49\textwidth]{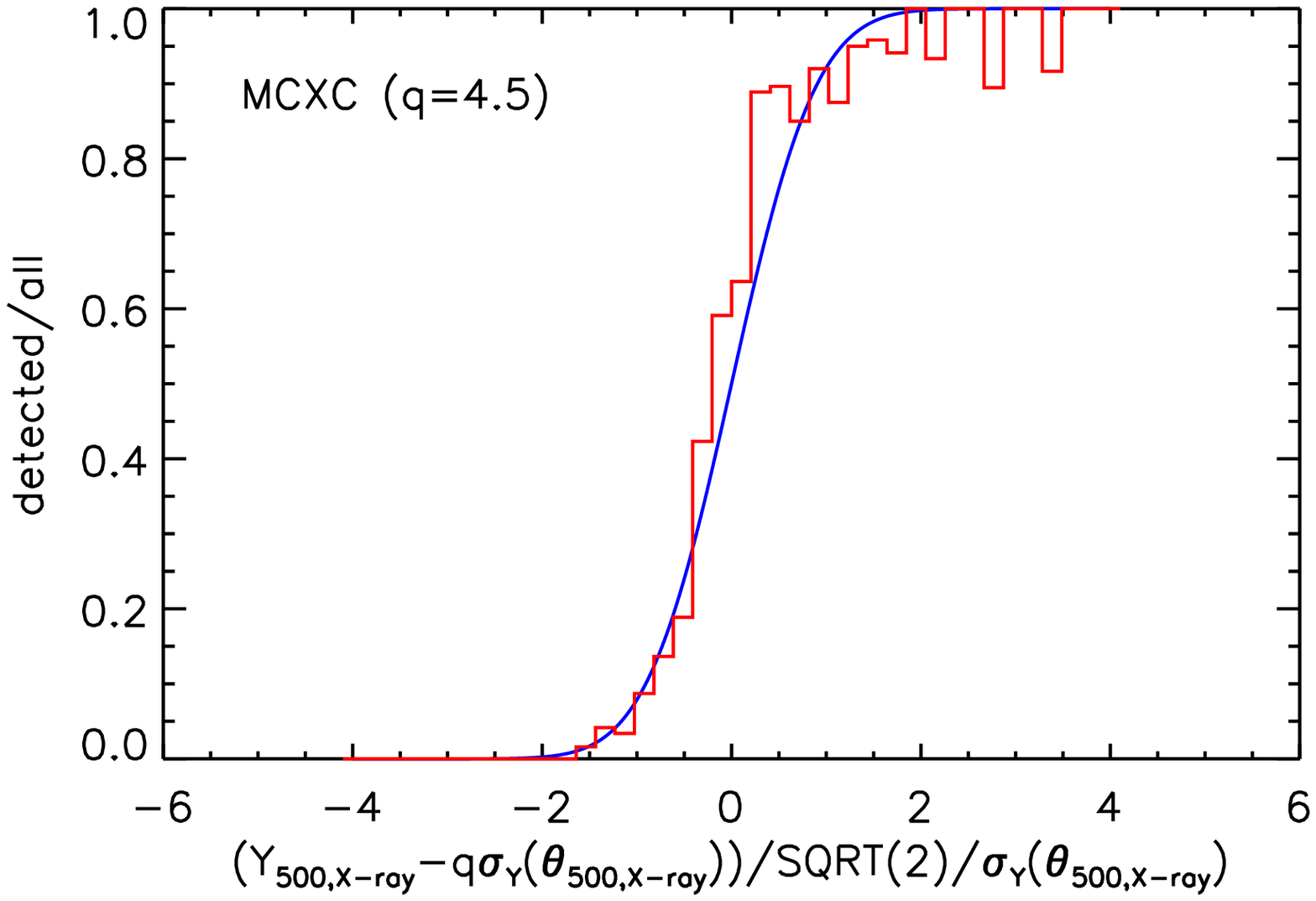}
\includegraphics[angle=0,width=0.49\textwidth]{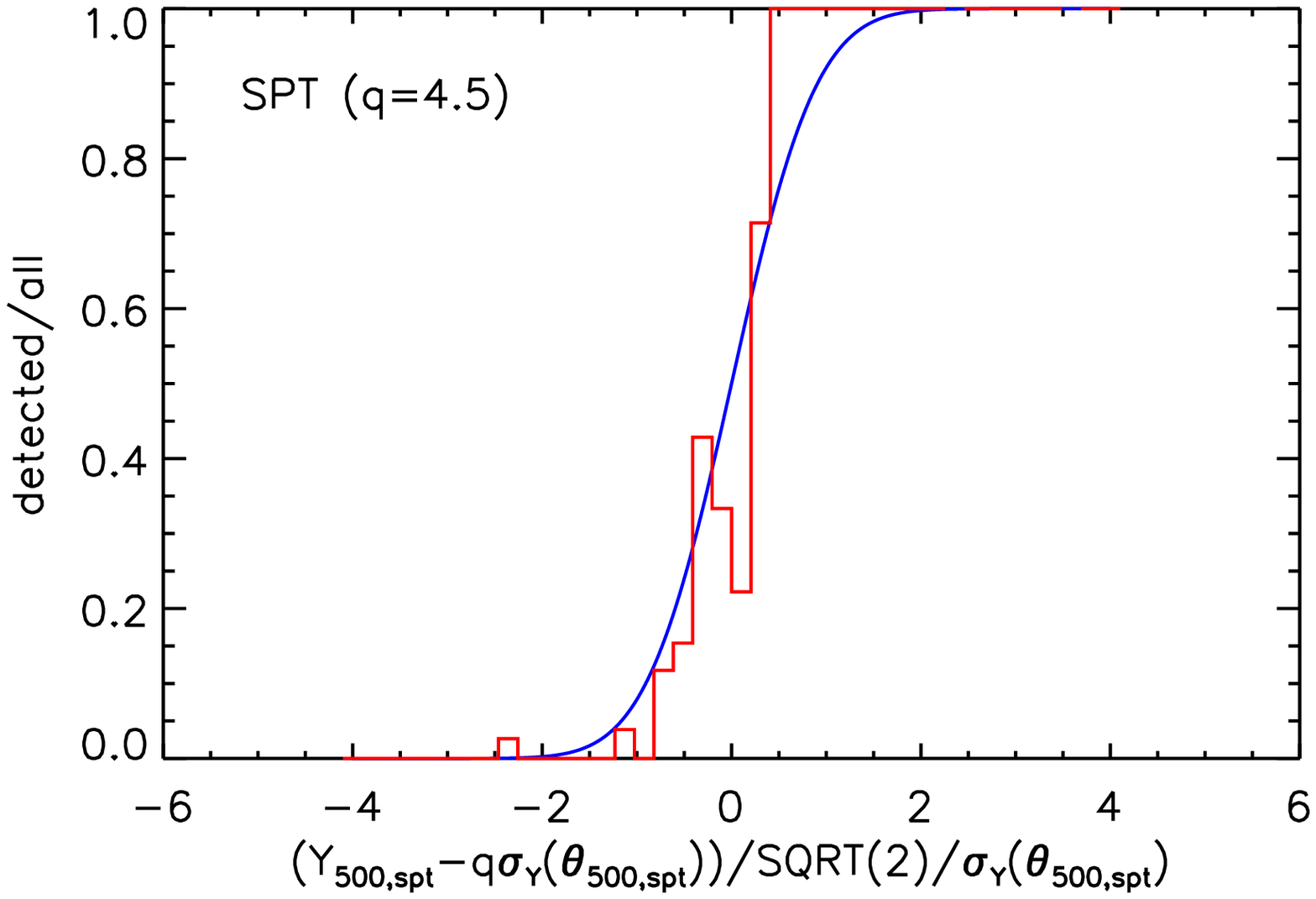}
\caption{ \mmfthree~completeness for the \psztwo~catalogue
  (\snr~threshold $q>4.5$) determined from the MCXC (left) and SPT
  (right) catalogues.  This external estimate (red histogram) is in good
  agreement with the analytic ERF calculation (solid blue line),
  except for SPT at the high probability end (see text).}
\label{fig:spt_mcxc_completeness}
\end{center}
\end{figure*}

We validated our Monte-Carlo completeness calculation and our simple
analytical ERF model by using the MCXC~\citep{pif11} and
SPT~\citep{blee14} catalogues. The \Planck\ detection threshold passes
across the cluster distributions of these two samples. This is
illustrated in Fig.~4 of~\cite{cham12} for the MCXC. This allows us to
characterize our completeness by checking if the fraction of detected
clusters follows the expected probability distribution as a function
of their parameters.  For each cluster of the MCXC catalogue, we use
the \mmfthree~algorithm to extract its flux \yfive~and associated
error $\sigma_Y$ at the location and for the size given in the
x-ray catalogue. We then build the quantity $(Y_{500} - q
\sigma_{\text{Y}})/{\sqrt 2}/\sigma_{\text{Y}}$, q being the detection
threshold (here 4.5) and $\sigma_\text{Y}$ the noise of the filtered
maps.  We make the corresponding histogram of this quantity for all
the clusters and for the clusters detected by \mmfthree. The ratio of
the two histograms is an empirical estimate of the completeness.
Results are shown in Fig.~\ref{fig:spt_mcxc_completeness} for the MCXC
(left) and the SPT (right) catalogues.
For MCXC, the estimation is in good agreement with the expected simple
analytical ERF model ($0.5 \, (1+{\rm ERF})$).  For SPT, the estimated
completeness is also in good agreement except for $(Y_{500} - q
\sigma_\text{Y})/{\sqrt 2}/\sigma_\text{Y}>1$ where it is higher than
the analytic expectation. We attribute this behaviour to the
correlation between SPT and \Planck\ detections. The SPT catalogue is
a SZ-based, so a cluster detected by SPT will have a higher than random
probability to be detected by \Planck. This leads to an overestimation
of the completeness at the high probability end.


\subsection{Position estimates}
\label{sec:qa_position}

We characterise the positional recovery of the \Planck\ detections
using injection into real data, including pressure profile and beam
variation.  We draw input clusters from a realistic distribution of
(\yfive,$\theta_{500}$), the same as used for the reliability in
Sect.~\ref{sec:reliability}.

\begin{figure}
\begin{center}
\includegraphics[angle=90,width=0.48\textwidth]{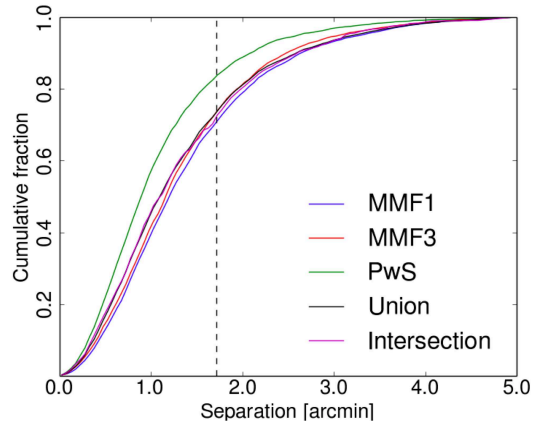}
\caption{Cumulative distribution of angular separations between
  estimated and input positions.  The dashed vertical line denotes the
  \Planck\ pixel size.}
\label{fig:qa_position}
\end{center}
\end{figure}

Fig.~\ref{fig:qa_position} shows the comparative performance of the
individual detection codes, and of the reference position chosen for
the union catalogue.  \pws~produces the most accurate positions, with
67\% of detected positions being within 1.18 arcmin of the input
position.  For \mmfone~and \mmfthree, the 67\% bound is 1.58 arcmin
and 1.52 arcmin respectively.  The union and intersection accuracy
follow that of the MMFs, with 67\% bounds of 1.53 arcmin.  We observe
that our inter-code merging radius of 5 arcmin is conservative given 
the expected position uncertainties.

\subsection{Impact of cluster morphology}
\label{sec:astrophysics_completeness}

\begin{figure}
\begin{center}
\includegraphics[angle=0,width=0.48\textwidth]{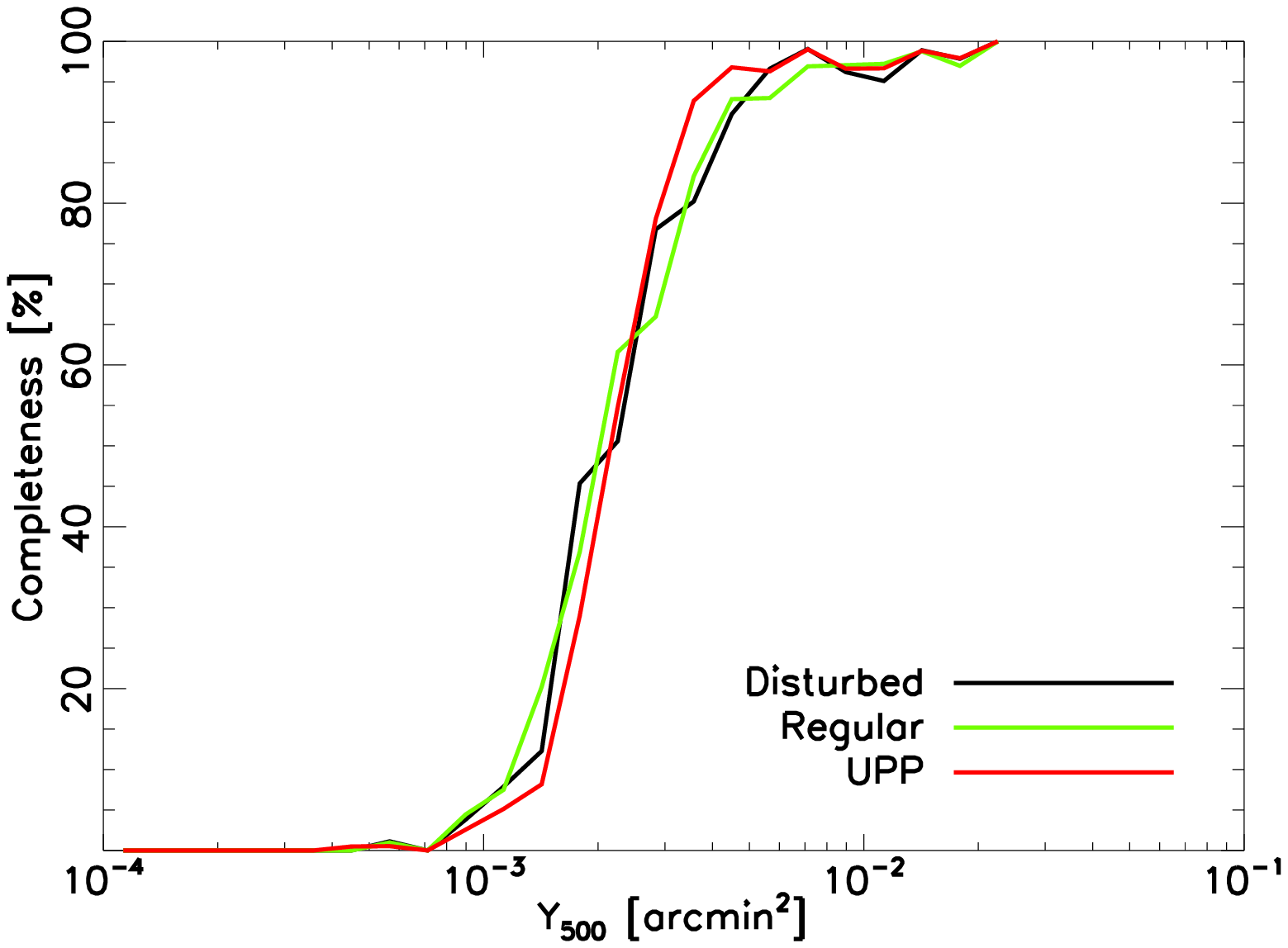}\\
\includegraphics[angle=0,width=0.48\textwidth]{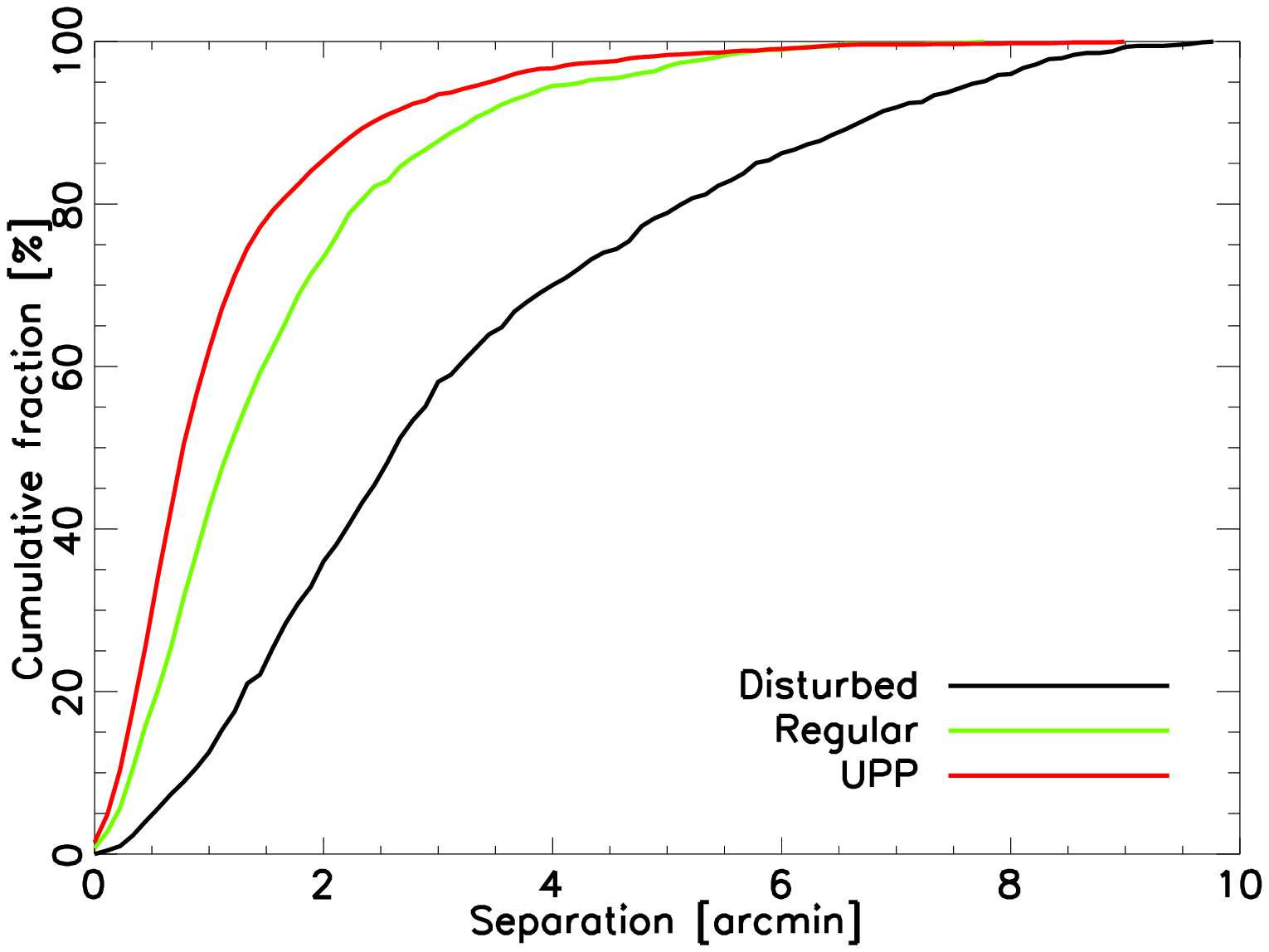}\\
\caption{Impact of cluster morphology on the completeness (top panel) and position estimates (bottom panel) for resolved clusters.  The simulated clusters are all injected with $\theta_{500} = 20'$, and all curves are for the union catalogue.  Cluster morphology has no impact on the completeness, but a significant impact on the position estimation.}
\label{fig:morphology_qa}
\end{center}
\end{figure}

Clusters are known to possess asymmetric morphologies and a wide range
of dynamical states, from irregular merging clusters to regular
relaxed clusters.  While the completeness simulations have included
some morphology variations through variation of the injected radial
pressure profile, this ignores the effects of the sub-structures and
asymmetries, which may induce detection biases for large clusters at
low redshift resolved by the \Planck\ beams, FWHM$\approx 7$ arcmin.

Neither of the external samples used in Sect.~\ref{sec:ext_comp} to
validate the completeness allow us to properly probe resolved,
irregular clusters at low-redshift. The MCXC is biased towards regular
clusters due to X-ray selection effects and the \Planck\ completeness
drop-off lies substantially beneath the SPT mass limit at low
redshift, so the drop-off is not sampled.

We address the effects of realistic morphology by injecting into the
\Planck\ maps the raw 2D projected Compton-y signal from a sample of
hydro-dynamically simulated cosmoOWLs clusters.  The clusters were
injected with a large enough angular extent, $\theta_{500}=20$ arcmin,
that they were resolvable in the \Planck\ data, and with a range of
\yfive~that encompassed the expected completeness drop-off.  20
candidate clusters were chosen from the sub-sample of cosmoOWLS
clusters selected by the mass cuts discussed in
Sect.~\ref{sec:MC_injection} based on their dynamical state.  The ten
clusters in the sub-sample with highest kinetic-to-thermal energy
ratio within $\theta_{500}$ constituted our disturbed sample, while
the regular sample comprised the ten clusters with the lowest
kinetic-to-thermal energy ratio within $\theta_{500}$.  These clusters
were injected 200 times, randomly distributed across the sky.  We also
created simulations injecting symmetric clusters with the UPP with the
same parameters and locations as the hydro-dynamic projections.  In
all cases, the signals were convolved with Gaussian beams to separate
the effects of beam asymmetries.

The completeness for regular, disturbed and UPP clusters is shown for
the union catalogue in the top panel of Fig.~\ref{fig:morphology_qa}.
There are no significant differences between the completeness
functions for the regular and disturbed clusters.  Both sets of
hydro-dynamic clusters show a slight widening effect in the
completeness caused by the variation in the effective pressure profile
away from the UPP assumed for extraction (the same effect as discussed
in Sect.~\ref{sec:completeness}).

Morphology has a clear impact on the estimation of cluster position.
The bottom panel of Fig.~\ref{fig:morphology_qa} shows the cumulative
distribution of angular separation between union and input positions
for the regular, disturbed and UPP clusters.  The disturbed clusters
show a significant reduction in positional accuracy.  Part of this is
physical in origin.  The clusters centres were defined here as the
position of the 'most-bound particle', which traces the minimum of the
gravitational potential and is almost always coincident with the
brightest central galaxy.  For merging clusters this position can be
significantly offset from the centre of the peak of the SZ
distribution.  A matching radius of $10$ arcmin, which is used in
Sect.~\ref{sec:ancillary_info}, ensured correct identification of
detected and injected positions.

\subsection{Reliability}
\label{sec:reliability}

The statistical reliability is the probability that a detection with
given detection characteristics is a real cluster.  We determine the
reliability using simulations of the \Planck\ data.  Clusters are
injected following the prescription in Sect.~\ref{sec:MC_injection},
except that the the clusters are injected such that cluster masses and
redshifts are drawn from a \citet{tin08} mass function and converted
into the observable parameters $(Y_{500},\theta_{500})$ using the
\Planck\ ESZ \yfive--$M_{500}$ scaling relation
\citep{planck2011-5.2a}.  The other components of the simulations are
taken from FFP8 simulation ensemble \citep{planck2014-a14}.  The
components include a model of galactic diffuse emission, with thermal
dust (including some emission from cold-clumps), spinning dust,
synchrotron and CO emission, and extra-galactic emission from the far
infra-red background.  The diffuse components are co-added to a set of
Monte-Carlo realisations of CMB and instrumental noise.  In addition
to the cluster signal, we also inject point sources drawn from a
multi-frequency model from the \Planck\ sky model
\citep{planck2014-a14}.  These point sources are mock detected, using
completeness information from the PCCS2 \citep{planck2014-a35}, and
harmonically infilled using the same process as for the real maps
prior to SZ detection.  This leaves a realistic population of residual
sources in the maps.  After detection, candidates that lie within the
simulated expanded source mask, or which match with the cold-core or
IR source catalogues from the real data, have their \snr~set to zero.

\begin{figure}
\begin{center}
\includegraphics[angle=0,width=0.48\textwidth]{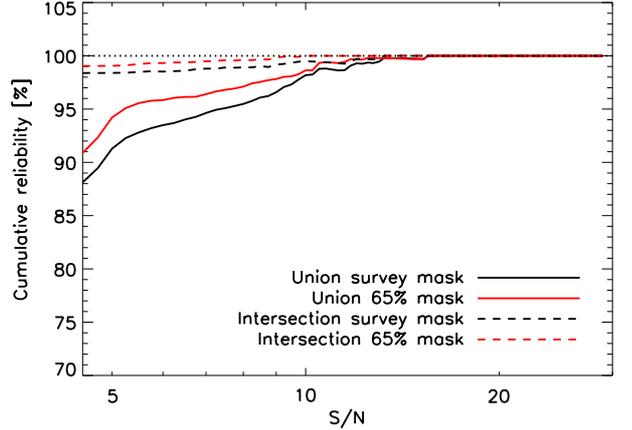}
\caption{Cumulative reliability as a function of \snr.}
\label{fig:reliability1}
\end{center}
\end{figure}

Fig.~\ref{fig:reliability1} shows the reliability as a function of
\snr~for the union and intersection samples across the whole survey
area and outside the 65\% galactic mask used to define the cosmology
samples.  Relative to the \pszone, the reliability of the union has
improved by 5\%, the lower noise levels have revealed more real
simulated clusters than spurious detections.  As was the case with the
\pszone, the reliability is improved significantly by removing more of
the galactic plane, where diffuse and compact galactic emission induce
extra spurious detections.


\subsection{Neural network quality assessment}

We supplement the simulation-based reliability assessment with an
a-posteriori assessment using an artificial neural-network.  The
construction, training and validation of the neural network is
discussed fully in \cite{agh14}.  The network was trained on nominal
mission \Planck\ maps, with a training set composed of three elements:
the positions of confirmed clusters in the \pszone~as examples of good
cluster signal; the positions of PCCS IR and radio sources as examples
of point-source induced detections; and random positions on the sky as
examples of noise-induced detections.  We provide for each detection a
neural network quality flag, Q\_NEURAL $=1-Q_\text{bad}$, following
the definitions in \cite{agh14}, who also tested the network on the
unconfirmed detections in the \pszone.  They showed that this flag
definition separates the high quality detections from the low quality
detections, as validated by the the \pszone~external validation
process, such that Q\_NEURAL $<0.4$ identifies low-reliability
detections with a high degree of success.

459 of the 1961 raw detections possess Q\_NEURAL $<0.4$ and may be
considered low-reliability.  This sample is highly correlated with the
IR\_FLAG, with 294 detections in common.  After removal of IR spurious
candidates identified by the IR\_FLAG, as discussed in
Sect.~\ref{sec:ir_spurious}, we retain 171 detections with bad
Q\_NEURAL, of which 28 are confirmed clusters.  This leaves 143
unconfirmed detections considered likely to be spurious by the neural
network.

The Q\_NEURAL flag is sensitive to IR induced spurious: detections
with low Q\_NEURAL quality flag are clustered at low galactic
latitudes and at low to intermediate \snr.  This clustering is not
seen for realisation-unique spurious detections in the reliability
simulations, which are identifiable as noise induced.
The reliability simulations underestimate the IR spurious populations
relative to the Q\_NEURAL flag.  Conversely, the neural network flag
by construction does not target noise-induced spurious detections:
$Q_\text{bad}$ is the parameter trained to indicate IR-induced
spurious.  The neural network flag also has some sensitivity to the noise 
realisation and amplitude in the data: the assessment is different to that applied to
the nominal mission maps in \citealt{planck2015-XXXVI}.

To place a lower limit on the catalogue reliability, we combine the
Q\_NEURAL information with the noise-induced spurious detections from
the reliability simulations.  For each reliability simulation
realisation, we remove the simulated IR spurious detections, which can be
identified either as induced by the FFP8 dust component, and thus
present in multiple realisations, or as induced by injected IR point
sources.  We replace these spurious counts with the unconfirmed low
Q\_NEURAL counts, smoothed so as to remove the steps due to small
number statistics.

\begin{figure}
\begin{center}
\includegraphics[angle=0,width=0.48\textwidth]{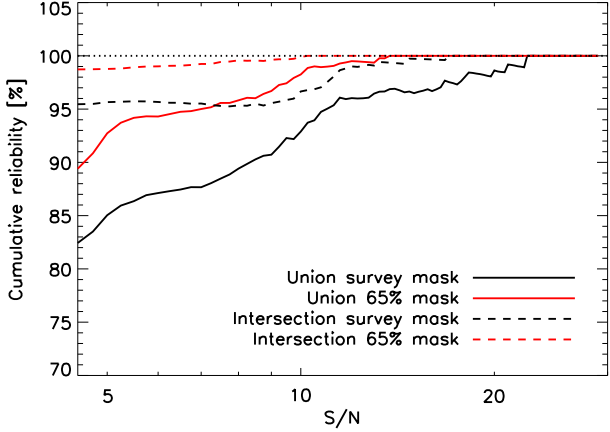}
\caption{Lower limits on the catalogue reliability, estimated by
  combining the reliability simulations with the Q\_NEURAL information
  (see text).}
\label{fig:reliability3}
\end{center}
\end{figure}

The combined lower limit of the reliability is shown in
Fig.~\ref{fig:reliability3}.  The lower limit tracks the simulation
reliability well outside the 65\% galactic dust mask.  For the whole
survey region, the lower limit is typically 6\% lower than the
simulation estimate, due either to over-sensitivity of the neural
network to dusty foregrounds, or shortcomings in the FFP8 galactic
dust component.

\section{Parameter Estimates}
\label{sec:param_est}

The SZ survey observable is the integrated Comptonisation parameter,
$Y_{sz}$.  As was the case for the \pszone, each of the extraction
codes has an associated parameter estimation code that evaluates, for
each detection, the two dimensional posterior for the integrated
Comptonisation within the radius $5R_{500}$, \yfiver, and the scale
radius of the GNFW pressure $\theta_{S}$.  The radius 5R500 is chosen
as it provides nearly unbiased (to within a few percent) estimates of
the total integrated Comptonisation, while being small enough that
confusion effects from nearby structures are negligible.

We provide these posteriors for each object and for each code, and
also provide \yfiver~in the union catalogue, defined as the expected
value of the \yfiver~marginal distribution for the reference detection
(the posterior from the code that supplied the union position and
\snr).

Below we also discuss the intricacies of converting the posteriors to
the widely used X-ray parameters \yfive$-\theta_{500}$.

\subsection{\yfiver~estimates}

\begin{figure}
\begin{center}
\includegraphics[angle=0,width=0.48\textwidth]{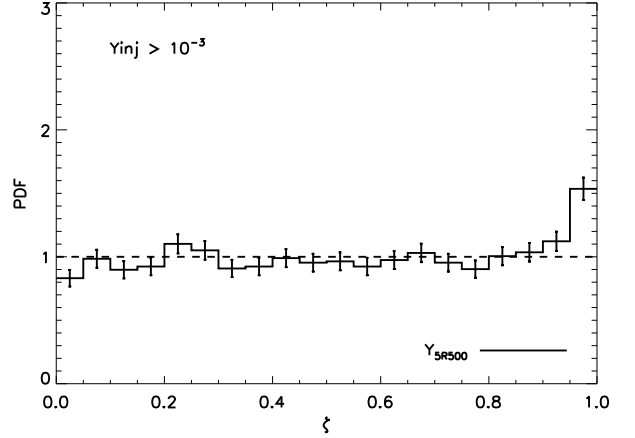}
\includegraphics[angle=0,width=0.55\textwidth]{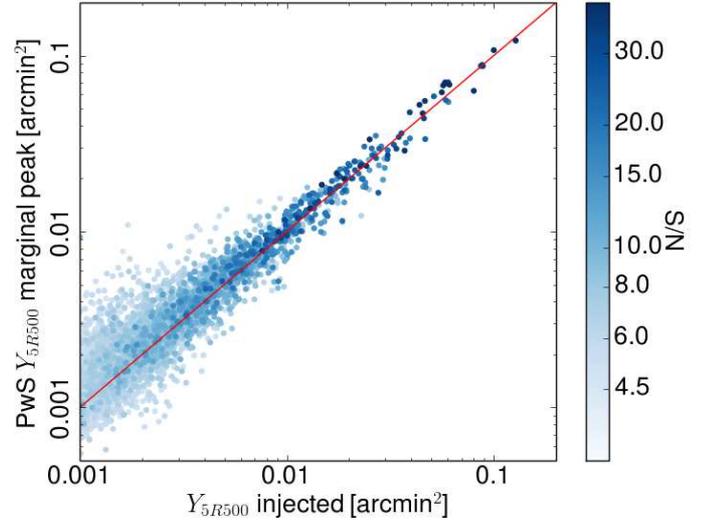}
\caption{\textit{Top panel:} Results of the posterior validation for
  \yfiver.  The histogram of the posterior probability, $\zeta$,
  bounded by the true \yfiver~parameter is almost uniformly
  distributed, except for a small excess in the tails of the
  posteriors, at $\zeta > 0.95$.  The histogram has been normalised by
  the expected counts in each $\zeta$ bin.  \textit{Bottom panel:}
  Comparison of recovered peak \yfiver~to the injected \yfiver.  The
  estimates are unbiased, though asymmetrically scattered, with a
  scatter that decreases as \snr~increases.}
\label{fig:y5r500_pws}
\end{center}
\end{figure}

To validate the \yfiver~estimates, we apply the posterior validation
process introduced in \cite{har14} to the \yfiver~marginal
distributions.  In brief, this process involves simulating clusters
embedded in the \Planck\ maps and evaluating the $Y-\theta$ posteriors
for each (detected) injected cluster.  For each posterior, we
determine the posterior probability, $\zeta$, bounded by the contour
on which the real underlying cluster parameters lie.  If the
posteriors are unbiased, the distribution of this bounded probability
should be uniformly distributed between zero and one.

This process allows us to include several effects that violate the
assumptions of the statistical model used to estimate the posteriors.
Firstly, by injecting into real sky maps, we include the non-Gaussian
contributions to the noise on the multifrequency-matched-filtered maps
that come from galactic diffuse foregrounds and residual point
sources.  Secondly, we include violations of the `signal' model that
come from discrepancies between the cluster pressure profile and the
UPP assumed for parameter estimation, and from sky-varying and
asymmetric effective beams that vary from the constant Gaussian beams
assumed for estimation.  The clusters are injected using the process
discussed in Sect. \ref{sec:MC_injection}, drawing injected pressure
profiles from the set of cosmoOWLs simulated profiles.

The top panel of Fig.~\ref{fig:y5r500_pws} shows the histogram of
$\zeta$ for the \pws~\yfiver~marginals.  The distribution is flat,
except for a small excess in the $0.95-1.0$ bin, which indicates a
small excess of outliers beyond the 95\% confidence region, in this
case 52\% more than statistically expected.  This suggests the
posteriors are nearly unbiased, despite the real-world complications
added to the simulations.  Note that we have considered only
posteriors where the injected \yfiver$~> 0.001$ arcmin$^2$, a cut that
removes the population effects of Eddington bias from consideration:
we focus here on the robustness of the underlying cluster model.

The bottom panel of Fig.~\ref{fig:y5r500_pws} shows the peak recovery
from the \pws~\yfiver~marginals compared to the true injected values.
The peak estimates are unbiased relative to the injected parameters.

\subsection{Conversion to \yfive}

The (\yfiver,$\theta_S$) estimates can be converted into
(\yfive,$\theta_{500}$) estimates using conversion coefficients
derived from the UPP model that was assumed for extraction.  However,
when the underlying pressure distribution deviates from this model,
the conversion is no longer guaranteed to accurately recover the
underlying (\yfive,$\theta_{500}$) parameters: variation of the
pressure profile can induce extra scatter and bias in the
extrapolation.

We demonstrate this by applying the posterior validation process to
the \yfive~posteriors, defined as the \yfiver~posteriors scaled with
the UPP conversion coefficient, as estimated from injected clusters
whose pressure profiles are drawn from the cosmoOWLS pressure profile
ensemble.  We validate posteriors for \yfive~calculated in two ways:
firstly by marginalising over the $\theta_{500}$ parameter, referred
to in previous publications as `Y blind'; and secondly by slicing the
(\yfive,$\theta_{500}$) posteriors at the true value of
$\theta_{500}$, equivalent to applying an accurate, externally
measured delta-function radius prior.

\begin{figure}
\begin{center}
\includegraphics[angle=0,width=0.48\textwidth]{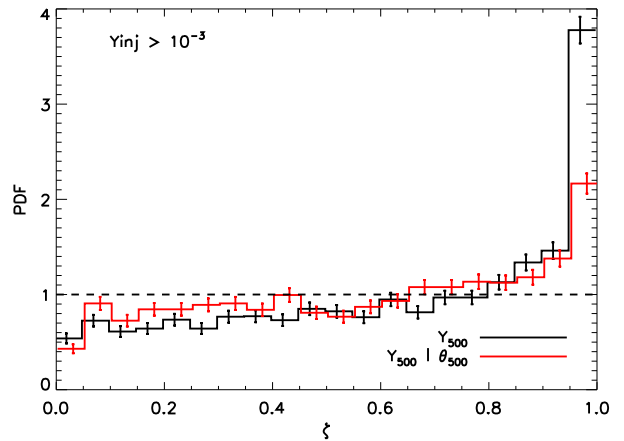}
\caption{Bounded probability histograms, as in the top panel of
  Fig.~\ref{fig:y5r500_pws}, but for the converted $p($\yfive$)$
  marginal and $p($\yfive$|\theta_{500})$ sliced posteriors.}
\label{fig:\yfive_HPDs}
\end{center}
\end{figure}

\begin{figure}
\begin{center}
\includegraphics[angle=0,width=0.55\textwidth]{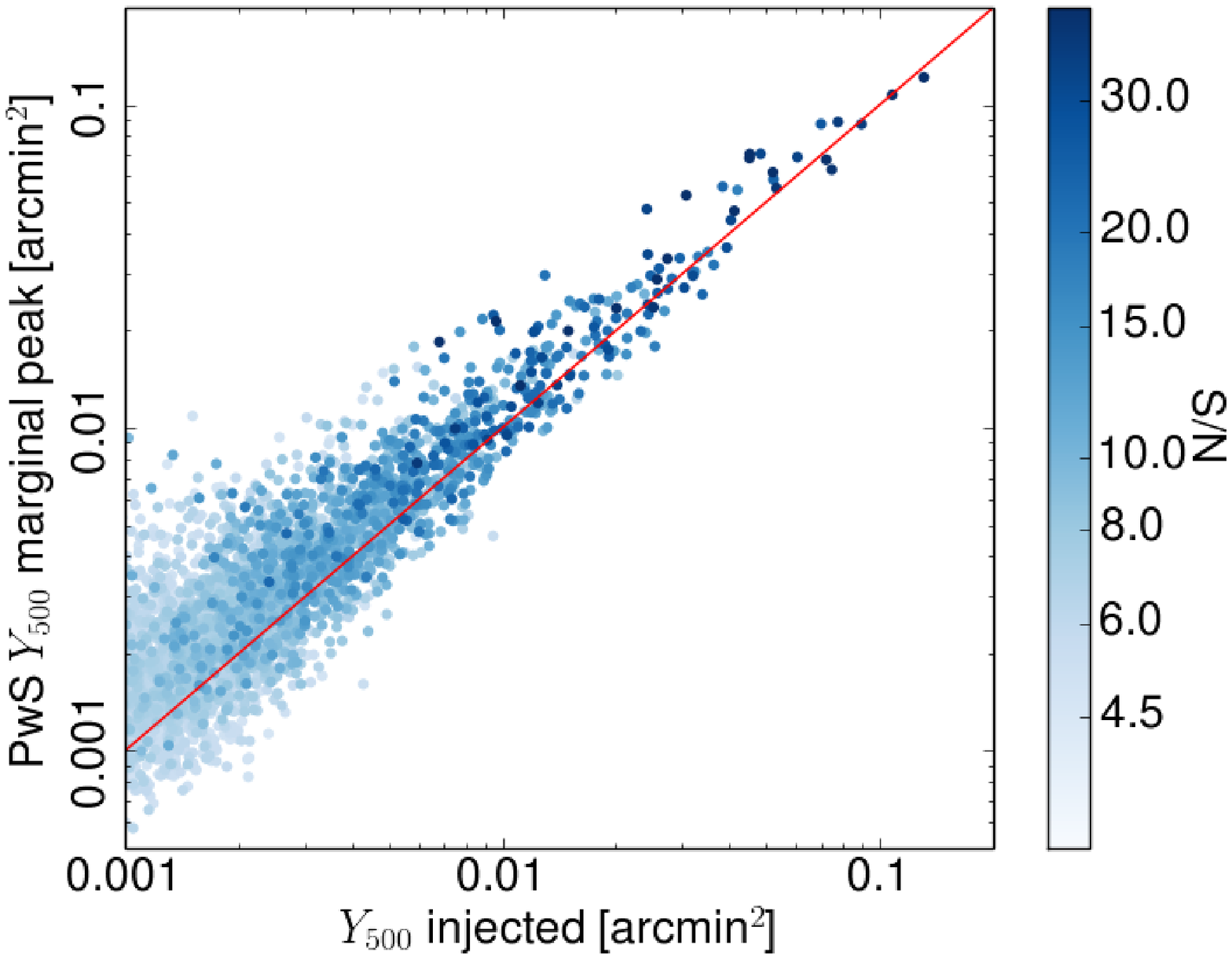}
\includegraphics[angle=0,width=0.55\textwidth]{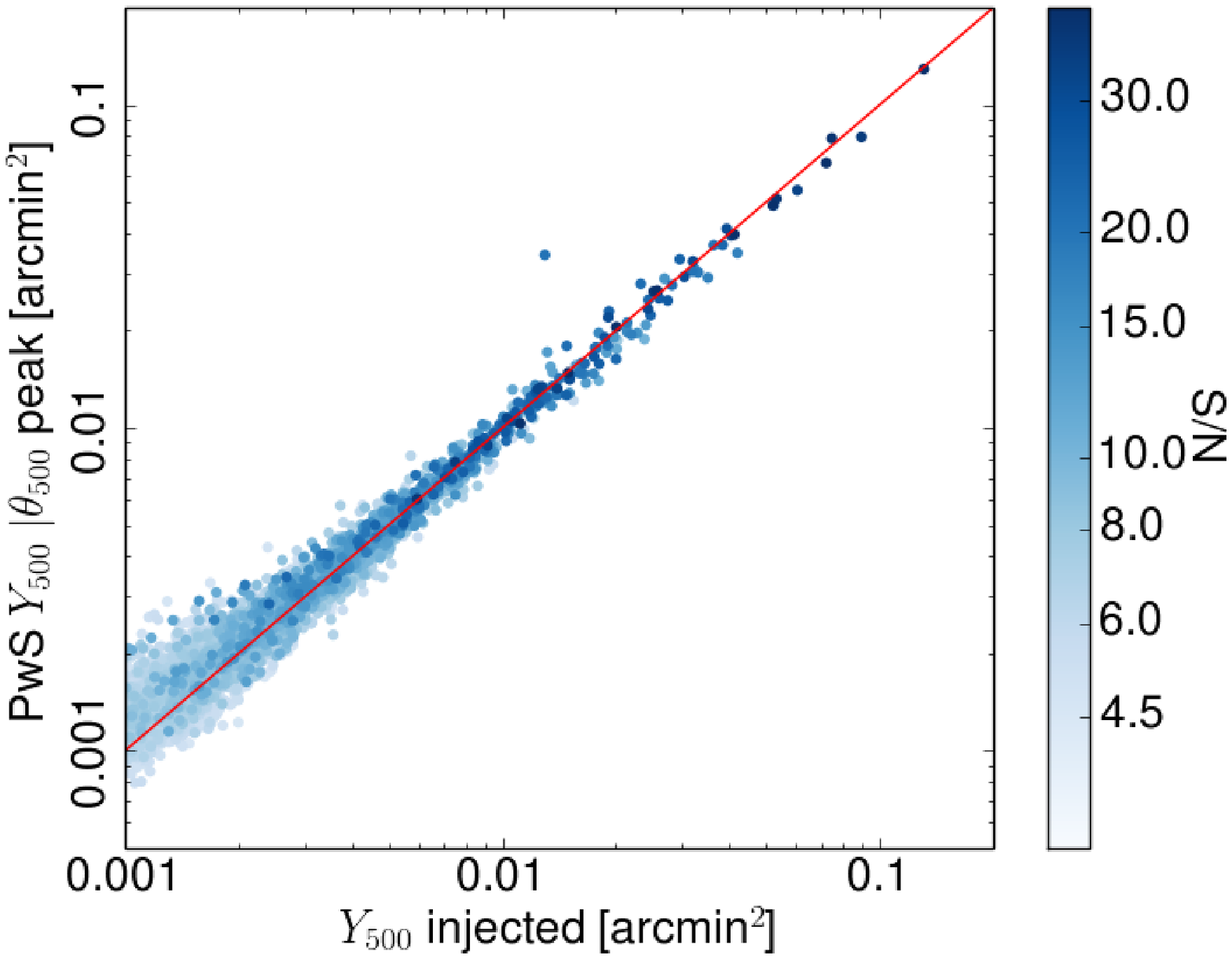}
\caption{Scatter of the recovered estimates of \yfive~with the input
  \yfive.  \textit{Top panel}: for the marginalised \yfive~posterior,
  `Y blind'.  \textit{Bottom panel}: for the sliced posterior
  $p($\yfive$|\theta_{500})$, assuming an accurate radius prior.}
\label{fig:\yfive_scatter}
\end{center}
\end{figure}

Fig.~\ref{fig:\yfive_HPDs} shows the bounded probability histograms
for the two \yfive~posteriors and Fig.~\ref{fig:\yfive_scatter} shows
the scatter of the peak of the posteriors with the input values of
\yfive.  The marginal \yfive~posteriors are poor, with histograms
skewed towards the tails of the distribution and large numbers of $>2\sigma$ outliers.  The scatter plot
reveals the peak estimates to possess a large scatter and to be
systematically biased high.  In contrast, the peak
$p($\yfive$|\theta_{500})$ estimates have much better accuracy and
precision and are distributed around the input values with low
scatter.  The bounded probability histogram of $p($\yfive$|\theta_{500})$ shows that while there is
a noticeable excess of detections in the wings, the posteriors are
reasonably robust.  If the posteriors were Gaussian, the skewness of
the $p($\yfive$|\theta_{500})$ histogram towards the tails would be
consistent with an underestimate of the Gaussian standard deviation of
21\%

We therefore recommend that, to estimate $Y_{500}$ accurately from
\Planck\ posteriors, prior information be used to break the
(\yfive,$\theta_{500})$ degeneracy.  However, we note that the
uncertainties on such $Y_{500}$ estimates will be slightly
underestimated.

\subsection{Mass and \yfive~estimates using scaling priors}
\label{section:Msz}

The key quantity which can be derived from SZ observables is the total
mass of the detected clusters within a given overdensity (we used
$\Delta=500$). To calculate the mass from \Planck\ data it is
necessary to break the size-flux degeneracy by providing prior
information, as outlined in the previous section. We used an approach
based on Arnaud et al. (in prep.), where the prior information is an
expected function relating $Y_{500}$ to $\theta_{500}$ that we
intersect with the posterior contours. We obtained this relation by
combining the definition of $M_{500}$ (see Eq. 9 in
\citealt{planck2013-p15}, connecting $M_{500}$ to $\theta_{500}$, for
a given redshift $z$) with the scaling relation $Y_{500}-M_{500}$
found in \citet{planck2013-p15}.  A similar approach was also used in
\citet{planck2013-p05a}, but in this work we use the full posterior
contours to associate errors to the mass value.
\begin{figure}
\begin{center}
\includegraphics[angle=0,width=0.48\textwidth]{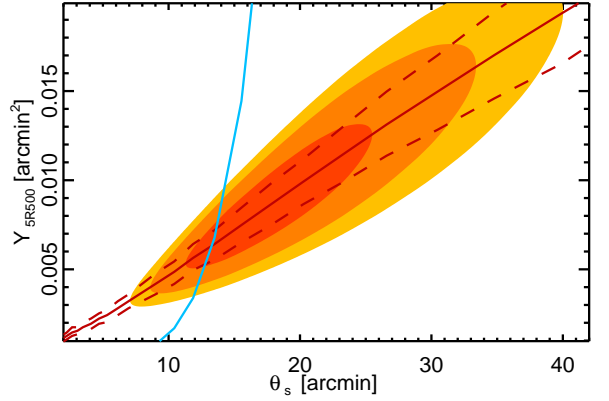}
\caption{Illustration of the posterior probability contours in the
  $Y_{5R500}-\theta_s$ plane for a cluster detected by \Planck: the
  contours show the 68, 95 and 99 percent confidence levels. The red
  continuous line shows the ridge line of the contours while the dashed
  lines the 1-$\sigma$ probability value at each $\theta_s$. The cyan
  line is the expected $Y-\theta$ relation at a given redshift that we
  use to break the degeneracy. }
\label{fig:ex_cont_mass}
\end{center}
\end{figure}

We illustrate our method in Fig. \ref{fig:ex_cont_mass}. At any fixed
value of $\theta_{S}$, we study the probability distribution and
derive the $Y_{5R500}$ associated to the maximum probability,
i.e. the ridge line of the contours (red continous line in
Fig. \ref{fig:ex_cont_mass}). We also derive the $Y_{5R500}$ limits
enclosing a $68\%$ probability and use them to define a upper and
lower degeneracy curve (dashed lines). From the intersection of these
three curves with the expected function (cyan line), we derive the
$M_\text{SZ}$ estimate and its $1\sigma$ errors, by converting
$Y_{5R500}$ to $Y_{500}$ and then applying the $Y_{500}-M_{500}$
scaling relation prior at the redshift of the counterpart.

$M_\text{SZ}$ can be viewed as the hydrostatic mass expected for a
cluster consistent with the assumed scaling relation, at a given
redshift and given the \Planck ($Y-\theta$) posterior information.  We
find that this measure agrees with external X-ray and optical data
with low scatter (see Sect.~\ref{sec:ancillary_info}).  For each
$M_\text{SZ}$ measurement, the corresponding \yfive~from the scaling
relation prior can be calculated by applying the relation.

We underline that the errors bars calculated from this method
consider only the statistical uncertainties in the contours, not the
uncertainties on the pressure profile nor the errors and scatter in
the $Y_{500}-M$ scaling relation, and should thus be considered a
lower limit to the real uncertainties on the mass.

We used the masses for the confirmation of candidate counterparts (see
Sect. \ref{sec:ancillary_info}) and we provide them, along with their
errors, in the \psztwo~catalogue for all detections with confirmed
redshift. We compared them with the masses provided in \pszone~for the
detections where the associated counterpart (and thus the redshift
value) has not changed in the new release (see Appendix B). We find
very good agreement between the two values which are consistent within
the error bars over the whole mass range.

In the individual catalogues, we provide for all entries an array of
masses as a function of redshift ($M_\text{SZ}(z)$), which we obtained
by intersecting the degeneracy curves with the expected function for
different redshift values, from $z=0$ to $z=1$. The aim of this
function is to provide a useful tool for counterpart searches: once a
candidate counterpart is identified, it is sufficient to interpolate
the $M_\text{SZ}(z)$ curve at the counterpart redshift to estimate its
mass.



\section{Consistency with the \pszone}
\label{sec:consistency}

\begin{table}
\caption{Results of fits between \snr~from the \pszone~and \psztwo,
  using the fitting function in Eq. \ref{equ:snr_fitfunc}. The assumed
  correlation of the uncertainties of $s_{1}$ and $s_{2}$ was 0.72.}
\begin{center}
\begin{tabular}{ccccc}
\hline\hline
$s_{1}$ & $s_{2}$& A & $\alpha$ & $\sigma$ \\ 
\hline
\psztwo~& \pszone~& $0.76 \pm 0.08$ & $0.72 \pm 0.01$ & $0.53 \pm 0.02$ \\ 
\hline
\end{tabular}
\end{center}
\label{tab:consistency_snr_surveys}
\end{table}%

The extra data available in the construction of the \psztwo~improves
the detection \snr~and reduces statistical errors in the parameter and
location estimates.  Here we assess the consistency between the two
catalogues, given the matching scheme discussed in
Sect.~\ref{section:psz1_cross_match}.

\subsection{Signal-to-noise}
\label{sec:psz1_psz2_snr}

\begin{figure}[h!]
\begin{center}
\includegraphics[angle=0,width= 0.55\textwidth]{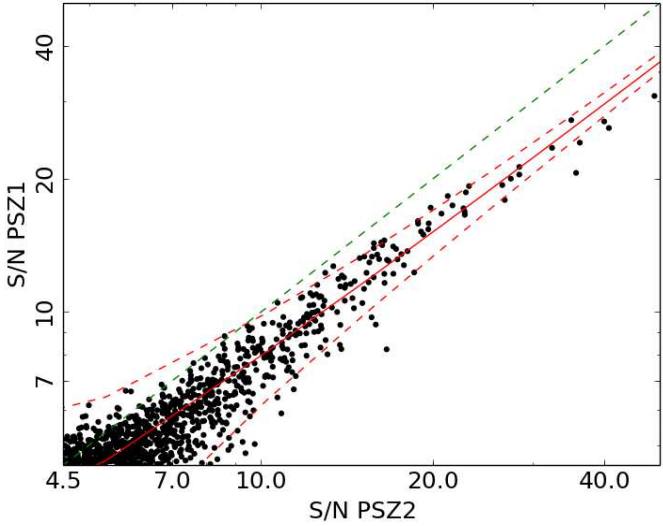}
\caption{Comparison of \snr~values for common \pszone~and
  \psztwo~detections.  The best fit relation is plotted in red, with
  $2\sigma$ scatter plotted by dashed red lines.  The green dashed
  line denotes the 1-1 relation.}
\label{fig:psz1_psz2_snr_consistency}
\end{center}
\end{figure}

We fit the relation between \snr~for common \pszone~and \psztwo~using
the the approach and model discussed in
Sect. \ref{sec:code_consistency}.  For the \pszone~and \psztwo, the
likelihoods for $s_1$ and $s_2$ have a strong
covariance, as more than half of the \psztwo~observations were used in
the construction of the \pszone.  We therefore assign a covariance of
$0.72$ between the two \snr~estimates, as is appropriate for Gaussian
errors sharing 53\% of the data.  As the errors are not truly
Gaussian, we allow for an intrinsic scatter between the \snr~estimates
to encapsulate any un-modelled component of the \snr~fluctuation.

The consistency of the \snr~estimates between the \pszone~and
\psztwo~are shown in Fig. \ref{fig:psz1_psz2_snr_consistency} and 
 the best fitting model in is shown in Table
\ref{tab:consistency_snr_surveys}.  Detections with \psztwo\
 $S/N>20$ are affected by changes
in the \mmfthree~\snr~definition.  For the \pszone, the
empirical standard deviation of the filtered patches was used to
define the \snr~in this regime, while the theoretical standard
deviation of \cite{melin2006} was used for lower \snr.  \mmfthree~now
uses the theoretical standard deviation for all \snr, consistent with
the ESZ and the definitions in the other detection codes.  For this
reason, the best fit model ignores detections at $S/N > 20$ in either
catalogue.  The \mmfthree~\snr~show a flat improvement
  relative to the ESZ \snr~(which was produced solely by \mmfthree), consistent
  with the reduced noise in the maps.

If the Compton-Y errors are entirely Gaussian in their behaviour, we
should expect the \snr~to increase by 37\% between the \pszone~and
\psztwo, ie: $\alpha=0.73$.  This is consistent within $1\sigma$ with
the fit, which describes the \snr~behaviour well to \snr$<20$.

\subsection{Position estimation}

\begin{figure}[h!]
\begin{center}
\includegraphics[angle=90,width= 0.48\textwidth]{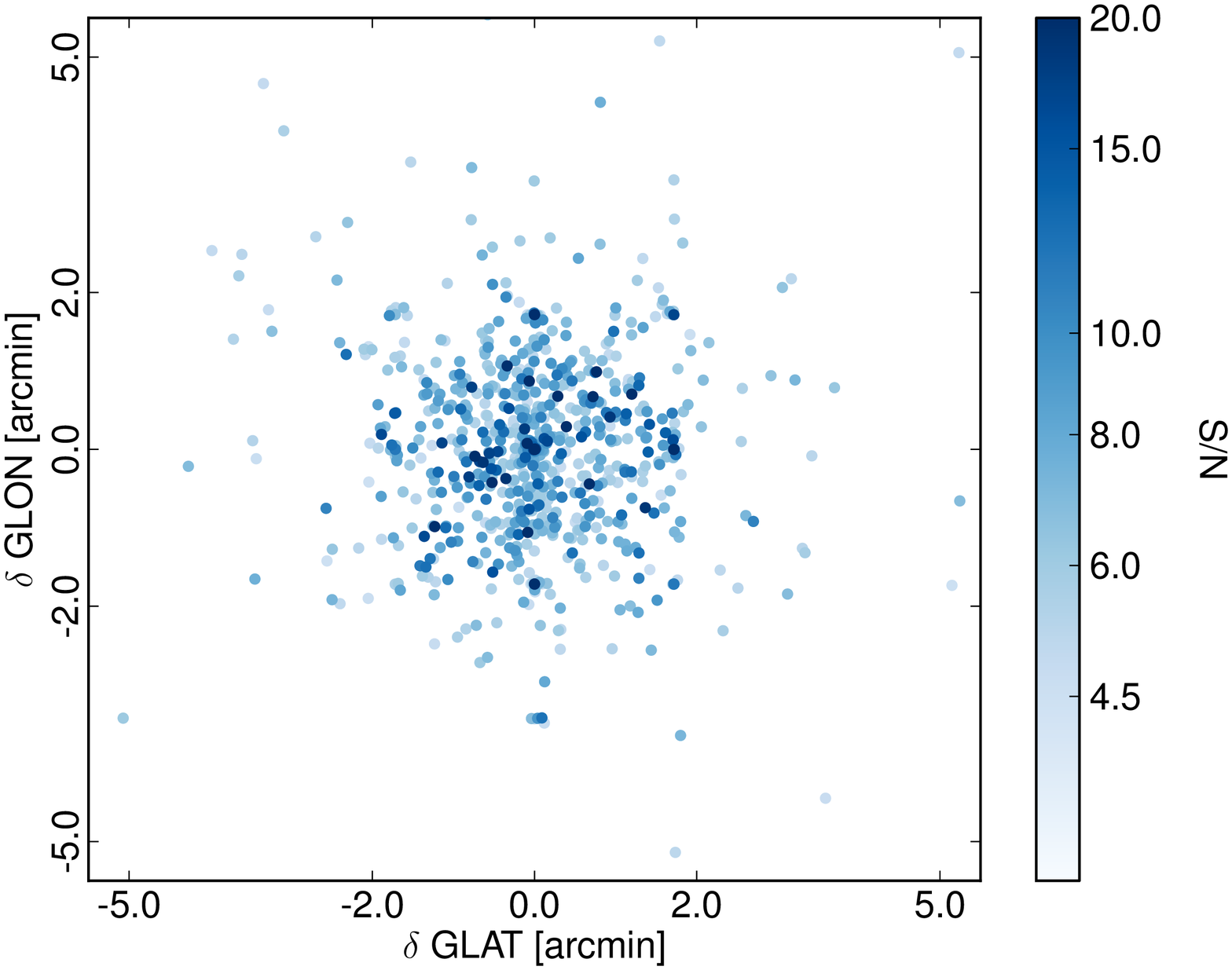} \\
\includegraphics[angle=0,width= 0.48\textwidth]{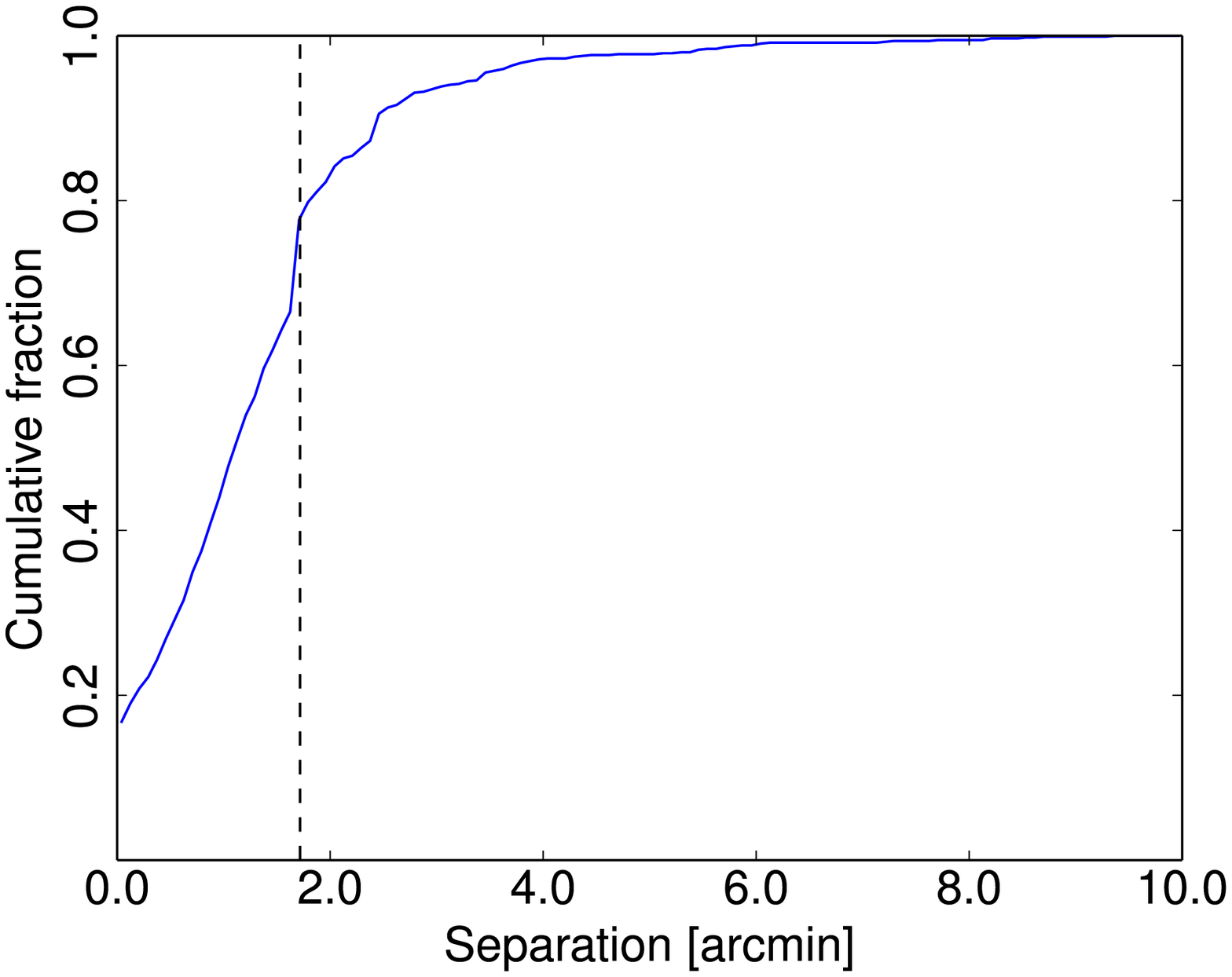}
\includegraphics[angle=0,width= 0.48\textwidth]{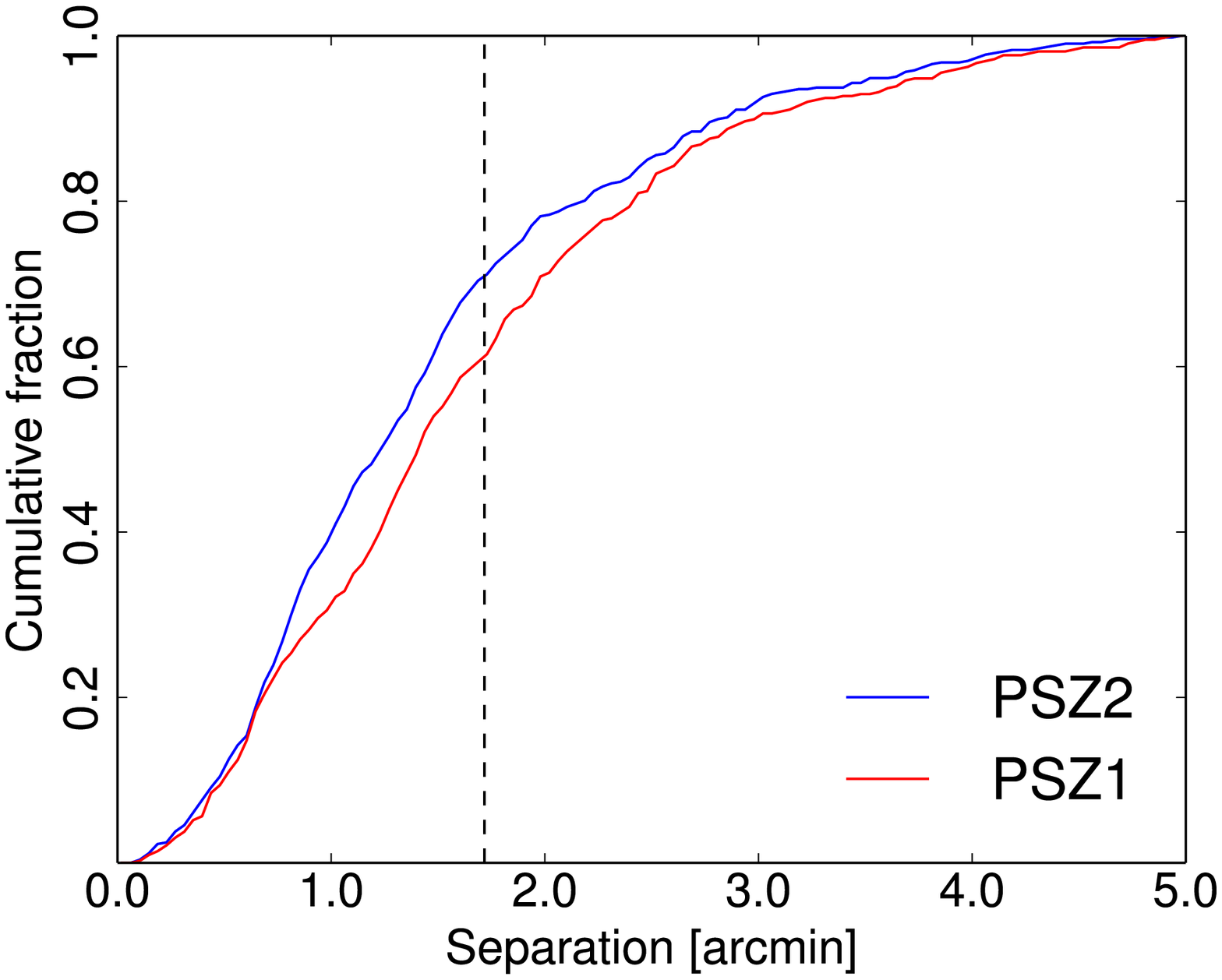}
\caption{\emph{Top panel}: Separation between \psztwo~and
  \pszone~positions for common detections.
\emph{Middle panel}: Cumulative distribution of the angular separation
between \pszone~and \psztwo~positions, with the \Planck\ pixel width
indicated by a dashed vertical line.  \emph{Bottom panel} Cumulative
distributions of angular separation to MCXC x-ray centres, for all
\pszone~and \psztwo~MCXC matches.  The vertical dashed line denotes
the \Planck\ Healpix pixel size.}
\label{fig:psz1_psz2_separation}
\end{center}
\end{figure}

The distribution of angular separations between the \psztwo~and
\pszone~position estimates is shown in
Fig. \ref{fig:psz1_psz2_separation}.  Of the common detections, 80\%
of the \psztwo~positions lie within one \Planck\ map pixel width, 1.7
arcmin, of the \pszone~position.  \mmfthree~does not allow for
sub-pixel positioning, so if the \mmfthree~position was used for the
union in both the \pszone~or \psztwo, the angular separation will be a
multiple of the pixel width.  This is evident in the cumulative
distribution of angular separations as discontinuities at 0, 1 and
$\sqrt{2}$ pixel widths.

We also compare the position discrepancy between the SZ detection and
the X-ray centres from the MCXC \citep{pif11}.
The bottom panel of Fig. \ref{fig:psz1_psz2_separation} shows the distributions of these
angular separations for the \psztwo~and \pszone.  The distributions
are calculated from the full MCXC match for each catalogue: the
\psztwo~includes 124 new detections.  The \psztwo~position estimates
are clearly closer to the X-ray centres than the \pszone: for the
\pszone, the 67\% error radius is 1.85 arcmin, while for the
\psztwo~this reduces to 1.6 arcmin.



\subsection{Missing \pszone~detections}
\label{sec:missing_detections}

The \pszone~produced 1227 union detections.  While the numbers of
detections has increased by 35\% in the \psztwo~to 1653, the number of
common detections is 936: 291 (23.7\%) of the \pszone~detections
disappear.  The high-purity intersection sample loses 44 detections,
of which 20 are lost entirely, and 24 drop out of the intersection
after one or two codes failed to detect them.  In this section, we
discuss these missing detections.  Table \ref{table_lost_dets1}
details each of the missing detections and provides an explanation for
why each is missing.

The first type of missing detection are those that fall under the new
survey mask, due to the increase in the number of point sources
being masked.  The masked areas sare pre-processed with harmonic infilling to prevent spurious
detections induced by Fourier ringing.  The increase of the mask area is
driven by \snr~improvements for IR sources in the high frequency
channels.  While the increase in the masked area is small (0.1\% of
the sky), the correlation between IR point sources and galaxy clusters
leads to a larger percentage of detections being masked.  In the
\pszone, these detections were contaminated by point source emission,
but the emission was just beneath the point source masking threshold.
21 \pszone~union detections fall behind the new mask.  Of these, three
were confirmed clusters, none received the highest validation quality
flag of 1 (denoting probable clusters) in the \pszone~validation
process\footnote{The \pszone~validation process produced three quality
  flags for unconfirmed clusters.  These were based on a combination
  of SZ signal quality, X-ray signal in the RASS maps and IR signal in
  the WISE maps.  Class 1 candidates satisfied good quality in all
  three measures and were high reliability candidates.  Class 2
  satisfied at least one measure with good quality, while class 3
  failed all three measures and so were considered probably
  spurious.}, four received the intermediate validation quality flag
2, and 14 received the lowest validation quality flag of 3, denoting
probable spurious.

The second type of missing detection is one which has a matching detection in
the full-mission data, but where the detection was rejected either by the
infra-red spurious cuts or by \pws~internal consistency cuts, both of which are 
discussed in Sect. \ref{sec:ir_spurious}. The IR cuts are responsible
for cutting 33 unconfirmed \pszone~detections, of which six were in
the intersection sample.  In the \pszone~validation process, none of
these received validation quality flag of 1, seven received quality
flag 2 and 26 received quality flag 3.  These were all $S/N<7$
detections. Five detections were lost because \pws~was the only
detecting code in the \psztwo~and they failed \pws~consistency
criteria: two of these were confirmed clusters.

\begin{figure}[h!]
\begin{center}
\includegraphics[angle=0,width= 0.48\textwidth]{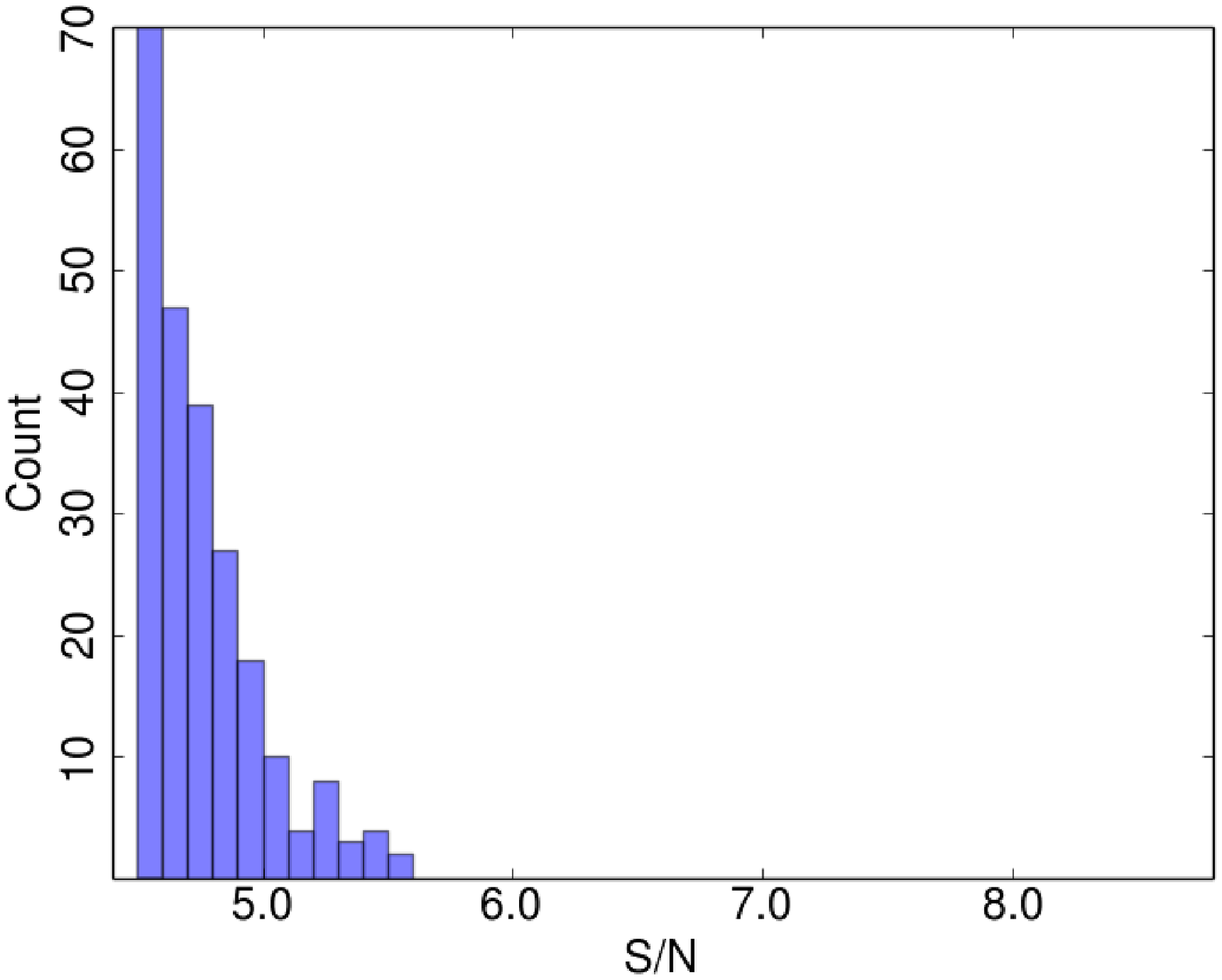}\\
\includegraphics[angle=0,width= 0.48\textwidth]{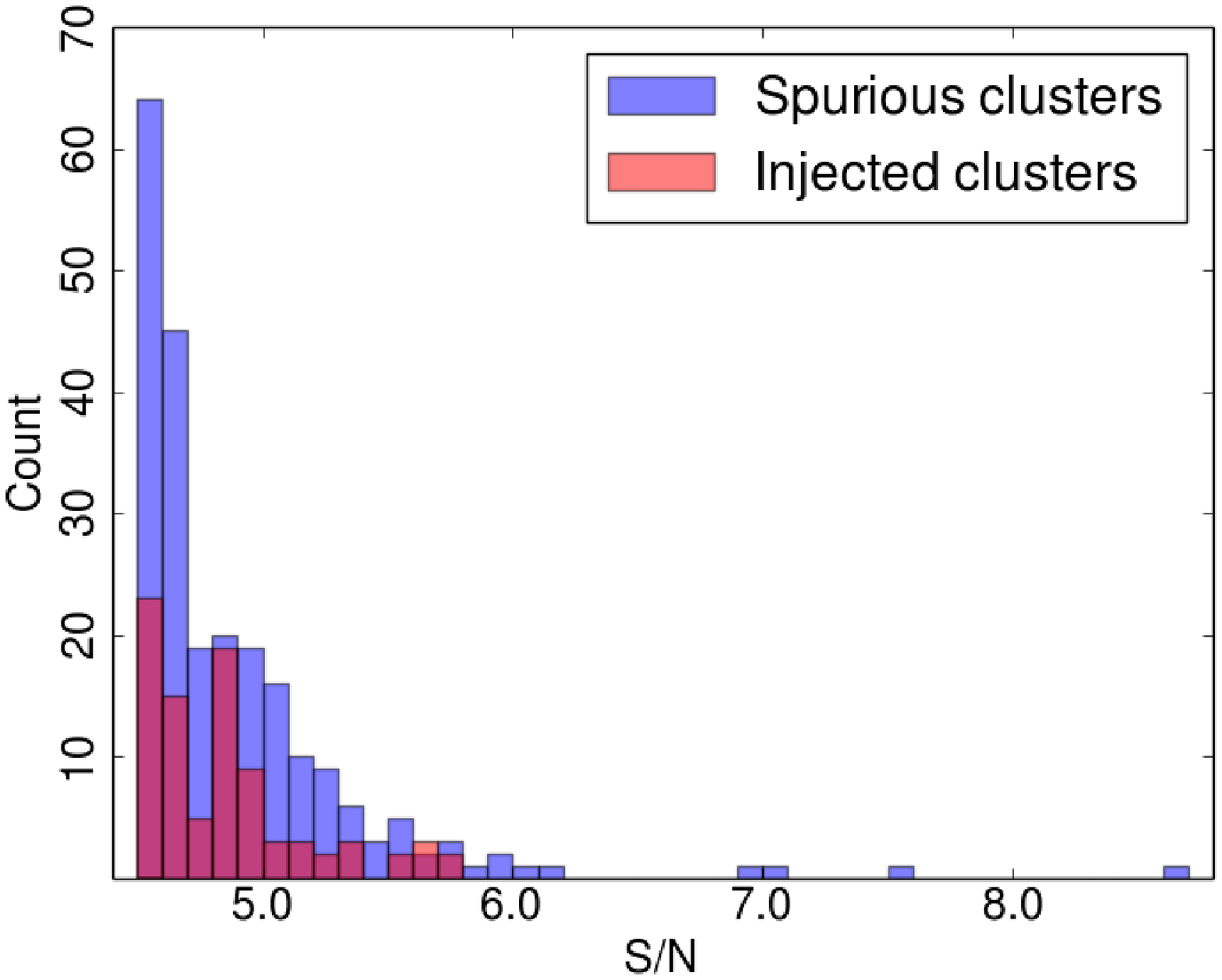}
\caption{Distribution of \snr~for missing nominal mission detections lost due
to downward fluctuation of the \snr\ rather than because of spurious rejection
  cuts or changes in the survey mask. \emph{Top panel} Detections lost from the
  \pszone. \emph{Bottom panel} Detections lost in simulations of the
  transition from the nominal to full mission.}
\label{fig:churn_losses}
\end{center}
\end{figure}

\begin{figure}[h!]
\begin{center}
\includegraphics[angle=0,width= 0.48\textwidth]{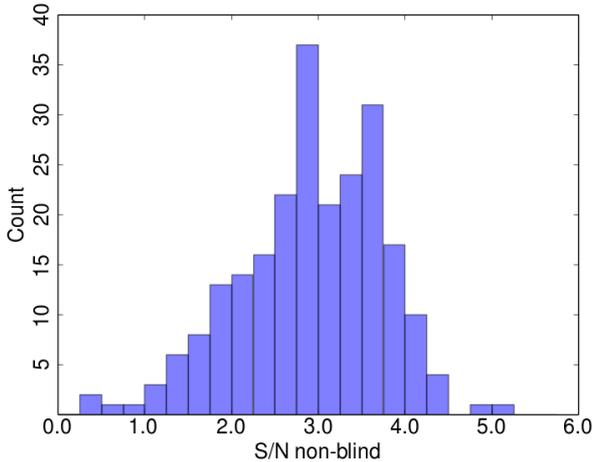}
\caption{Non-blind \pws~\snr~for the SZ signal at the location of
  missing \pszone~detections that were not masked out or cut for IR
  contamination.}
\label{fig:missing_SNR}
\end{center}
\end{figure}

The final type contains the majority (232) of the missing detections.
These are low-significance detections close to the \pszone~threshold
that have downward-fluctuated with the full mission data and are now
beneath the \psztwo\ threshold.  This occurs for some detections
despite the fact that the \snr~improves for most.  The top panel of
Fig.~\ref{fig:churn_losses} shows the \pszone\ \snr~distribution of
the downward-fluctuated detections. These were weak detections: 87\%
were within $0.5\sigma$ of the detection threshold and 82\% of them
were single-code detections.  While many of these may be spurious
detections, 81 confirmed clusters have been lost.  61 of these were
single-code detections and 70 of them were within $0.5\sigma$ of the
threshold.  Based on \Planck\ data alone, these clusters were weak SZ
detections and were likely to be Eddington biased above the threshold in the
\pszone.  We have estimated the \snr~for these lost \pszone\ detections
in the full-mission maps using \pws~in a non-blind analysis at the
\pszone~positions.  Fig. \ref{fig:missing_SNR} shows the distribution
of these non-blind \snr: for most an apparently significant signal
still exists in the maps, but it is now too weak to exceed the
detection threshold, typically lying between $2<S/N<4$.  The non-blind \snr~for
this category is shown per detection in Table \ref{table_lost_dets1}.
Two detections have a non-blind \snr~above the selection threshold.
For these detections, the noise level for the non-blind analysis
(centred on the PSZ1 location) was lower than for any of the patches
in the mosaic used for the cluster detection.

To verify that this sample of missing detections is consistent with
the change in data, we simulated the transition from the \pszone~to
the \psztwo~using FFP8 half-mission noise realisations to approximate
the nominal mission: this produces a pair of data-sets with
appropriately correlated noise characteristics.  A common sample of
clusters and point sources were injected into the simulations and the
full pipeline was applied to construct catalogues from both simulated
datasets.  The simulations produced a total loss of 353 detections, of
which 24 were lost due to the expansion of the point source mask, ten
were lost due to changes to the \pws~spurious rejection criteria, and
319 were lost due to downward fluctuation of the \snr~beneath the
detection threshold.

The \snr~distribution of the latter group is shown in the bottom panel
of Fig.~\ref{fig:churn_losses}.  75\% lie within $0.5\sigma$ of the
detection threshold and 85\% were single-code detections.  While this
group was primarily composed of 230 spurious detections, 89 injected
clusters were lost.  The loss of these injected clusters illustrates
that, as a statistical process, cluster detection is dependent on the
realisation of the noise in the filtered patch-maps and we should
expect that substantial numbers of confirmed but weak cluster
detections will be lost due to noise fluctuations.

These simulations over-estimate the loss-rate of nominal mission
detections.  This may be due in part to unsimulated changes in the
sample selection applied to the real data.  We were unable to simulate systematic changes
in the IR spurious rejection that may, had they been incorporated, have resulted in some spurious
detections from the nominal mission simulation being cut from the comparison.

\subsection{Compton Y estimates}
\label{sec:consis_Y}

\begin{figure*}
\begin{center}
\includegraphics[angle=90,width= 0.48\textwidth]{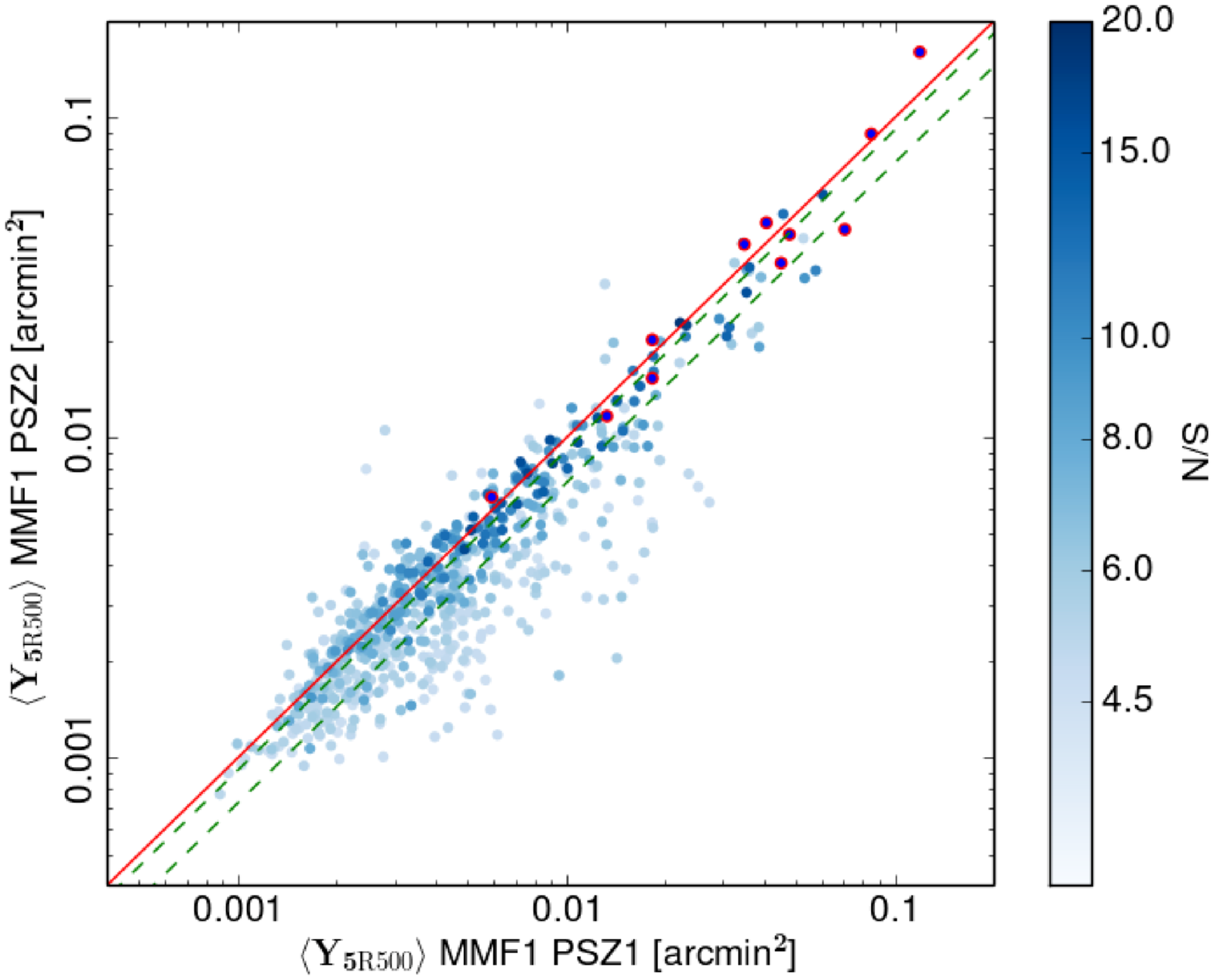}
\includegraphics[angle=90,width= 0.48\textwidth]{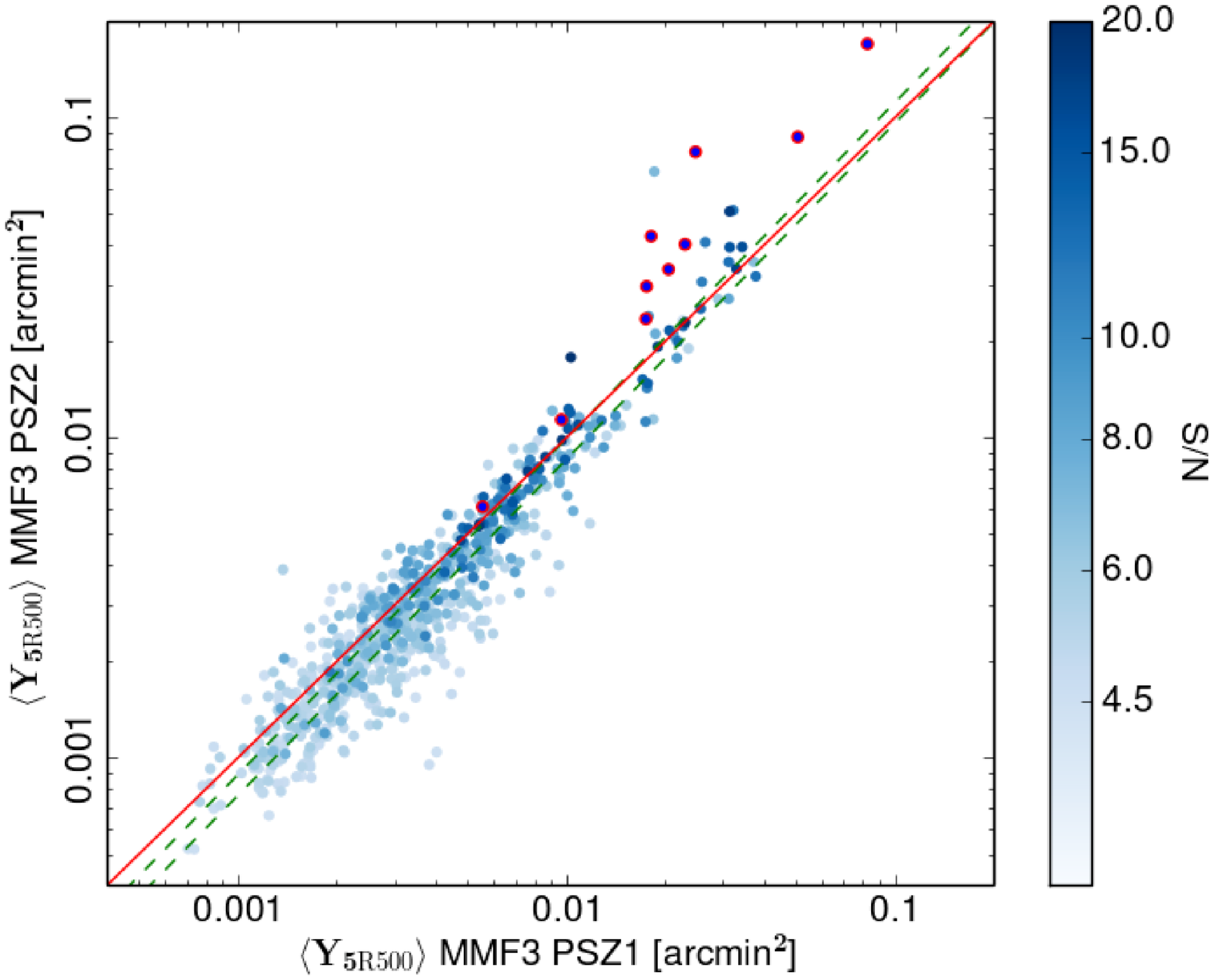}\\
\includegraphics[angle=90,width= 0.48\textwidth]{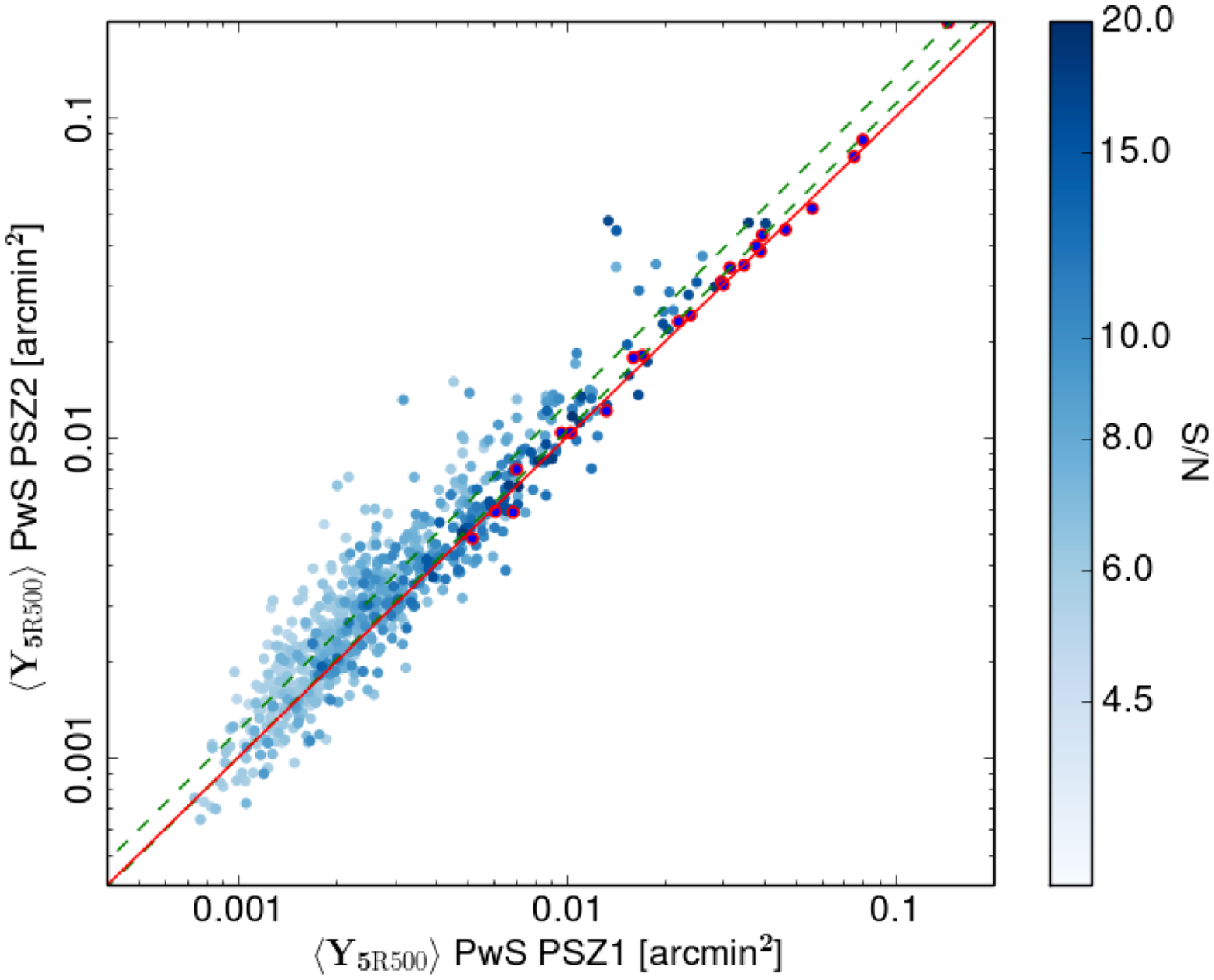}
\includegraphics[angle=90,width= 0.48\textwidth]{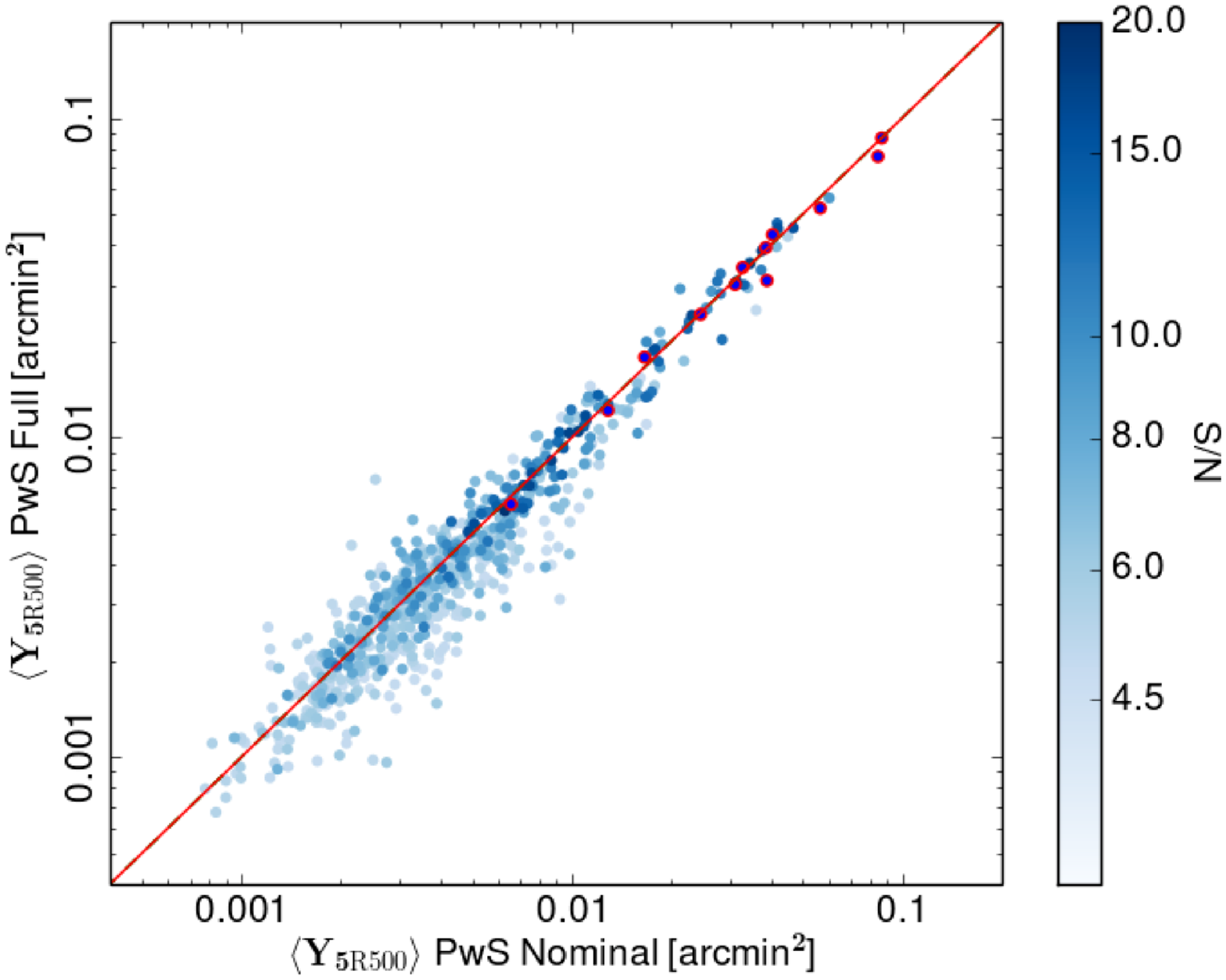}
\caption{Comparison of \yfiver~estimates from individual codes in the
  \pszone~and \psztwo.  The \yfiver~estimator is the mean of the
  $Y-\theta$ posteriors, marginalised over $\theta_{S}$ (`Y blind').
  The circled red points denote sources with $S/N > 20$.  The dashed
  green lines show the $1-\sigma$ envelope of the best-fit relations
  shown in Table~\ref{tab:consistency_Y5R500}.  \mmfone~estimates are
  shown top left, \mmfthree~top right.  \pws~estimates are shown
  bottom left.  The bottom right panel compares \pws~estimates having
  re-analysed \pszone~data using uninformative priors on \yfiver~and
  $\theta_S$.}
\label{fig:psz1_psz2_comptonY}
\end{center}
\end{figure*}

The Compton \yfiver~estimates from each code are compared to the
\pszone~estimates in Fig. \ref{fig:psz1_psz2_comptonY}.  The
\yfiver~estimates that we consider here are the mean estimates of the
\yfiver~marginal posteriors, having marginalised over the scale
radius $\theta_{s}$.

The best fit relations between the \psztwo~and \pszone~values for each
code are shown in Table \ref{tab:consistency_Y5R500}.  These were fit
using a similar procedure to the \snr~estimates discussed in the
previous section.  We assume a log-linear relation between the
estimates of the form

\begin{equation}
\log \frac{Y_{2}}{Y_{\text{piv}}} = A + \alpha \log \frac{Y_{1}}{Y_{\text{piv}}}
\label{eq:y_fits},
\end{equation}
with a log-normal intrinsic scatter $\sigma_\text{int}$ and $Y_{\text{piv}}=
3\times 10^{-4}$ arcmin$^{2}$.  We again assume a bivariate Gaussian
likelihood for the estimates, with a correlation of 0.72.

Fig.~\ref{fig:psz1_psz2_comptonY} compares the \yfiver~estimates for
each of the three detection codes.  High \snr~detections have more
consistent estimates of \yfiver.  For \mmfthree, detections at \snr$>
20$ are significantly changed due to the changes in the treatment of
these detections discussed in
Appendix~\ref{appendix:extraction_refine}.  These points are excluded
from the fit to the relation.  The scatter on the high \snr~estimates
is determined by the robustness of the noise power spectrum estimation
to small changes in the data.  For \pws, the high \snr~estimates have
particularly low scatter, due to the robust nature of the noise
estimation that accounts for Compton-Y `noise' contributed by
neighbouring clusters.

The low \snr~detections show systematic deviations for each of the
codes.  For the MMFs, these are caused by the correction of
\pszone~Eddington bias in the \psztwo~data, which is visible in
Fig.~\ref{fig:psz1_psz2_comptonY} as clouds of faint points where the
\yfiver~estimate reduces in the \psztwo.  The opposite is the case for
\pws~estimates, where the faint detections show upward deviation in
the \psztwo.  This is caused by a change in the priors: for the PSZ1,
\pws~used a power-law prior in \yfiver, which was replaced in the
\psztwo~with the uninformative flat prior, as this produced more
robust \yfiver~estimates in the posterior validation process discussed
in the previous section.  We have confirmed that \pws~behaves in the
same way as the MMFs when uninformative priors are used for both
\pszone~and \psztwo~parameter estimates (see the bottom right panel of
Fig.~\ref{fig:psz1_psz2_comptonY}).

\begin{table*}
\caption{Results of fits between \yfiver~from the \pszone~and \psztwo, following Eq.~\ref{eq:y_fits}.}
\begin{center}
\begin{tabular}{llrrr}
\hline\hline
\multicolumn{1}{c}{$Y_{1}$} & \multicolumn{1}{c}{$Y_{2}$}& \multicolumn{1}{c}{A} & \multicolumn{1}{c}{$\alpha$} & \multicolumn{1}{c}{$\sigma_\text{int}$} \\
\hline
\pszone~\mmfone~& \psztwo~\mmfone~& $-0.087 \pm 0.006$ & $1.00 \pm 0.02$ & $0.083 \pm 0.003$ \\
\pszone~\mmfthree~& \psztwo~\mmfthree~& $-0.054 \pm 0.002$ & $1.05 \pm 0.01$ & $0.054 \pm 0.003$ \\
\pszone~\pws~& \psztwo~\pws~& $0.056 \pm 0.005$ & $1.02 \pm 0.01$ & $0.068 \pm 0.003$ \\
\hline
\end{tabular}
\end{center}
\label{tab:consistency_Y5R500}
\end{table*}

To verify that the bias effects seen in
Fig.~\ref{fig:psz1_psz2_comptonY} are within expectations, we extracted
\yfiver~estimates from the half to full-mission transition simulations
described in Sec.~\ref{sec:missing_detections}.  We confirm the same behaviour
in these simulations as in the real data: low \snr~detections from the
MMFs show a correction of Eddington bias in the full mission while
\pws~low \snr~detections are affected by change from power-law to
uninformative priors in the posterior estimation and are typically
higher in the full mission.



\section{Ancillary Information}
\label{sec:ancillary_info}

\subsection{Cross-match with PSZ1}
\label{section:psz1_cross_match}

We begin the search for counterparts by conducting a cross-match with
the well-validated \pszone.  All matches within 5 arcmin of a
\pszone~detection are accepted as a true match.  Both catalogues used
this radius as the merging limit to define unique detections, both in
the merge of Cartesian patch catalogues to form an all-sky catalogue
and in the formation of the union).  This step produced no multiple
matches.

Several of our detections are clear matches with \pszone~detections at
higher radii than this, so we consider matches out to 10 arcmin, as is
the case with the X-ray and optical counterpart searches described
below.  This step produced 18 potential matches, two of which were
non-unique.  We apply a further condition to accept these
high-separation matches: that the \psztwo~\snr~be greater than the
\pszone~\snr~and be consistent with the \snr~relation determined in
Sect. \ref{sec:psz1_psz2_snr}.  For the two non-unique matches, the
nearer match was chosen both times and this match also better fit the
\snr~relation.


\subsection{X-ray information}
\label{subsec:MCXC}
We use the \textit{Meta-Catalogue of X-ray detected Clusters of
  galaxies} (MCXC, \citealt{pif11}) for the association of \Planck\ SZ
candidates with known X-ray clusters, as was done in
\citet{planck2013-p05a}. MCXC is based on the \textit{ROSAT} All Sky
Survey and complemented with other serendipitous catalogues and with
the \textit{Einstein} Medium Sensitivity Survey. It includes 1743
clusters distributed over the whole sky and provides coordinates,
redshifts and X-ray luminosity measured within $R_{500}$,
$L_{\text{X,}500}$.  The association of \Planck\ SZ candidates with
MCXC clusters follows two steps: first a positional matching between
the catalogues then a verification of the association using the
$L_{\text{x,}500}-M_{500}$ relation \citep{pra09}.  In the first step,
we looked for possible counterparts of \Planck\ SZ candidates in the
MCXC within a searching radius of 10 arcmin around the
\Planck\ position. We found one counterpart for $537$ candidates and
multiple matches for another 16 objects.  In the second step, we
verified our associations by looking at their position in the
$L_{\text{x,}500}-M_{500}$ plane (Fig.~\ref{fig:lm_plane}). For the
X-ray luminosity, we use the $L_{\text{x,}500}$ value provided in the
MCXC, while we calculate the mass from our own data, as described in
Sect.~\ref{sec:param_est}.  In Fig. \ref{fig:lm_plane}, we compare our
results with the expected $L_{\text{x,}500}-M_{500}$ relation
\citep{pif11}: we consider as good associations those whose position
in the $L_{\text{x,}500}-M_{500}$ does not differ from the expected
one by more than twice the intrinsic scatter in the relation
($\sigma_\text{int}=0.183$ \citealt{pra09}). Based on this criterion
we discarded the association with an MCXC cluster for two objects,
PSZ2 G$086.28+74.76$ and PSZ2 G$355.22-70.03$,
 both new \psztwo\ detections. \\ 

We performed a further check of the candidate counterparts, by
studying the separation between the \Planck\ and the MCXC
positions. Indeed, the relatively large search radius (10 arcmin) may
have led to spurious associations, which might have escaped our
selection on the $L_{\text{x,}500}-M_{500}$ relation.  In
Fig.~\ref{fig:sep}, we compare the separation between the \Planck\ and
the MCXC positions with two relevant angular scales: the positional
uncertainty of the \Planck\ detections $\theta_\text{err}$ ($90\%$
confidence level, provided in the catalogue) and the cluster size as
quantified by $\theta_{500}$\footnote{We calculated $\theta_{500}$
  from the mass proxy $M_{500}$, using the redshift of the MCXC
  counterpart.}. Ideally, one would keep as good counterparts those
systems where the angular separation is smaller than both
$\theta_{500}$ and $\theta_\text{err}$ (lower left quadrant in
Fig.~\ref{fig:sep}) but this would lead to a large number of rejected
matches including many objects in the \pszone. Therefore, we chose a
less conservative criterion: we excluded only those associations where
the separation is larger than both $\theta_{500}$ and
$\theta_\text{err}$ (upper right quadrant in Fig.~\ref{fig:sep}). We
thus allow the separation to exceed $\theta_{500}$ if the MCXC
counterpart falls within the \Planck\ accuracy (upper left quadrant)
and to exceed $\theta_{err}$ if it is smaller than the cluster expected
size (lower right quadrant).  We noticed that the most deviant
clusters in the latter case (with $\theta>2\theta_{500}$) are
associated with nearby clusters ($z<0.14$) with $\theta_{500}>7$
arcmin, resolved by \Planck. In this phase of the analysis, we thus
discarded the three associations to PSZ2 G$247.97+33.52$, PSZ2
G$212.93-54.04$, and PSZ2 G$209.79+10.23$
in the upper right quadrant of Fig.\,\ref{fig:sep}: for these systems, the
separations between the \Planck\ and MCXC are always
larger than 5 arcmin, which would correspond to a physical distance of
$\simeq 1 $Mpc at the redshift of the MCXC objects (all at $z>0.2$). \\

We also used the position in the $L_{\text{x,}500}-M_{500}$ and in the
separation plane (Fig.~\ref{fig:sep}) to select the most likely
counterparts for the objects where two or more MCXC clusters were
found within our search radius of 10 arcmin.  For seven out of 16
objects, one counterpart does not match the criteria described above
and we are thus left with only one good counterpart.
For six \Planck\ detections, both MCXC counterparts fullfill our
requirements. We thus rank them based on their distance from the
$L_{\text{x,}500}-M_{500}$ scaling relation and their separation in
terms of $\theta_{500}$ and $\theta_\text{err}$, and select as most
likely counterpart the one with smaller values for at least two out of
three indicators. We provide details of the other possible
counterparts in the Comments file.  In two cases, the same MCXC
cluster can be associated with two \Planck\ detections and we used the
procedure described above to select the most likely associations.

In the last step of our analysis, we checked our matching with MCXC with
the matches made in the \pszone\ catalogue: in most cases the MCXC
counterparts in the two catalogues coincide. We examined in detail the
cases where, following our selection criteria, we would have broken
the association with the MCXC counterpart which was chosen in
\pszone. This led to three restorations:
 \begin{itemize}
  \item PSZ2 G$247.97+33.52$ (\pszone\ index 842) association
    with RXCJ$0956.4-1004$ lies in the forbidden area in the
    separation plane. However, RXCJ$0956.4-1004$ (also known as A901)
    is a multi-component cluster \citep{bosch13}, and the \psztwo\
    position lies close to the position of one of the components. We
    thus decided to keep the association. 
\item PSZ2 G$302.41+21.60$(\pszone\ index 1054) and \psztwo\
  G$332.29-23.57$(\pszone\ index 1158) are both associated very
  low-redshift clusters (RXC J$1248.7-4118$ and RXC J$1847.3-6320$, both  $z < 0.015$) which are marginal outliers in the L-M plane. However,  our mass proxy estimate may be less reliable
  for local objects due to the large cluster extent and we thus decided to
  keep these associations.
 \end{itemize}


\subsubsection{Comparison to $L-M$ relation}

\begin{figure}[h]
\includegraphics[angle=0,width=0.48\textwidth]{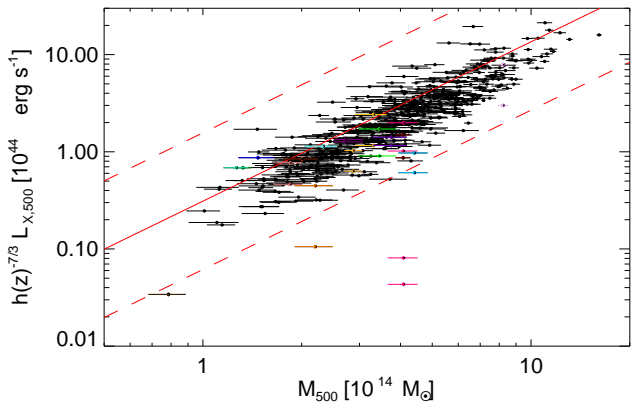}
\caption{Comparison of candidates associated with the MCXC catalogue
  with the expected $L_{\text{x,}500}-M_{500}$ scaling relation (red
  line). The parallel dashed lines identify the region of the plane
  within $2\sigma_\text{int}$ from the expected scaling relation,
  where $\sigma_\text{int}$ is the logarithmic intrinsic scatter of
  the relation we used. Black points are confirmed MCXC associations,
  while magenta squares mark the associations discarded by the L-M
  criterion. Pairs of coloured diamonds mark the two possible
  counterparts for objects with multiple associations.}
\label{fig:lm_plane}
\end{figure}

\begin{figure}[h]
\includegraphics[angle=0,width=0.48\textwidth]{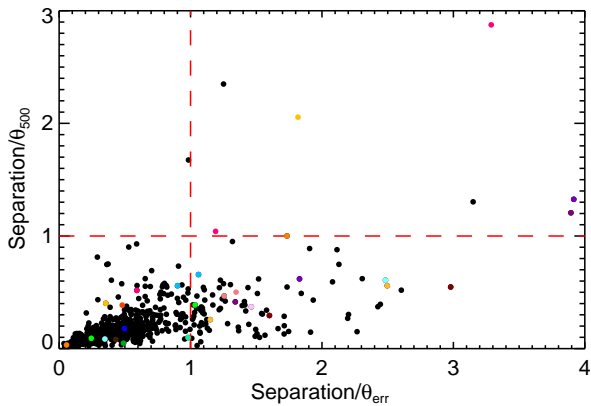}
\caption{Separation between the \Planck\ and MCXC positions in
  terms of the positional uncertainty of the \Planck\ detection and of
  the cluster $R_{500}$. The horizontal and vertical dashed lines mark
  our acceptance threshold.}
\label{fig:sep}
\end{figure}

It is interesting to note in Fig.~\ref{fig:lm_plane} that most points
lie below the expected scaling relations of \citet{pra09}, although
well within the intrinsic scatter, meaning that clusters in our
subsample are systematically under-luminous (by about 21\%, or $-0.41\sigma_\text{int}$) in X-rays
at a given mass. We recall here that this subsample (intersection of
\psztwo\ with MCXC) is not representative and thus cannot be
quantitatively compared with a well defined representative sample such
as REXCESS, for which the $L_{\text{x,}500}-M_{500}$ was derived by
\citet{pra09}.  The systematic offset observed in
Fig.~\ref{fig:lm_plane} does not contradict the good agreement between
X-ray predictions and \Planck\ measurements found with a statistical
approach in \citet{planck2011-5.2a}. It can be explained taking into
account selection effects and the scatter in the $Y-L_\text{X}$
scaling relation: when cutting in SZ \snr, clusters
with a high SZ signal (and thus a high mass) for a given X-ray
luminosity are preferentially selected. Another effect, which could
also partly contribute to the offset in Fig.~\ref{fig:lm_plane}, is
the presence of a cluster population with different X-ray properties
in the \psztwo\ sample which will be discussed in Sect.~$8.2$.

\subsection{Optical information}
\label{sec:optical_info}

We benefit from a wealth of publicly available data over the northern
sky, principally thanks to the Sloan Digital Sky Survey
\citep[SDSS,][]{york2000} that covers most of the northern
extragalactic sky with imaging in five optical bands (\emph{ugriz}).
A number of cluster catalogues have been extracted from these data
using different finding algorithms \citep{koe07, hao10, wen12}.  Among
these, the \redmapper\ catalogue \citep{rykoff2014}, published since
the \pszone\ release and containing many more clusters, has proven to
be the most useful for identifying counterparts to \Planck\ SZ
sources.  We also supplement \redmapper\ with other optically
information.

\subsubsection{RedMAPPer}
\label{sec:redmapper}

The \redmapper\ algorithm detects clusters by looking for spatial
over-densities of red-sequence galaxies.  It provides accurate
photometric redshift estimates for all sources, spectroscopic
redshifts for the brightest central galaxy (BCG) when available, and
richness estimates.  We used the proprietary \redmapper\ catalogue
(v5.10) provided by the authors and containing over 400,000 objects.

In our procedure, detailed further in Bartlett et al. 2014 (in prep.),
each \Planck\ SZ source is first matched to a maximum of three
\redmapper\ clusters falling within a radius of 10 arcmin.  They are
subsequently ranked by richness and labeled first-, second-, and
third-ranked matches.  We then calculate the \Planck\ mass proxy,
$M_{\rm sz}$, for each SZ source at the redshifts of its matched
\redmapper\ clusters.  The best \redmapper\ counterpart is then
selected based on cuts in angular separation and richness.  The
angular cuts incorporate both the \Planck\ positional uncertainty and
the physical extent of the cluster estimated from the calculated
$M_\text{sz}$.

\begin{figure}
\begin{center}
\includegraphics[angle=0,width=0.48\textwidth]{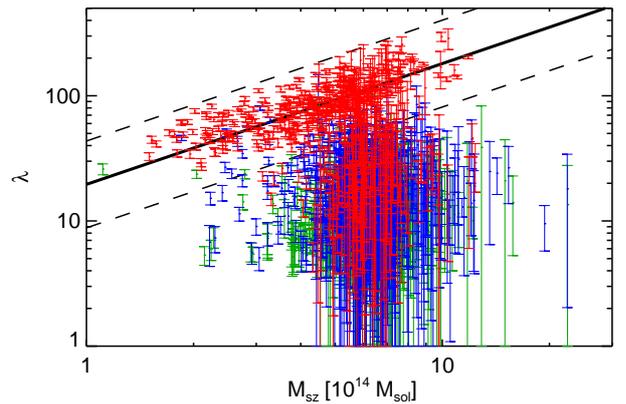}
\includegraphics[angle=0,width=0.48\textwidth]{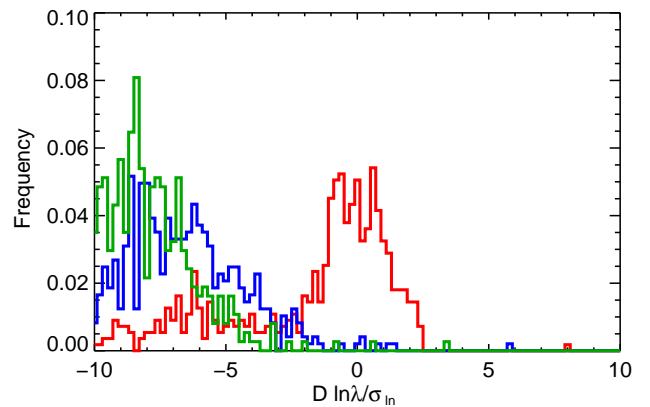}
\caption{Distribution of positional matches within a 10 arcmin radius
  in the richness-$M_{\rm sz}$ plane.  The red points in the upper
  panel represent the highest richness match, blue the second (when
  present) and green points the third-ranked richness match (when
  present).  The mean scaling law from \cite{rozo2014a} is shown as
  the solid line, with the dashed lines delineating the $\pm 3\sigma$
  band.  In the lower panel, we show the distribution of these points
  relative to the mean relation, normalized to the logarithmic
  scatter.  The red, blue and green histograms refer to the first-,
  second-, and third-ranked matches, respectively.}
\label{fig:richdist}
\end{center}
\end{figure}

These angular criteria alone would leave multiple possible
counterparts in many cases, given the high surface density of
\redmapper\ clusters.  Any ambiguity is efficiently reduced by the
richness cut, which is based on the existence of a well-defined
relation between richness, $\lambda$, and $M_{\rm sz}$.  The relation
was established by \citet{rozo2014a} on the \Planck\ 2013 SZ cluster
catalogue and is expressed as \begin{equation} \langle \ln\lambda |
  M_{\rm sz}\rangle = a + \alpha\ln(\frac{M_{\rm sz}}{M_{\rm p}}),
\end{equation}
with $a=4.572\pm0.021$, $\alpha=0.965\pm0.067$, and $M_{\rm p}=5.23\times 10^{14}\, \Msolar$.  The measured dispersion at given $M_{\rm sz}$ is $\sigl= 0.266\pm0.017$.  

In Fig.~\ref{fig:richdist} we compare the distribution of the first,
second-, and third-ranked \redmapper\ matches in the $\lambda-M_{\rm
  SZ}$ plane to this scaling relation.  The quantity
$\Dln\lambda\equiv [\ln(\lambda)-\langle\ln\lambda\rangle]/\sigl$ is
the deviation of measured richness from the expected mean.  We see
that the first-ranked matches (in red) display a prominent peak with
$\pm 3\sigl$.  This reaffirms the existence of the scaling relation
and motivates its use in defining the final \redmapper\ counterparts
for the \psztwo.

\begin{figure}
\begin{center}
\includegraphics[angle=0,width=0.48\textwidth]{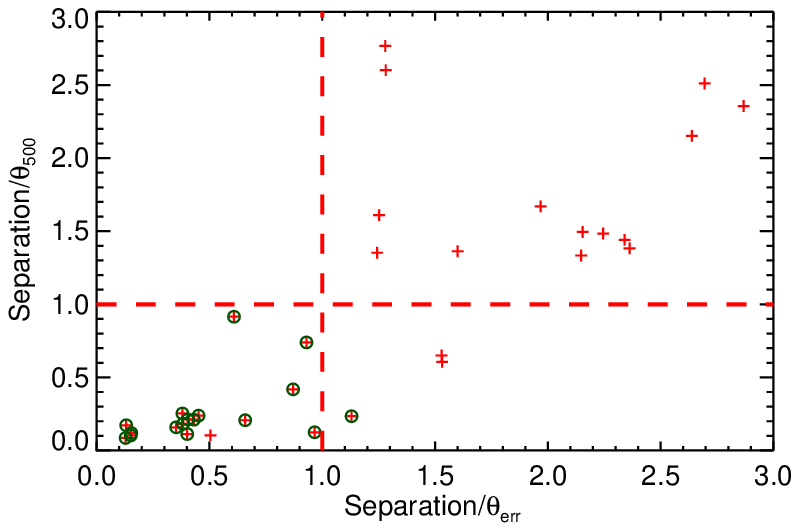}
\includegraphics[angle=0,width=0.48\textwidth]{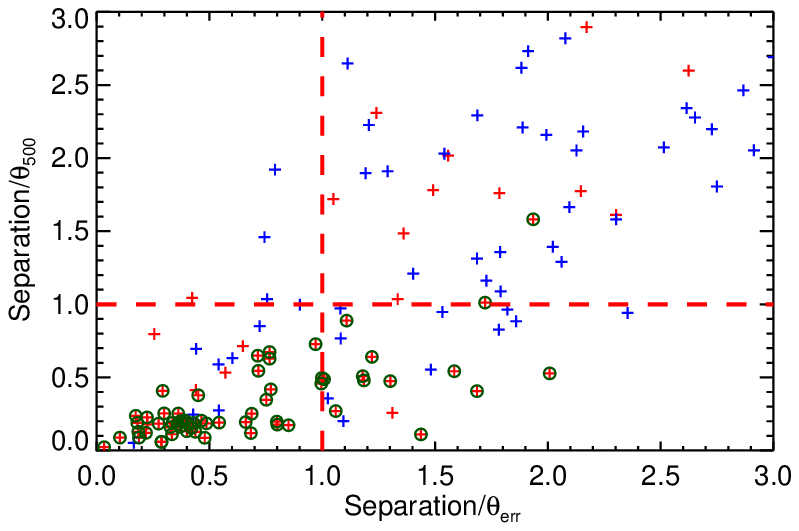}
\caption{Selection criteria for \redmapper\ counterparts.  Objects are
  plotted following the same colour scheme as in the previous figure,
  i.e., first-, second- and third-ranked matches represented by red,
  blue and green symbols, respectively.  The bands delineate the
  acceptable region defined by the angular criterion, and circled
  points indicate objects that also satisfy the richness cut
  (Criterion 1).  {\em Upper panel --} Single matches within a 10
  arcmin radius.  {\em Lower panel --} Double matches within a 10
  arcmin radius.  Note that in both panels, only first-ranked matches
  appear as good counterparts by satisfying both criteria.}
\label{fig:cuts}
\end{center}
\end{figure}

We define the best \redmapper\ counterparts with the following cuts:
\begin{enumerate}
\item $(\theta/\poserr) \le 1$ OR $(\theta/\thetafive) \le 1$;
\item $\left|\Dln\lambda\right| \le 3$;
\item When more than one object remains a possible counterpart, we
  choose the highest ranked match.
\end{enumerate}
The criteria on angular separation, $\theta$, allow objects with
centres either within \Planck's positional uncertainty or 
within the estimated size of the cluster (or both).  The second
criterion imposes the richness requirement based on the scaling
relation.  If there remain more than one possible counterpart
satisfying these two criteria, then we choose the one with the largest
richness.  These latter cases, however, deserve closer examination, in
particular for potential projection effects.

Fig.~\ref{fig:cuts} shows the distribution of \redmapper\ objects for
\Planck\ sources with single and double matches.  The dashed lines
delineate the angular cuts, while open circles identify those objects
that also satisfy the richness cut.  We see that the richness cut
effectively eliminates objects that would be accepted on angular
criteria alone.  For the double matches, there are no good
second-ranked \redmapper\ counterparts because only the highest
richness object satisfies the richness cut.

Table~\ref{tab:counterparts} summarizes the distribution of the 375
counterparts found with the above cuts.  It is grouped into sets of
columns for single-, double-, and triple-matched \Planck\ objects.
For each grouping, the total number of \Planck\ sources is given in
parentheses in the heading.  The table entries list the number of
matches in matrix format as follows: the element $(i,j)$ of a matrix
gives the number of \Planck\ sources with both $i$- and $j$-ranked
\redmapper\ counterparts; for example, six of the 438 triple matches
have a good second-ranked counterpart only, while eight have both good
first- and second-ranked counterparts.

\begin{table}[tmb]
\begingroup
\newdimen\tblskip \tblskip=5pt
\caption{Distribution of the 375 good counterparts.}
\label{tab:counterparts}
\nointerlineskip
\vskip -3mm
\footnotesize
\setbox\tablebox=\vbox{
\newdimen\digitwidth
\setbox0=\hbox{\rm 0}
\digitwidth=\wd0
\catcode`*=\active
\def*{\kern\digitwidth}
\newdimen\signwidth
\setbox0=\hbox{+}
\signwidth=\wd0
\catcode`!=\active
\def!{\kern\signwidth}
\halign{\hbox to 0.85in{#\leaderfil}\tabskip 2em&
   \hfil#\hfil\tabskip=3em&
   \hfil#\hfil\tabskip=1em&
   \hfil#\hfil\tabskip=3em&
   \hfil#\hfil\tabskip=1em&
   \hfil#\hfil\tabskip=1em&
   \hfil#\hfil\tabskip=0pt\cr
\noalign{\doubleline}
\omit&\multispan6\hfil M{\sc atches}\hfil\cr
\noalign{\vskip -3pt}
\omit&\multispan6\hrulefill\cr
\noalign{\vskip 3pt}
\omit&Single&\multispan2\hfil Double\hfil&\multispan3\hfil Triple\hfil\cr
\noalign{\vskip 3pt}
\omit\hfil R{\sc ank}\hfil&(/40)&\multispan2\hfil (/85)\hfil&\multispan3\hfil (/438)\hfil\cr
\noalign{\vskip 3pt\hrule\vskip 5pt}
1&17&58&0&283&8&0\cr
2&&&0&&6&0\cr
3&&&&&&2\cr
\noalign{\vskip 3pt\hrule\vskip 3pt}}}
\endPlancktable
\tablenote {{}} One object with three possible good matches.\par
\endgroup
\end{table}

\subsubsection{Other optical information}
\label{sec:other_optical}
 
We perform targeted searches for counterparts within the SDSS
footprint for all \Planck\ sources without good \redmapper\ matches by
applying the \redmapper\ algorithm on a case-by-case basis.  This
yielded an additional 17 counterparts and associated redshifts.
 
We add optical confirmations and redshifts from several other
sources. These include optical counterparts for \pszone\ clusters
published recently from PanSTARRs \citep{liu14} and from
\Planck\ collaboration optical follow-up observations
\citep{PIP26,planck2015-XXXV}.  We also search for counterparts in the
NED\footnote{The NASA/IPAC Extragalactic Database (NED) is operated by
  the Jet Propulsion Laboratory, California Institute of Technology,
  under contract with the National Aeronautics and Space
  Administration.} database, again removing any negated duplicate
matches from the systematic searches.  We compared the NED redshifts for
\psztwo\ matches to the redshifts from all the other ancillary
catalogues we have studied.  The NED redshifts have 88\% agreement with
these sources within $\Delta z < 0.02$.  We therefore caution that NED
redshifts should be considered the least reliable of our counterpart
assignments.  The NED associations are dominated by optical
associations with Abell \citep{abe58} and Northern Optical Cluster
Survey clusters \citep{gal03}.

Finally, we add four high-z counterparts confirmed using SDSS data and
which are discussed in Appendix~\ref{appendix:high-z}.
 
\subsection{IR information}
\label{sec:WISE}

At \Planck\ detection positions, we have searched for galaxy
overdensities in the AllWISE mid-infrared source catalogue
\citep{cutr13}. The AllWISE source catalogue includes the combined
cryogenic and NEOWISE \citep{mainzer11} observations from the
Wide-field Infrared Survey Explorer mission (WISE;
\citealt{Wri10}). The data cover the entire sky and we used the
deepest bandpasses, the 3.4~$\mu$ (W1) and 4.6~$\mu$ (W2) channels. We
predicted galaxy (W1-W2) colours from \cite{bruz03} stellar population
models, and searched for galaxy over-densities of the same colour in
successive redshift ranges from z=0.3 to z=1.5
(e.g. \citealt{papo10,mei12,mei14,stanf14}). At redshift $z<0.3$, the
contrast between red mid-infrared galaxies and the background is not
efficient for galaxy cluster detection, so we searched only in the
fields of \Planck\ detections already validated at redshift $z>0.3$,
and detections not yet confirmed or with unknown redshift.

We estimated a significance of the over-densities by comparing the
number of galaxies found in a region of co-moving diameter of 1 Mpc,
with the background galaxies found in regions of the same area. To
estimate the background density, we calculated for each redshift range
both a local background for each candidate, in a region within 15
arcmin from the \Planck\ detection, and a master background derived
from the estimators for all the \Planck\ fields.  A substantial
percentage of the \Planck\ detections ($\sim37\%$) are affected by
artefacts from bright stars in the WISE data, which compromise
meaningful assessment of the galaxy over-densities.  The bulk of these
are at low galactic latitude ($|b| < 20$).  This means that we do not
expect to reach a detection completeness of better of $60-70\%$.

After visual inspection of all detections, we have flagged our
detections in regions not affected by bright star artefacts with the
following classification: 3 - Significant galaxy overdensity detected
; 2 - Probable galaxy overdensity; 1 - Possible galaxy overdensity; 0
- No significant galaxy overdensity.  We also include these
classifications for detections in regions affected by bright star
artefacts: -1 Possible galaxy overdensity; -2 - No significant galaxy
overdensity; -3 No assessment possible.

To test our classification, and evaluate our completeness and purity,
we blindly apply our automated and visual inspection to 100 objects:
50 confirmed $z>0.5$ clusters and 50 random positions in the sky. We
show the results of this validation test in Table~\ref{wisetest}:
$59\%$ of the fields have images with bright star artefacts, including
17\% which are class -3. In class 3 and 2, we obtain a $96\%$ and
$80\%$ purity, respectively, for the validated clusters. Given the
high purity of the class 3 detections, we classify these objects as
confirmed infra-red clusters.

In Table~\ref{wiseplanck}, we show the number of WISE
\Planck\ detections in each class for the 935 \Planck\ detections with
$z>0.3$ or unknown redshift. A detailed study of the WISE detections
will be published in a separate paper (Mei et al. 2015, in prep.). 73 new clusters
have been confirmed (class 3 and not validated by other methods) by
our WISE image analysis.  A further 54 probable new clusters are
identified (class 2 and unvalidated by other methods).


\begin{table}[tmb]
\begingroup
\newdimen\tblskip \tblskip=5pt
\caption{Total, validated, and spurious detections in a blind test of
  our WISE detection classification using 50 real and 50 spurious
  fields.}
\label{wisetest}
\nointerlineskip
\vskip -1mm
\footnotesize
\setbox\tablebox=\vbox{
\newdimen\digitwidth
\setbox0=\hbox{\rm 0}
\digitwidth=\wd0
\catcode`*=\active
\def*{\kern\digitwidth}
\newdimen\signwidth
\setbox0=\hbox{+}
\signwidth=\wd0
\catcode`!=\active
\def!{\kern\signwidth}
\halign{\hbox to 0.85in{$#$\leaderfil}\tabskip 2em&
   \hfil#\hfil\tabskip=1em&
   \hfil#\hfil\tabskip=1em&
   \hfil#\hfil\tabskip=0pt\cr
\noalign{\doubleline}
\omit& \multispan3\hfil T{\sc est} D{\sc etections}\hfil\cr
\noalign{\vskip -3pt}
\omit&\multispan3\hrulefill\cr
\noalign{\vskip 3pt}
\omit\hfil C{\sc lass}\hfil&Total&Validated&Spurious\cr
\noalign{\vskip 3pt\hrule\vskip 5pt}
!3&24&23&*1\cr
!2&10&*8&*2\cr
!1&*3&*0&*3\cr
!0&*4&*0&*4\cr
-1&14&*2&12\cr
-2&28&*8&20\cr
-3&17&*9&*8\cr
\noalign{\vskip 3pt\hrule\vskip 3pt}}}
\endPlancktable
\endgroup
\end{table}


\begin{table*}[tmb]
\begingroup
\newdimen\tblskip \tblskip=5pt
\caption{WISE \Planck\ detection classification. }
\label{wiseplanck}
\nointerlineskip
\vskip -1mm
\footnotesize
\setbox\tablebox=\vbox{
\newdimen\digitwidth
\setbox0=\hbox{\rm 0}
\digitwidth=\wd0
\catcode`*=\active
\def*{\kern\digitwidth}
\newdimen\signwidth \setbox0=\hbox{+} \signwidth=\wd0
\catcode`!=\active \def!{\kern\signwidth} \halign{\hbox to
  0.85in{$#$\leaderfil}\tabskip 2em& \hfil#\hfil\tabskip=2em&
  \hfil#\hfil\tabskip=1em& \hfil#\hfil\tabskip=1em&
  \hfil#\hfil\tabskip=0pt\cr \noalign{\doubleline} \omit&
  \multispan4\hfil WISE Planck D{\sc etections}\rlap{$^{\rm
      a}$}\hfil\cr \noalign{\vskip -3pt}
  \omit&\multispan4\hrulefill\cr \noalign{\vskip 3pt}
  \omit&&&Previously\cr \omit\hfil C{\sc lass}\hfil&Total&\% of
  Sample&Confirmed&Unconfirmed\cr \noalign{\vskip 3pt\hrule\vskip 5pt}
  !3&374& 40& 301& *73\cr !2&*68& *7& *14& *54\cr !1&*55& *6& **7&
  *48\cr !0&*88& *9& *15& *73\cr -1&*42& *4& **5& *37\cr -2&*97& 10&
  *15& *82\cr -3&211& 23& *55& 156\cr \noalign{\vskip 3pt\hrule\vskip
    3pt}}} \endPlancktablewide \tablenote {{a}} For each WISE
detection class, we show the total number of \Planck\ detections and
their percentage with respect to the 935 objects with known redshift
$z>0.3$ or unknown redshift, the number of previously confirmed
\Planck\ clusters and the number of unconfirmed
\Planck\ detections.\par \endgroup
\end{table*}

\subsection{SZ information}
\label{subsec:ACT_SPT}
We searched for counterparts of our \Planck\ detections using
catalogues obtained with other SZ surveys, such as the South Pole
Telescope (SPT) and the Atacama Cosmology Telescope (ACT), and update
the list of SZ confirmations by direct follow-up with the Arc-minute
Micro-kelvin Interferometer (AMI). \\

\subsubsection{SPT}
For SPT, we refer to the recently released catalogue \citep{blee14},
extracted from the full 2500 deg$^2$ SZ survey.  It contains 677 SZ
detections, of which 516 have been confirmed as clusters, through
optical and near-IR observations. The catalogue contains the
photometric redshifts (spectroscopic when available) and mass estimate
of the confirmed clusters. More specifically, we chose to use the
``Fiducial Cosmology'' Catalogue provided by the SPT collaboration, to
be consistent with the cosmological parameters used in the present
paper.  We performed a two step matching process as described in
Sect.~\ref{subsec:MCXC}: a positional match within 10 arcmin, leading
to 89 single and five double matches, which we verified by comparing
our mass estimate with the one provided in the SPT catalogue. The mass
estimates are usually consistent at better than $3 \sigma$, except for
one detection, PSZ2 G$249.87-21.65$, where they differ at
$3.5\sigma$. We note, however, that the error bars on the
\Planck\ mass reflect only the statistical error on the probability
contours and do not consider the uncertainties nor the intrinsic
scatter in the scaling relation used to break the
degeneracy. Moreover, as discussed also in the SPT case by
\citet{blee14}, the use of a fixed scaling relation and of a fiducial
cosmology results in an underestimation of the statistical and
systematic uncertainties in both datasets. Therefore, we decided to
keep the SPT counterparts for PSZ2 G$249.87-21.65$, which is also
associated in the SPT catalogue to the same MCXC cluster (RXC
J$0628.8-4143$) as in our matching.  We checked the position of the
matches in the separation plane: none of the single matches have been
discarded in this way, while in four out of five multiple matches, one
of the counterparts was excluded following this criterion. In the
remaining match, two possible counterparts are allowed and we selected
as the most likely the one with a smaller mass difference and a
smaller separation in terms of $\theta_{500}$. \\

We observe a systematic difference between the masses we derived from
\Planck\ data and those provided in the SPT catalogue.  The mass is a
derived quantity which requires scaling information to be assumed
before it can be calculated from the SZ signal measured by either
instrument.  Comparison of the SZ observables is complicated by the
different scales probed by each instrument: \Planck\ is sensitive to
the cluster outskirts while SPT is sensitive to the core regions.  Any
comparison necessarily requires model extrapolation, which is
complicated further by the different pressure models used in the two
measurements.  A robust comparison will require a joint analysis of
the data, which is beyond the scope of this paper.




  \subsubsection{ACT}
 
For ACT, we use the catalogue published in~\cite{has13}, which
contains both the most recent ACT detections and an update of the 23
\cite{mar11} detections.  32 \psztwo\ detections match with ACT
clusters in a 10 arcmin radius. 28 have ACT UPP-based SZ masses
consistent with \Planck\ $M_\text{SZ}$ (less than $3 \sigma$
deviation). Four have more than $3 \sigma$ deviation: PSZ2 G053.44-36.25
(ACT-CLJ2135.1-0102)  and PSZ2 G130.21-62.60 (ACT-CLJ0104.8+0002)
are actually considered as good matches by our
 \redmapper\ association, PSZ2 G262.27-35.38 (ACT-CLJ0516-5430)
  is also a good match for
 our MCXC association. The last one, PSZ2 G265.86-19.93 (ACT-CLJ0707-5522), is a good
 match in the \pszone.  We decided to leave blank the ACT field of the
 \psztwo\ catalogue for these four clusters but, given the
 uncertainties on the mass determination and associated errors, we did
 not break the corresponding \redmapper, MCXC and
 \pszone\ association.

\subsubsection{AMI}
Following the ongoing follow-up observations of the \Planck\ cluster
candidates, a total of $161$ clusters with $4.5<$\snr $<20$ were
observed with AMI. The detection significance is then characterised by
calculating of the natural logarithm of the Bayes factor, $\ln B_{10}$,
\begin{equation}\label{eq:logbayesfactor}
{\rm ln}B_{10}=\Delta {\rm ln} {\mathcal{Z}_{10}}={\rm ln} {\mathcal{Z}_1}-{\rm ln}
{\mathcal{Z}_0},
\end{equation}
where ${\rm ln} {\mathcal{Z}_1}$ and ${\rm ln} {\mathcal{Z}_0}$ are
the natural logarithm of the Bayesian evidence for model $H_1$ and
$H_0$ respectively.  Model $H_1$ accounts both for the cluster signal
and the contribution from radio sources while $H_0$ only takes into
account the radio source environment. Further details of the AMI
observations, of the Bayesian methodology and of the modelling of
interferometric SZ data, primordial CMB anisotropies, and resolved and
unresolved radio point sources, as well as of the criteria used to
categorise clusters are given in \cite{perr14} and references therein.

In this context the detection significance for the $161$ clusters is
described in Table~\ref{table:ami}.  132 of the 161 AMI-confirmed
\Planck\ clusters are included in the \psztwo.
%
%

\begin{table}[tmb]
\begingroup
\newdimen\tblskip \tblskip=3pt
\caption{AMI scale for an interpretation of the detection significance of the
Planck cluster candidates.}
\label{table:ami}
\nointerlineskip
\vskip -3mm
\footnotesize
\setbox\tablebox=\vbox{
\newdimen\digitwidth
\setbox0=\hbox{\rm 0}
\digitwidth=\wd0
\catcode`*=\active
\def*{\kern\digitwidth}
\newdimen\signwidth
\setbox0=\hbox{+}
\signwidth=\wd0
\catcode`!=\active
\def!{\kern\signwidth}
\halign{\hbox to 1.6in{#\leaderfil}\tabskip 2em&
   \hfil$#$\hfil\tabskip=1em&
   \hfil#\hfil\tabskip=0pt\cr
\noalign{\doubleline}
\omit&\cr
\noalign{\vskip -3pt}
\omit\hfil Category\hfil&\ln{B_{10}}&$N$\cr
\noalign{\vskip 3pt\hrule\vskip 5pt}
Clear detection&     {\rm ln}B_{10}\ge3&     102\cr
\noalign{\vskip\tblskip}
Moderate detection&  0\le {\rm ln}B_{10}<3&  *30\cr
\noalign{\vskip\tblskip}
Non-detection&       -3 \le {\rm ln}B_{10}<0&*25\cr
\noalign{\vskip\tblskip}
Clear non-detection& {\rm ln}B_{10}\le-3&    **4\cr
\noalign{\vskip 3pt\hrule\vskip 3pt}}}
\endPlancktable
\endgroup
\end{table}

\subsection{Redshift compilation}
\label{sec:redshift_compilation}

We provide a single redshift estimate for each detection where at
least one redshift is known for a matched counterpart.  In many cases,
we have multiple estimates per detection.  This section discusses the
compilation of redshift information and how redshifts are assigned in
the final catalogue.  Confirmation statistics and final assigned
redshift numbers are summarised in Table~\ref{val_table}.

Initial redshift estimates are taken from the \pszone\ catalogue
redshift compilation, given the matching in
Sect.~\ref{section:psz1_cross_match}.  We include new follow-up
results from the Planck collaboration \citep{PIP26, planck2015-XXXV},
which include spectroscopic updates to \pszone\ photometric redshifts and new
confirmations of \pszone\ detections.  We also include external
updates to \pszone\ counterpart redshifts from the NED and SIMBAD
databases \citep{planck2015-XXXVI}.

After these steps, we cycle through priority levels in our systematic
counterpart searches, in the following order of priority: MCXC,
\redmapper\, ACT and SPT.  We compare the updated \pszone\ and MCXC
redshifts with the redshifts from \redmapper\, where available, and
prioritise \redmapper\ redshifts highest amongst available photometric
redshifts.  We test spectroscopic redshifts at $z>0.1$ for consistency
within $\Delta z < 0.03$ and $\Delta z / z <0.1$ of the
\redmapper\ photo-z.  Any discrepancies are considered on an
individual basis.  In a small number of cases, we choose the
\redmapper\ redshift.  We also reject a small number of counterpart
assignments where that counterpart is a bad match in \redmapper.
These cases are discussed in Appendix
\ref{appendix:redshift_differences}.

The common sample between the \psztwo\ and each of the external
samples is denoted in the catalogue.  After the systematic searches,
we assign any remaining unconfirmed clusters to database counterparts
where available.  If a \pszone\ match assigns to the same counterpart
as one of the negated counterparts from the systematic searches, where
the possible counterpart violated the consistency criteria, then the
\pszone\ match is also negated.

\begin{table*}[tmb]
\begingroup
\newdimen\tblskip \tblskip=5pt
\caption{Summary of ancillary information.  The highest available
  priority redshift source, following the ordering in the Priority
  column, provides the reference confirmation and redshift.  When two
  priorities are given, the first number pertains to spectroscopic
  redshifts and the second number to photometric redshifts.  The
  \psztwo\ contains 1203 confirmed clusters, of which 289 are
  Planck-discovered. 87 of these are clusters newly identified in this
  paper: 73 are confirmed by WISE, eight are new identifications in
  SDSS data 
  and six are confirmed by AMI.}
\label{val_table}
\nointerlineskip
\vskip -1mm
\footnotesize
\setbox\tablebox=\vbox{
\newdimen\digitwidth
\setbox0=\hbox{\rm 0}
\digitwidth=\wd0
\catcode`*=\active
\def*{\kern\digitwidth}
\newdimen\signwidth
\setbox0=\hbox{+}
\signwidth=\wd0
\catcode`!=\active
\def!{\kern\signwidth}
\halign{\hbox to 1.25in{#\leaderfil}\tabskip 2em&
     \hfil #\hfil\tabskip=1em&
   \hfil#\hfil\tabskip=2em&
   \hfil#\hfil\tabskip=1em&
   \hfil#\hfil\tabskip=1em&
   \hfil#\hfil\tabskip=2em&
        #\hfil\tabskip=0pt\cr
\noalign{\doubleline}
\omit \hfil Confirmation \hfil&&&Joint&Reference&Planck-&\omit\hfil Redshift\hfil\cr
\omit  \hfil source \hfil& Validation &Priority&sample size&confirmations&discovered&\omit \hfil reference\hfil\cr
\noalign{\vskip 3pt\hrule\vskip 5pt}
 ENO follow-up&   10&            1/5& \dots& *22& *18& \citealt{planck2015-XXXV}\cr
RTT follow-up&  11&             1/5& \dots& *45& *31& \citealt{PIP26}\cr
PanSTARRs&      12&        6& \dots& *16& *16& \citealt{liu14}\cr
 \redmapper\  non-blind& 13& \dots& \dots& *17& **5& This paper: Sect.~\ref{sec:other_optical}\cr
SDSS high-z&    14&    \dots& \dots& **4& **4& This paper: Appendix \ref{appendix:high-z}\cr
AMI fu&     15&        \dots& \dots& *10& *10& \dots\cr
WISE&      16&        \dots& \dots& *73& *73& \dots\cr
PSZ1 2013&  20&          1&   782& 348& 125& \citealt{planck2013-p05a}\cr
MCXC&    21&               2&   551& 447& **0& \citealt{pif11}\cr
SPT&       22&           4/5&   *94& *39& **4& \citealt{blee14}\cr
ACT&         23&         4/5&   *28& **1& **0& \citealt{has13}\cr
 \redmapper\  &   24&           3&   374& 122& **2& \citealt{rykoff2014}\cr
Updated \pszone\ & 25& 1& \dots & *19& **0& \citealt{planck2015-XXXVI} \cr
NED&          30&          7& \dots& *40& **1& Various\cr
\noalign{\vskip 3pt\hrule\vskip 3pt}}}
\endPlancktablewide
\endgroup
\end{table*}


\section{Sample Properties}
\label{sec:sample_prop}

\subsection{Mass and redshift properties}
\begin{figure*}
\begin{center}
\includegraphics[angle=0,width=0.48\textwidth]{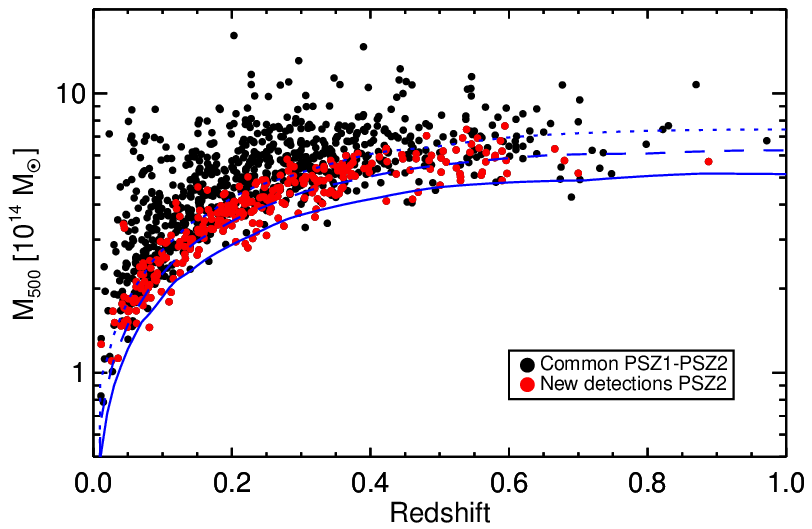} 
\includegraphics[angle=0,width=0.48\textwidth]{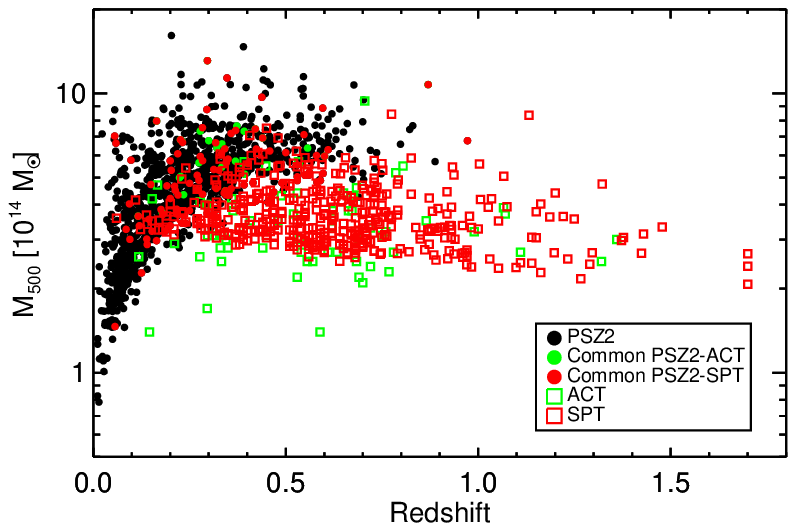} 
\caption{\emph{Left panel:} Distribution of the 1094 \psztwo\ clusters
  with counterparts with known redshift in the $M_{500} - z$ plane.
  New \psztwo\ detected clusters are indicated with red dots, while
  commmon \pszone\ and \psztwo\ clusters are indicated by black
  dots.  The solid, dashed and
  dotted lines indicate respectively the 20\%,50\%
  and 80\% survey completeness contours for the \psztwo. \emph{Right panel:}
  Distribution of the \psztwo\ clusters with associated redshift in
  the $M_{500}-z$ plane compared to the SPT (Bleem et al 2014) and ACT
  (Hasselfield et al 2013) catalogues. Black circles represent
  \psztwo\ clusters, while red and green filled circles mark common
  SPT/\psztwo\and ACT/ \psztwo\ clusters, respectively. The remaining
  SPT and ACT clusters not detected by \Planck\ are shown with red and
  green empty squares. }
\label{fig:msz_z_distr}
\end{center}
\end{figure*}

We discuss here the distribution of \Planck\ SZ-selected clusters in
the mass-redshift ($M_{500}-z$) plane, using the mass proxy derived
with scaling relations as discussed in
Sect.~\ref{sec:ancillary_info}. For 1094 detections with known
redshifts in the \psztwo\ catalogue (Sect.~\ref{sec:ancillary_info}),
we show in Fig.~\ref{fig:msz_z_distr} their position in the
$M_{500}-z$ plane, compared with the expected completeness function $C(M_{500},z)$
 of our survey (we show the 20\%, 50\% and 80\% completeness levels). 
 These curves indicate the points in the $M_{500}-z$ plane at which clusters have
$C$\% chances to be detected.  They were computed for the full survey area.  The red points
in Fig.~\ref{fig:msz_z_distr} show the 298 new \psztwo\ confirmed
detections, with redshifts, that were not found in the previous version of the
catalogue.  The black points show the common \pszone-\psztwo\ detections. 

We stress that the $M_{500}-z$ distribution in
Fig.~\ref{fig:msz_z_distr} cannot be considered as fully representative
of the \Planck\ SZ selection, since it reflects the biases due to the
non-uniform knowledge of redshifts over the sky in the ancillary
information we used (Sect.~\ref{sec:ancillary_info}). For instance, we
have an extensive redshift information in the sky area covered by the
SDSS survey thanks to the \redmapper\ catalogue \citep{rykoff2014} but
not in the remaining part of the sky.  The incomplete redshift
information can also explain the rarity of new detections in the
\psztwo\ catalogue with respect to \pszone\ at high redshift: at
$z>0.6$ we have 36 objects, but only four new \psztwo\ detections.  We
note however that most of the \pszone\ clusters in this redshift range
were not present in existing catalogues but they were confirmed as
clusters and their redshift was measured thanks to the massive follow
up campaign with optical and X-ray telescopes which was undertaken by
the \Planck\ Collaboration for \pszone\ candidates
(\citealt{planck2013-p05a} and references therein) and which continued
also after the 2013 release (\citealt{PIP26,planck2015-XXXV}). Since
a similar observational campaign has not yet been possible for new
\psztwo\ detections, we could not populate further the high-mass
high-z part of the $M_{500}-z$ plane.

The new \psztwo\ confirmed detections (red points in
Fig.~\ref{fig:msz_z_distr}) are mostly low-mass objects close to the
detection limit of the survey.  The mean mass of confirmed clusters
over the whole redshift range in the \psztwo\ is $4.82 \times
10^{14}\ M_\odot$, which is lower than in the \pszone\ ($5.12\times
10^{14}\ M_\odot$). The common sample of 795 objects contains the higher
mass clusters detected by both surveys, with mean mass $5.16\times 10^{14}\ M_\odot$. 
This is expected, since the common sample contains none of the new low-mass \psztwo\
 detections and none of the missing low \snr\ \pszone\ detections, discussed in Sect.~\ref{sec:missing_detections},
  which were likely to have been low mass.

\begin{figure}
\begin{center}
\includegraphics[angle=0,width=0.48\textwidth]{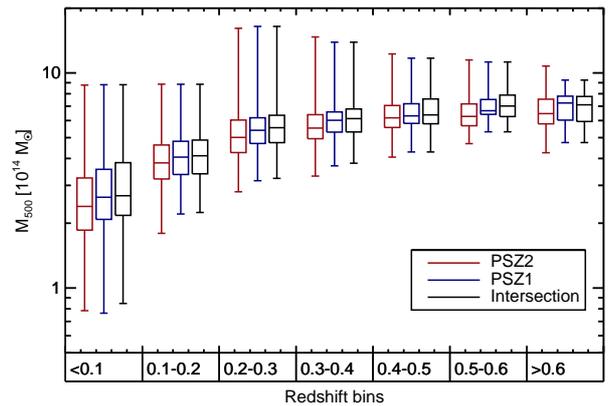} 
\caption{Box-and-whisker diagrams showing the mass distribution of the \psztwo\ (red), \pszone\ (blue) and their intersection (black) sample in seven redshift bins. The bottom and the top of the boxes represent the first and third quantile of the data, while the band inside the box shows the median (i.e. the second quantile). The ends of the whiskers mark the minimum and maximum of the data. }
\label{fig:msz_boxplots}
\end{center}
\end{figure}

This is also shown in
Fig. \ref{fig:msz_boxplots}, where we compare the mass distribution of
the confirmed clusters in the \psztwo, the \pszone\ and their common
sample, for several redshift bins. The median mass and the first and
third quantiles are always lower for the \psztwo\ than for the \pszone\ and
 the common sample, showing that we are significantly expanding the
sample towards lower masses.

Fig.~\ref{fig:msz_z_distr} also shows a comparison of the SZ selected
samples from the \Planck, ACT and SPT surveys.  \Planck~tends to detect the
rarest high-mass clusters observed at high-redshift in these
partial-sky surveys and provides a complementary clean SZ selection at
lower redshifts, where the \Planck~frequency range provides sufficient
information to disentangle the SZ signal of large clusters from the
background.

\subsection{X-ray underluminous clusters}
\label{sec:xray_underlum}

The presence of a bright cool-core, characterised by a peaked surface
brightness profile, has been shown to bias X-ray flux selected cluster
samples in favour of peaked, relaxed objects with respect to
morphologically disturbed systems \citep{eck11}.
In contrast, SZ selected samples have produced more disturbed systems
than expected, with SZ discovered clusters typically lying on the
lower end of the mass-luminosity relation \citep{planck2011-5.1b}.
There has also been much interest in the possible existence of
severely X-ray under-luminous clusters, with several authors
identifying potential systems (eg: \citealt{bow97,pop07,die09,tre14}
and references therein) and suggesting a model where these clusters
are dynamically young objects, still undergoing accretion and mergers
and yet to reach equilibrium.  An alternative suggestion is that line-of-sight
structures may bias mass and richness estimates high relative to the X-ray luminosity \citep{bow97,gil15}.

However, the reported under-luminosity
of these objects is disputed. \cite{andr11} note that the
under-luminosity is often claimed relative to biased scaling
relations, and that the significance of the under-luminosity is
amplified due to underestimation of the true scatter in the relation.

In the SDSS area, the majority of \Planck\ detections possess
counterparts in the \redmapper\ catalogue, with redshifts and optical
richness estimates.  We construct a test sample from the
\Planck-\redmapper\ intersection at low redshift, $z<0.2$.  This
sample of 148 clusters can be expected to be detectable in the
\emph{ROSAT} maps.  While \redmapper\ is not complete for the
\Planck\ mass ranges at these redshifts, it allows us to construct a
sample that is independent of any X-ray selection effects and
therefore well-suited for finding under-luminous clusters, as any
biases in selection will affect both normal and under-luminous X-ray
cluster alike.

For each of these clusters, we calculate the X-ray count-rate using a
growth-curve analysis on the \emph{ROSAT} $0.5-2.4$ keV band maps
following \cite{boh00}.  We derived the count-rate and its upper and
lower limit from the growth curve at $\theta_{500}$, which we calculated
from the \Planck\ mass proxy. We then converted the RASS count rates
into flux in the $0.1-2.4$ keV energy range, using an absorbed thermal
model, where we used the galactic absorption at the \Planck\ position,
the redshift of the \redmapper\ counterpart and the temperature
derived from the mass proxy through the M-T scaling relation by
\cite{arn07}. We then converted fluxes into luminosity in the
$0.1-2.4$ keV channel.  We found good agreement with the reported MCXC
values of $L_{500}$ for those clusters also present in the MCXC.


\begin{figure*}
\begin{center}
\includegraphics[angle=0,width=0.48\textwidth]{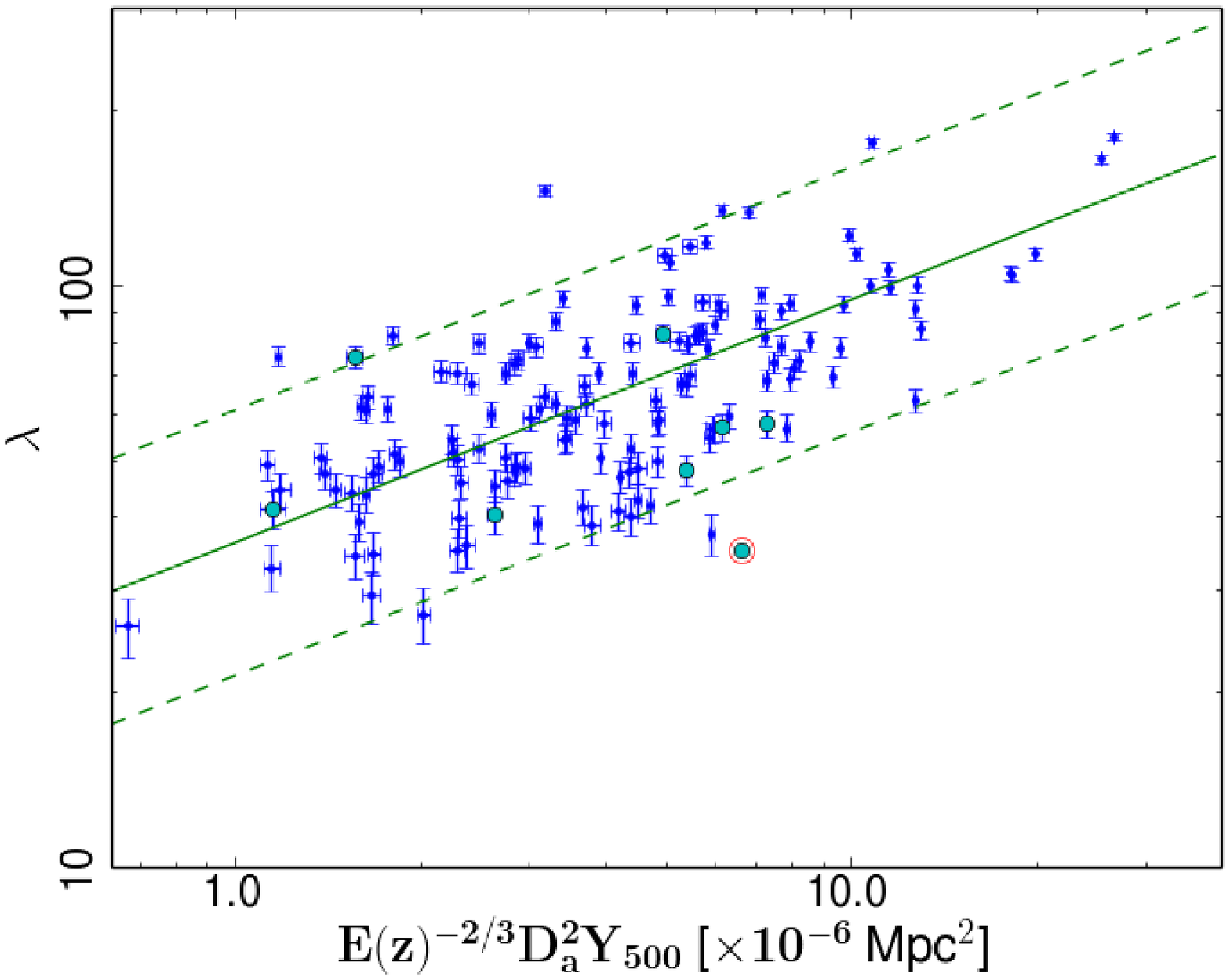}
\includegraphics[angle=0,width=0.48\textwidth]{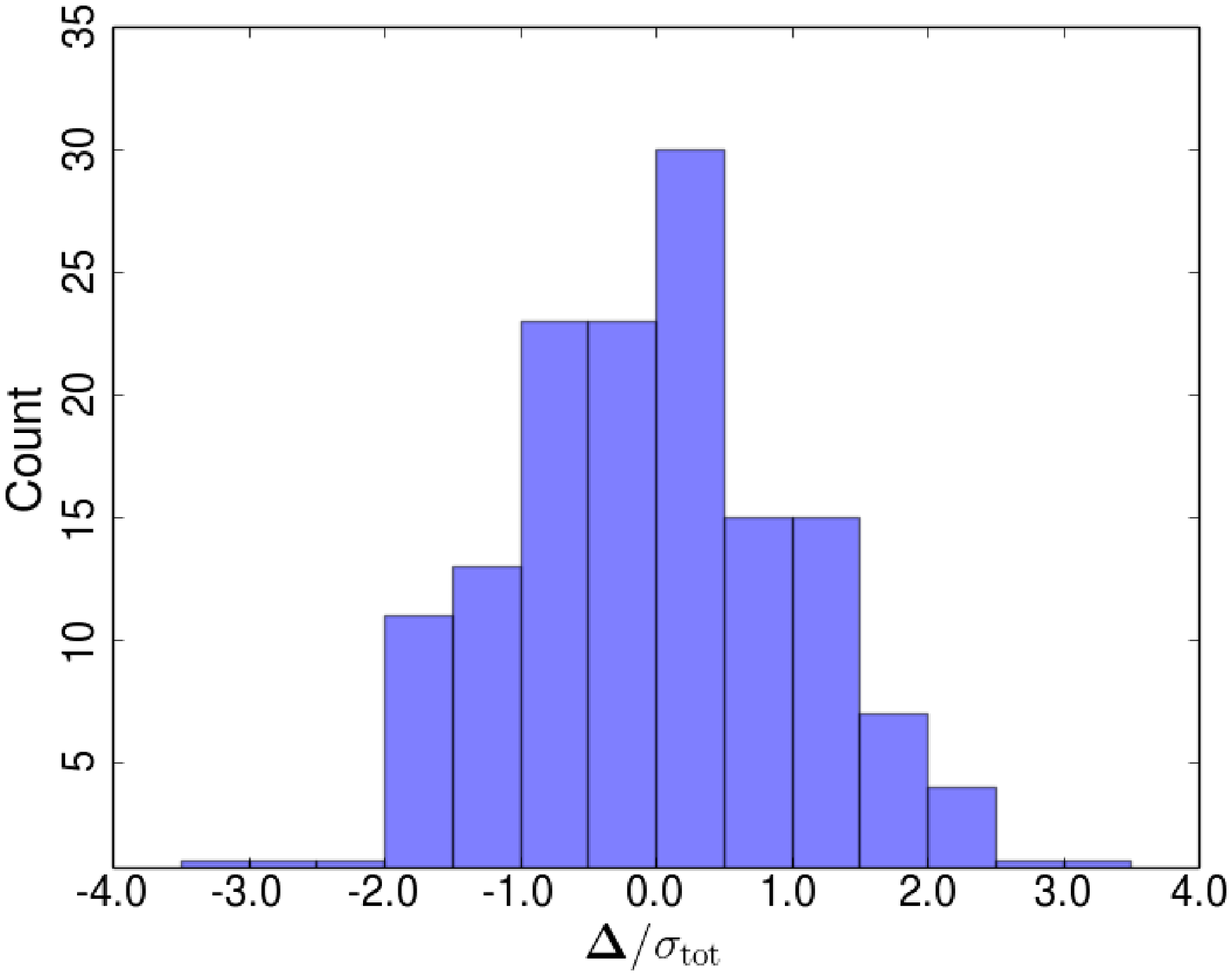}\\
\includegraphics[angle=0,width=0.48\textwidth]{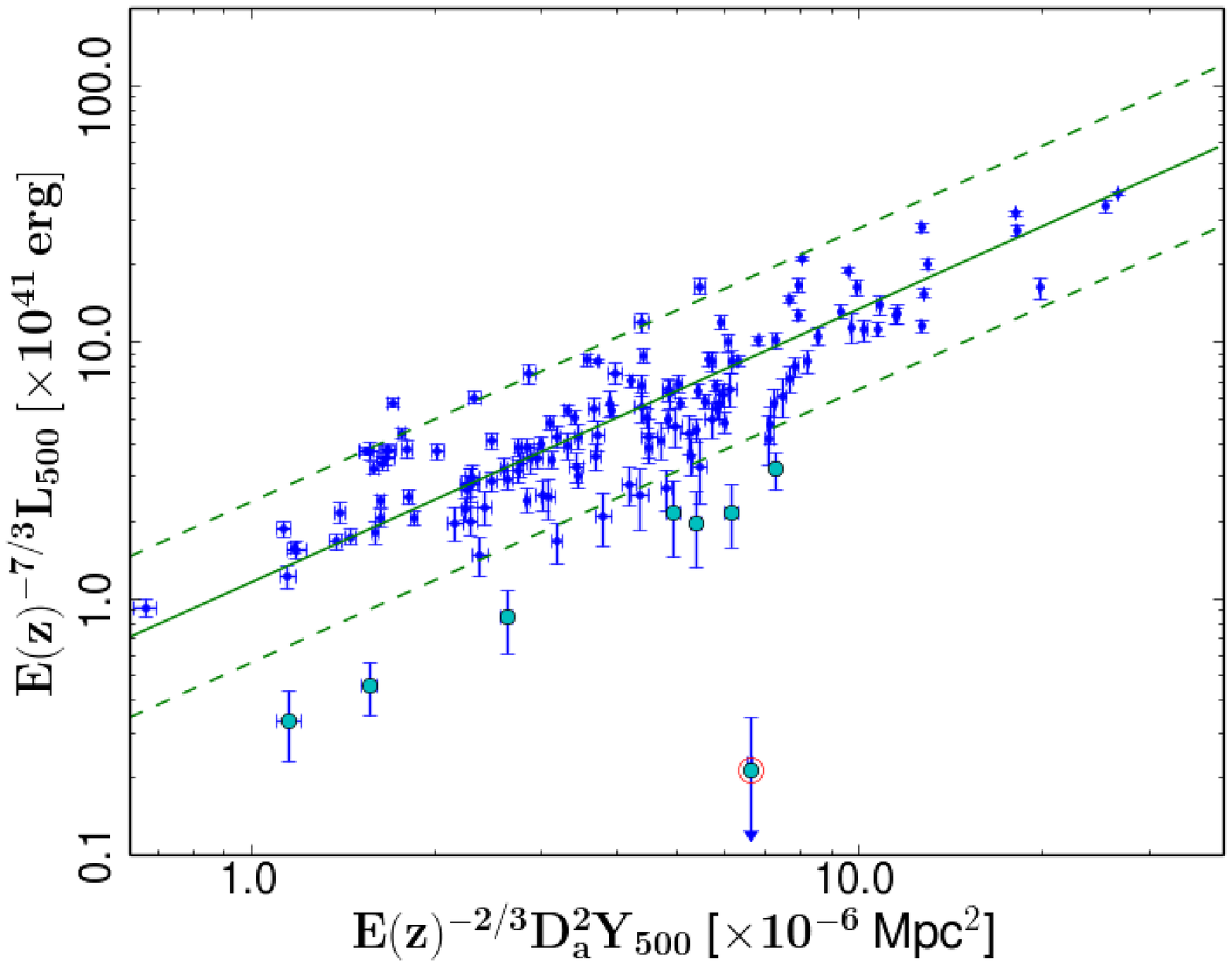}
\includegraphics[angle=0,width=0.48\textwidth]{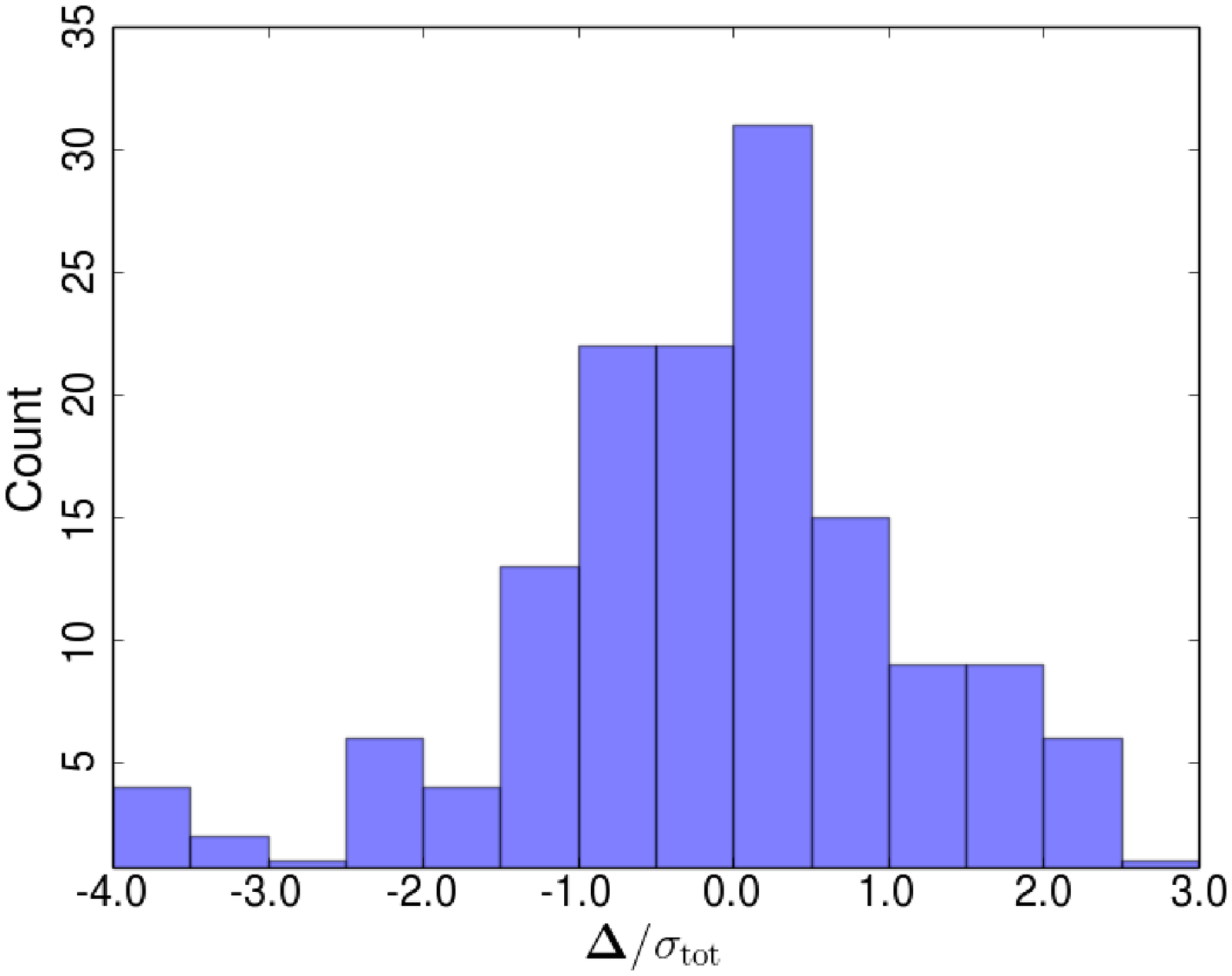}
\caption{Properties in the \yfive$-\lambda$ (top panels) and
  \yfive-$L_{500}$ (bottom panels) planes for the
  \Planck-\redmapper\ sample at $z<0.2$.  Under-luminous candidates
  are denoted with cyan circles in the scatter plots to the left,
  which also show the best fit relation and the dispersion
  $\pm2\sigma_\text{tot}$. The circled red point is a cluster with
  contaminated Y signal.  The right plots show the histograms of
  orthogonal deviation $\Delta_{\perp}$ for each relation.}
\label{fig:xray_underlum}
\end{center}
\end{figure*}

We then searched for outlier clusters from the \yfive-$L_{500}$ and
\yfive$-\lambda$ relations, using the \yfive\ calculated following
Sect.~\ref{section:Msz}.  In both cases, we find the best-fit relation
for our sample using the BCES algorithm.  To exclude outliers from the
fits, we clipped objects with orthogonal residual
$|r_{\perp}|>2.5\sigma_\text{tot}$ from the best-fit relations, where
$\sigma_\text{tot}$ is the raw scatter around the relation derived
from the median-absolute deviation.  We then iterated the
sigma-clipping process until converged.

The top panels of Fig.~\ref{fig:xray_underlum} show the best-fit
relations and their $\pm2\sigma_\text{tot}$ scatters between
\yfive\ and \redmapper\ richness $\lambda$, and between \yfive\ and
$L_{500}$, for the test sample of 148 \Planck-\redmapper\ clusters.
The bottom panels of Fig.~\ref{fig:xray_underlum} show the histograms
of $\Delta$, normalised by $\sigma_\text{tot}$.

The points highlighted with larger cyan circles denote clusters with
$L_{500}$ more than 2.5$\sigma_\text{tot}$ below the best-fit
relation.  These clusters are under-luminous in X-rays for their
\yfive.  However, their \yfive\ estimates are consistent with their
optical richness and do not lie preferentially beneath the relation.
The consistency of optical and SZ mass proxies suggest either that
these clusters are under-luminous for their mass, or that both the $Y_{500}$ 
and $\lambda$ estimates are biased high.



One exception to this is PSZ2 G127.71-69.55, 
discrepant with both relations: at $10\sigma_\text{tot}$ for
\yfive$-L_{500}$ and $3\sigma_\text{tot}$ for \yfive$-\lambda$.  It is
circled in red in Figure~\ref{fig:xray_underlum}.  This is the only
cluster in the \Planck-\redmapper\ sample with a poor Q\_NEURAL flag
and is either a failed \redmapper\ match (lying close to the matching
threshold in the $M_\text{sz}-\lambda$ plane) or a cluster with a
severely IR contaminated spectrum for which the
\Planck\ \yfive\ estimate is likely overestimated.  We therefore
remove this from the list of under-luminous candidates.

We expanded the search for under-luminous clusters across the whole
sky, testing all clusters with $z<0.2$ against the mean
\yfive$-L_{500}$ relation determined above.  This expanded the number of
under-luminous candidates to 22.  These objects warrant follow-up
observations to determine their dynamical state and to search for line-of-sight
structures that may bias high both the \yfive\ and optical richness. They are flagged in the
\textsc{COMMENT} field of the catalogue.  They should be of interest in
understanding the systematic differences between SZ and X-ray
selection.

\section{Summary}
\label{sec:conclusions}


\begin{table*}[tmb]
\begingroup
\newdimen\tblskip \tblskip=5pt
\caption{Counterpart summary for \psztwo\ compared to \pszone.  Common
  samples are defined as those \psztwo\ detections with the given
  property that has a counterpart with that property in the
  \pszone\ 2015.  The intersection comprises those detections common
  to all three detector codes.  Low reliability candidates possess a
  poor neural-network quality assessment flag.  In the \pszone, low
  reliability candidates possess the lowest external quality
  assessment flag.  SZ clusters denote clusters with SZ detections in
  ACT or SPT. \pszone\ 2013 refers to the 2013 release of the
  catalogue \citep{planck2013-p05a}, and \pszone\ 2015 to a recent
  addendum updating the counterpart information of the catalogue
  \citep{planck2015-XXXVI}.}
\label{tab:counterpart_summary}
\nointerlineskip
\vskip -1mm
\footnotesize
\setbox\tablebox=\vbox{
\newdimen\digitwidth
\setbox0=\hbox{\rm 0}
\digitwidth=\wd0
\catcode`*=\active
\def*{\kern\digitwidth}
\newdimen\signwidth
\setbox0=\hbox{+}
\signwidth=\wd0
\catcode`!=\active
\def!{\kern\signwidth}
\halign{\hbox to 1.2in{#\leaderfil}\tabskip 2em&
   \hfil#\hfil\tabskip=0.5em&
   \hfil#\hfil\tabskip=2em&
   \hfil#\hfil\tabskip=2em&
   \hfil#\hfil\tabskip=0.5em&
   \hfil#\hfil\tabskip=0pt\cr
\noalign{\doubleline}
\omit\hfil Sample\hfil&\pszone\ 2013&\pszone\ 2015&\psztwo&Common&New \psztwo\cr
\noalign{\vskip 3pt\hrule\vskip 5pt}
Union&          1227& 1227& 1653& 937& 716\cr
Intersection&   *546& *546& *827& 502& 325\cr
\noalign{\vskip 5pt}
Confirmed&      *861& *947& 1203& 820& 383\cr
Candidates&     *366& *292& *546& *99& 447\cr
Low reliability&*142& *131& *143& *39& 104\cr
\noalign{\vskip 5pt}
Total X-ray&    *501& *501& *603& 477& 126\cr
MCXC&           *455& *455& *551& 427& 124\cr
SZ clusters&    **82& **82& *110& *79& *31\cr
\noalign{\vskip 3pt\hrule\vskip 3pt}}}
\endPlancktablewide
\endgroup
\end{table*}

The \Planck~satellite is unique in providing broad frequency coverage
over the whole sky with good sensitivity to both the high frequency
spectral increment and the low frequency decrement of the thermal SZ effect. In this paper, we
have presented second \Planck\ catalogue of SZ sources (\psztwo).
This is based on data from the full 29 month mission and uses a
methodology that refines the one used to produce the \pszone\ from 15.5
months of data. The catalogue is based on the union of results from
three cluster detection codes \citep{planck2013-p05a}.  The
\psztwo\ contains 1653 cluster candidates distributed across 83.6\% of
the sky.  The catalogue was validated using external X-ray, optical,
SZ and near infra-red data, producing confirmation for 1203 candidates
with 1094 redshifts.  The catalogue contains 716 new detections including 366 confirmed
clusters with newly identified SZ signal.  87 of our confirmed clusters are newly identified
in this paper.  We have found good consistency with the
\pszone\ and re-detect 937 SZ sources from the \pszone\ sample of
1227. We have investigated the missed detections: the vast majority of
these were low-significance \pszone\ detections whose \snr\ has fluctuated
beneath the detection threshold. The majority of these are expected to be
spurious detections.

The current status of our knowledge of counterparts for our detections at various frequencies
is summarised in Table~\ref{tab:counterpart_summary} and compared to the \pszone.  Our optical validation scheme is
based on the newly released SDSS-based \redmapper\ catalogue
\citep{rozo14}. This produced 374 high-quality matches where the
counterparts are consistent with \Planck~mass and position information.
We reject 188 possible matches where the mass or position information
is inconsistent with the \Planck\ information. This underlines the importance of consistency checks when
matching with high density SDSS catalogues.  Our X-ray and SZ
counterpart searches implement similar consistency criteria leading to
tight control over mismatches.

Central to the counterpart search process is the understanding of the \Planck~SZ
parameter estimates.  We have validated the \Planck~Compton-Y
posteriors using detailed simulations that include an ensemble of
hydro-dynamically simulated pressure profiles that vary from the
pressure profile assumed by our extraction algorithms.  Our
\yfiver\ estimates are robust to mis-matches in the pressure profile.
When translating to \yfive, we have shown the importance of accurate
prior information about radius to break the $Y-\theta$ degeneracy and
produce accurate and precise estimates from the \Planck\ data.  Our
counterpart searches make extensive use of the \Planck\ mass-proxy;
this uses prior information about redshift and scaling relations to
derive mass constraints which show low scatter with respect to
external estimates.  We provide this $M_\text{sz}$ for all candidates
with a redshift, and provide $M_\text{sz}(z)$ in the range $0<z<1$ for
all other candidates.  We expect this information to be useful in
future comparisons with external data.

Central to any statistical use of a cluster sample is the survey
selection function.  We have estimated the catalogue completeness
using Monte-Carlo source injection and we provide this as a product
for the full survey and for various sub-samples as a function of
selection \snr.  We have validated the completeness through a
comparison with external X-ray data and high resolution SZ data from
SPT \citep{blee14}, which spans the redshift range and angular sizes
of the \Planck~data.  We estimate the catalogue to be 83-87\% pure,
based on simulations of the \Planck\ data and detection-by-detection
quality assessment utilising machine learning.  Higher reliability
sub-samples can be constructed easily: the main contaminant is
infra-red galactic emission and as such the reliability is a strong
function of galactic latitude.  Specifically, the cluster cosmology
zone that covers 65\% of the sky contains 1308 detections at $\sim
90$\% reliability, and full survey intersection catalogue (objects
detected by all three codes) contains 827 detections at $>95\%$
reliability.

Cosmology using the cluster counts is also dependent on external
observational data to provide cluster redshifts. \cite{planck2014-a30}
have produced cosmological constraints using samples drawn from the
\psztwo, containing 493 candidates from the intersection sample and
439 drawn from the single-code \mmfthree\ sample.  Utilising larger
samples from the \psztwo\ requires further redshift information.  We
also expect the \psztwo\ to contain many high-mass clusters at
$z>0.6$.  So far only 36 have been identified, of which 21 were
identified in targeted follow-up observations of \pszone\ candidates.
For these reasons, the \psztwo\ should motivate further follow-up
observations.  In particular, the catalogue contains 73 clusters
confirmed by WISE infra-red data that currently have no redshift
information but which are likely to be high-redshift.

Understanding the biases in cluster selection that affect samples
defined at different wavelengths will be important for interpreting
statistical results from existing surveys and those planned for the
near future.  Using a low-redshift overlap sample from \psztwo\ and
\redmapper, we have identified a population of low-z clusters with
`typical' optical and SZ properties, but which are underluminous for
their mass in \emph{ROSAT} X-ray data.  These clusters may be part of
a population of dynamically disturbed clusters that are
under-represented in X-ray selected surveys.  These objects will
be interesting targets for multi-wavelength follow-up to determine
their dynamical state.

In the near future, \Planck~all-sky SZ data can be combined with
observations of the large-scale structure by surveys such as
PAN-STARRS, LOFAR, Euclid, LSST, and RSG/e-ROSITA.  This will provide
an unprecedented multi-wavelength view of the evolution of large-scale
structure that will revolutionise our understanding of the physics
governing this process.

%
%
%

\begin{acknowledgements}
The Planck Collaboration acknowledges the support of: ESA; CNES and CNRS/INSU-IN2P3-INP (France); ASI, CNR, and INAF (Italy); NASA and DoE (USA); STFC and UKSA (UK); CSIC, MINECO, JA, and RES (Spain); Tekes, AoF, and CSC (Finland); DLR and MPG (Germany); CSA (Canada); DTU Space (Denmark); SER/SSO (Switzerland); RCN (Norway); SFI (Ireland); FCT/MCTES (Portugal); ERC and PRACE (EU). A description of the Planck Collaboration and a list of its members, indicating which technical or scientific activities they have been involved in, can be found at \url{http://www.cosmos.esa.int/web/planck/planck-collaboration}.
We thank Ian McCarthy for
providing images and profiles of simulated clusters from cosmo-OWLS.
This research has made use of the NASA/IPAC Extragalactic Database (NED), which is operated by the Jet Propulsion Laboratory, California Institute of Technology, under contract with the National Aeronautics and Space Administration, and the SIMBAD database, operated at CDS, Strasbourg, France
This research made use of data retrieved from SDSS-III. Funding for SDSS-III has been provided by the
Alfred P. Sloan Foundation, the Participating Institutions, the National Science Foundation, and
the U.S. Department of Energy Office of Science. The SDSS-III web site is http://www.sdss3.org/.
This research has made use of data processed by the Centre d'Analyse de DonnŽes Etendues \url{http://cade.irap.omp.eu/} and has made use of the HEALPix pixelisation software \url{http://healpix.sourceforge.net} \citep{gor05}.
This work was performed using the Darwin Supercomputer of the University of Cambridge High Performance Computing Service (http://www.hpc.cam.ac.uk/), provided by Dell Inc. using Strategic Research Infrastructure Funding from the Higher Education Funding Council for England and funding from the Science and Technology Facilities Council.
\end{acknowledgements}

\bibliographystyle{aat}
\bibliography{sz_catalogue,Planck_bib}

\appendix
\section{High-redshift SDSS confirmations}
\label{appendix:high-z}

Table \ref{table:high-z} gives optical information for four
high-redshift confirmations found using a search in SDSS data around
unmatched \Planck\ detections.


We use a multi-wavelength approach to confirm the clusters.  Each of
these candidates possess coincident high-redshift optical
over-densities in SDSS, firm infra-red confirmations from WISE, and
significant emission in the \emph{ROSAT} $0.5-2.4$ keV band.  We
estimate the X-ray luminosity from the \emph{ROSAT} maps using
growth-curve analysis, and confirm that the luminosity is consistent
with the measured $M_\text{sz}$ as discussed in
Sect.~\ref{subsec:MCXC}.

One interesting case is PSZ2 G097.52+51.70, which appears to be a near
line-of-sight projection with components at $z=0.7$ and $z=0.333$,
separated by $1.91$ arcmin.  Both systems may contribute to the
observed \Planck\ signal.  We have associated to the high-redshift
cluster because it is significantly closer to the SZ centre ($0.714$
arcmin vs. $2.47$ arcmin separation), and because it is coincident
($0.23$ arcmin) with the \emph{ROSAT} X-ray centre.  The $z=0.333$
system shows less significant hard-band emission.

\begin{table}
\caption{Optical information for our high-z SDSS confirmations.
  Alongside the redshift, we give the RA and DEC of the BCG and if the
  redshift is spectroscopic, $N_z$ gives number of cluster members
  with spectroscopic redshifts.}
\begin{center}
\begin{tabular}{lrrlc}
\hline\hline
\multicolumn{1}{c}{NAME} & $\alpha_{BCG}$ & $\delta_{BCG}$ & \multicolumn{1}{c}{$z$} & $N_z$ \\
\hline
PSZ2 G076.18-47.30 & 343.1475 & 4.5381 & 0.666 & 3 \\
PSZ2 G087.39+50.92 & 231.6383 & 54.1520 & 0.748 & 1 \\
PSZ2 G089.39+69.36 & 208.4382 & 43.4843 & 0.68 & \dots \\
PSZ2 G097.52+51.70 & 223.8374 & 58.8707 & 0.7 & \dots \\
\hline
\end{tabular}
\end{center}
\label{table:high-z}
\end{table}%

\section{Differences in PSZ1 and PSZ2 redshift assignments}
\label{appendix:redshift_differences}

The \psztwo\ contains 782 clusters which had redshift estimates in the
\pszone.  We assign the same redshift in all but 43 of these cases.
In 25 of these cases, there is no significant difference, defined by
$|\Delta z| > 0.03$ or $|\Delta z| / z > 0.1$, between the estimates
and we have updated \pszone\ photometric redshift from various sources
to new estimates from \redmapper\ or \Planck\ follow-ups.

We have updated a further seven \pszone\ photometric redshifts with recent \Planck\ ENO follow-up redshifts \citep{planck2015-XXXV} where the redshift has significantly changed.  Of these, one was from the \pszone\ SDSS search, two were from PanSTARRs and four were from earlier \Planck\ photometric follow-ups.  These updates are included in the 2015 update of the \pszone\ redshift compilation \citep{planck2015-XXXVI}.

We discuss the remaining 11 significant differences below:

\begin{itemize}
\item \textbf{PSZ2 G020.66+37.99} (\pszone\ INDEX 51):
  \\ \cite{rozo2014a} discuss this cluster in depth.  RedMAPPer finds
  two overlapping clusters, $z_\text{spec}=0.338, \lambda=85.4$ and
  $z_\text{spec}=0.443, \lambda=23.5$.  The PSZ1 redshift
  $z_\text{phot}= 0.39$ from WFI \Planck\ follow-up is likely to be
  biased by members from the less rich and more distant system.  For
  the \psztwo, we choose the higher richness match and quote the
  $z_\text{phot}= 0.345$.  The 2015 update to the \pszone\ adopts this change.
\item \textbf{PSZ2 G066.68+68.44} (\pszone\ INDEX 222): \\ This is an
  ambiguous system.  The \pszone\ used $z_\text{spec}= 0.1813$ from
  NORAS \citep{boh00}.  For the \psztwo\ we have quoted
  $z_\text{phot}=0.163$ from \redmapper\, which agrees with the
  estimate $z_\text{spec}=0.16$ from BCS follow-up \citep{stru99,
    craw95} and the SDSS BCG estimate $z_\text{spec}= 0.163$.  The
  decision between these two \emph{ROSAT} follow-up spectroscopic
  redshifts rests on the \redmapper\ information, which identifies a
  rich $\lambda=84.1$ system at $z_\text{phot}= 0.163$.
\item \textbf{PSZ2 G087.39+50.92} (\pszone\ INDEX 299): \\ The
  \psztwo\ position has moved closer to a clear high-redshift SDSS
  cluster at $z_ \text{spec}=0.748$ at separation
  $0.9\times\theta_\text{err}$, and away from the \pszone\ SDSS match
  which is now at $2.62\times\theta_\text{err}$.  There is also clear
  \emph{ROSAT} $0.5-2.4$ keV X-ray emission at the high-z location,
  whose strength is consistent with the SZ emission, while there is no
  significant emission at the \pszone\ SDSS match location.
\item \textbf{PSZ2 G090.66-52.34} (\pszone\ INDEX 308): \\ The
  \pszone\ redshift $z_\text{spec}= 0.1784$ came from a single galaxy
  spectrum \citep{stru99}.  \redmapper\ suggest this galaxy is likely
  to be in the foreground, with the rich $\lambda=85$ cluster at
  slightly high redshift $z_\text{phot}= 0.197$.  We note however that
  the difference is small (10.2\%).
\item \textbf{PSZ2 G113.91-37.01} (\pszone\ INDEX 416): \\ We adopt
  the \cite{rozo2014a} update of the NORAS redshift, which replaced a
  $\lambda=7.1$ group at $z=0.135$ with a rich $\lambda=159$ cluster
  at $z=0.371$ separated by $8$ arcmin.
\item \textbf{PSZ2 G121.13+49.64} (\pszone\ INDEX 443): \\ We note
  that this system is a probable projection.  We adopt the correction
  of the redshift from \citep{rozo2014a}, noting that the richness of
  the $z=0.22$ component is consistent with the SZ signal, while the
  $z=0.438$ system matched in the \pszone\ is insufficiently rich. The 2015 update to the \pszone\ adopts this change.
\item \textbf{PSZ2 G143.26+65.24} (\pszone\ INDEX 513): \\ The
  \pszone\ association with ACO 1430 is correct, but we update the
  redshift of $z_\text{spec}=0.211$ from two members \citep{stru91}
  with $z_{phot}= 0.363$ from \redmapper.  The X-ray and optical
  images show an E-W elongation and two possible galaxy
  concentrations, possibly a projection.  The high-redshift component
  has richness consistent with the SZ signal. The 2015 update to the \pszone\ adopts this change.
\item \textbf{PSZ2 G151.19+48.27} (\pszone\ INDEX 537): \\ The
  \pszone\ association with A0959 is correct. NED lists two literature
  redshifts for this cluster: 0.289 (which we adopt) and 0.353
  (adopted in the PSZ1). This object is bimodal in the optical and in
  the X-ray and is almost certainly a projection. \redmapper\ suggests
  association with 0.289 component based on consistency of richness
  with the SZ signal. The 2015 update to the \pszone\ adopts this change.
\item \textbf{PSZ2 G259.30+84.41} (\pszone\ INDEX 888): \\ We adopt
  the correction of \cite{rozo2014a} of the \pszone\ redshift
  $z_\text{phot}=0.4125$ from NSCS, instead matching to a clear and
  rich SDSS cluster within 1 arcmin of the Planck position at
  $z_\text{phot}=0.323$. The 2015 update to the \pszone\ adopts this change.
\item \textbf{PSZ2 G310.81+83.91} (\pszone\ INDEX 1093): \\ The
  \psztwo\ matched to an SDSS cluster at $z_\text{phot}= 0.446$.
  \redmapper\ finds this potential counterpart to be insufficiently
  rich to be detected by \Planck\ at this redshift and there is no
  X-ray emission in \emph{ROSAT}.  The \Planck\ detection may be
  spurious: there are point sources detected at 353 and 545GHz and the
  neural network quality assessment suggests a contaminated spectrum.
  We therefore break the association and leave the detection
  unconfirmed.
\item \textbf{PSZ2 G318.62+58.55} (\pszone\ INDEX 1123): \\ We adopt
  the correction of \cite{rozo2014a} of the \pszone\ redshift
  $z_\text{spec}=0.1144$.  This associates to a different cluster at
  $4.56$ arcmin separation, with \redmapper\ redshifts $z_\text{phot}=
  0.22$ ($z_\text{spec}= 0.233$), and richness $\lambda= 69.3$.
\end{itemize}

\section{Modifications of the extraction algorithms since the \pszone\ release}
\label{appendix:extraction_refine}

\subsection{\mmfone}

The \mmfone\ code used for the \psztwo\ is the same as for the
\pszone\, with the following changes:
\begin{itemize}
\item Positions estimates are now calculated with sub-pixel positioning using posteriors marginalised over all other parameters, rather than taking the pixel centre closest to the peak.
\item Position error radius estimation has been debugged.
\item The 2D contour grids for ($Y,\theta$) expand dynamically if the 91\% confidence region is not entirely contained within the grid.  This expansion is not applied to \snr$ < 5$ detections.
\end{itemize}

\subsection{\mmfthree}

For the \psztwo\ release, we made three improvements on our
\mmfthree\ code:
\begin{itemize}
\item Bright clusters impact the estimation of background. For the
  \pszone\, we have adopted two different estimators of the
  \mmfthree\ background depending on the \snr\ of the detections. If
  the \snr\ was below 20, the theoretical calculation was used (see
  Eq. 7 in~\cite{melin2006}) while, for \snr\ greater than 20, we used
  the standard deviation of the filtered map. When the cluster signal
  is subdominant in the map the two estimators return the same result
  but they differ if the cluster is bright (typically with \snr\ above
  20) due to the contamination of the background by the cluster
  itself. The choice of using two estimators has been made to make the
  \mmfthree\ background estimate compatible with \pws\ and
  \mmfone\ for the \pszone.  We tested it against the QA after the
  2013 release and found that it biases significantly \mmfthree\ two
  dimensional ($\theta_s$,\yfiver) contours with respect to injected
  cluster size and flux at high \snr. We thus decided to come back to
  the theoretical calculation of the background across all the
  \snr\ range as in the earlier version of \mmfthree\ used for the
  ESZ. This choice fixes the issue with high \snr\ cluster contours in
  the QA. But it increases the \snr\ value and shifts the
  ($\theta_s$,\yfiver) contours for \mmfthree\ detections in the
  \psztwo\ release with respect to \pszone\ for the detections with
  \snr\ above 20 more than the increase expected from the additional
  integration time.  For the \psztwo\, \pws\ now estimate the
  cross-channel covariance matrix under the `\textit{native}'
  prescription (see \pws\ section in this appendix). This improvement
  makes the background estimate compatible with the theoretical
  calculation from \mmfthree\ and also gives unbiased estimates for
  the \pws\ two dimensional in the QA.
\item \mmfthree\ two dimensional ($\theta_s$,\yfiver) contours in the
  \pszone\ were tested against the QA after the 2013 release. This
  could not be done in early 2013 for lack of time. The contours were
  found to be wider than expected. The code has been corrected and new
  contours have been produced and included in the \Planck\ Legacy
  Archive for the \pszone. The \psztwo\ relies on this new and fully
  tested estimate of the two dimensional contours.
\item The \mmfthree\ positional error for the \pszone\ was
  overestimated. The code has been corrected and tested against the
  QA. New estimates have been produced and included in the Planck
  Legacy Archive for the \pszone. The \psztwo\ uses the new estimate
  of the positional error.
\end{itemize}

\subsection{\pws}

The \pws\ code used for \psztwo\ is similar to the one used for
\pszone, with two modifications:
\begin{itemize}
\item The cross-channel covariance matrix is now always estimated
  using iterative recalibration but is parameterised in order to produce
  a smoother estimate. Using the QA, we have shown that the new
  calibration only impacts the \snr\ estimate ($\sim 12$\%) keeping
  unchanged all other parameter estimates.  The new \snr\ estimates are
  consistent with the other codes.
\item For ($\theta_s$,\yfiver), we now adopt non-informative priors,
  formulated using Jeffrey's method~\citep{car12}, instead of
  informative priors derived from a fiducial cosmology and mass
  function.
\end{itemize}

\onecolumn

\section{Description of the delivered products}
\label{appendix:products}

The data products comprise: (i) the main catalogue, which contains the characterised catalogue with ancillary information; (ii) individual algorithm catalogues produced by each of the codes prior to merging to create the main catalogue, which contain $Y-\theta$ parameter posteriors per cluster; (iii) selection function files containing the completeness and survey masks for various sample definitions.

\subsection{Main catalogue}

The table contains the following columns:

\begin{center}
\begin{tabular}{L{1.2in}llp{9cm}}
\hline\hline
Column Name & Data Type & Units & Description\\
\hline
INDEX  \dotfill        & Integer(4) & \hspace{8pt}\ldots         & Index of detection (see note 1)\\
NAME   \dotfill        & String     &  \hspace{8pt}\ldots        & Name of detection (see note 2)\\
GLON    \dotfill       & Real(8)    & degrees    & Galactic longitude ($0\deg \le l < 360\deg$)\\
GLAT     \dotfill      & Real(8)    & degrees    & Galactic latitude ($-90\deg \le b \le 90\deg$)\\
RA        \dotfill     & Real(8)    & degrees    & Right ascension (J2000)\\
DEC       \dotfill     & Real(8)    & degrees    & Declination (J2000)\\
POS\_ERR   \dotfill    & Real(4)    & arcmin     & Uncertainty in position (see note 3)\\
SNR     \dotfill       & Real(4)    & \hspace{8pt}\ldots         & \snr\ of detection\\
PIPELINE     \dotfill  & Integer(4) & \hspace{8pt}\ldots         & Pipeline from which information is taken: the reference pipeline (see note 4)\\
PIPE\_DET   \dotfill   & Integer(4) & \hspace{8pt}\ldots         & Information on pipelines making detection (see note 4)\\
PCCS2       \dotfill    & Boolean    & \hspace{8pt}\ldots         & Indicates whether detection matches with any in PCCS2 catalogues\\
PSZ \dotfill & Integer(4) & \hspace{8pt}\ldots & Index of matching detection in PSZ1, or -1 if a new detection \\ 
IR\_FLAG    \dotfill    & Integer(1)     & \hspace{8pt}\ldots         & Flag denoting heavy IR contamination\\
Q\_NEURAL \dotfill & Real(4) & \hspace{8pt}\ldots & Neural network quality flag (see note 5) \\
Y5R500  \dotfill  & Real(4) & $10^{-3}$ arcmin$^2$ & Mean marginal \yfiver\ as measured by the reference pipeline \\
Y5R500\_ERR   \dotfill & Real(4) & $10^{-3}$ arcmin$^2$ & Uncertainty on\yfiver\ as measured by the reference pipeline \\
VALIDATION    \dotfill & Integer(4) & \hspace{8pt}\ldots         & External validation status (see note 6)\\
REDSHIFT\_ID    \dotfill    & String     & \hspace{8pt}\ldots         & External identifier of cluster associated with redshift measurement (see note 7).\\
REDSHIFT    \dotfill   & Real(4)    &  \hspace{8pt}\ldots        & Redshift of cluster (see note 7)\\
MSZ  \dotfill  & Real(4) & $10^{14} M_\odot$ & SZ mass proxy (see note 8)\\
MSZ\_ERR\_UP   \dotfill& Real(4)  & $10^{14} M_\odot$ & Upper 1$\sigma$ SZ mass proxy confidence interval (see note 8)\\
MSZ\_ERR\_LOW  \dotfill & Real(4)  & $10^{14} M_\odot$ & Lower 1$\sigma$ SZ mass proxy confidence interval (see note 8)\\
MCXC  \dotfill    & String     & \hspace{8pt}\ldots         & ID of X-ray counterpart in the MCXC if one is present\\
REDMAPPER    \dotfill   & String     & \hspace{8pt}\ldots         &ID of optical counterpart in the \redmapper\ catalogue if one is present\\
ACT    \dotfill      & String    & \hspace{8pt}\ldots         & ID of SZ counterpart in the ACT catalogues if one is present\\
SPT   \dotfill       & String     & \hspace{8pt}\ldots         & ID of SZ counterpart in the SPT catalogues if one is present\\
WISE\_FLAG    \dotfill    & Integer(4)     & \hspace{8pt}\ldots         & Confirmation flag of WISE IR overdensity (see note 9)\\
AMI\_EVIDENCE   \dotfill     & Real(4)     & \hspace{8pt}\ldots         & Bayesian evidence for AMI counterpart detection (see note 9)\\
COSMO      \dotfill    & Boolean    & \hspace{8pt}\ldots         & Indicates whether the cluster is in the cosmology sample\\
COMMENT    \dotfill    & String    & \hspace{8pt}\ldots         & Comments on this detection \\
\hline
\end{tabular}
\end{center}

Notes:

\begin{enumerate}

\item The index is determined by the order of the detections in the union catalogue.  The matching entries in the individual catalogues have the same index to facilitate cross-referencing.

\item The names are in the format PSZ2~Gxxx.xx$\pm$yy.yy where xxx.xx is the Galactic longitude and $\pm$yy.yy is the Galactic latitude of the detection, both in degrees. The coordinates are truncated towards zero, not rounded.

\item The value given here is the 95\% confidence interval of the distribution of radial displacement.

\item The PIPELINE column defines the pipeline from which the values in the union catalogue are taken: 1 = \mmfone; 2 = \mmfthree; 3 = \pws.

The PIPE\_DET column is used to indicate which pipelines detect this object. The three least significant decimal digits are used to represent detection or non-detection by the pipelines. Order of the digits: hundreds = MMF1; tens = MMF3; units = \pws. If it is detected then the corresponding digit is set to 1, otherwise it is set to 0.

\item The neural network quality flag is 1-$Q_\text{bad}$, following the definitions in \citealt{agh14}.  Values Q\_NEURAL$<0.4$ denote low-reliability detections.

\item The VALIDATION column gives a summary of the external validation, encoding the most robust external identification: -1 = no known external counterpart; 10=ENO follow-up; 11= RTT follow-up; 12= PanSTARRs; 13= \redmapper\ non-blind; 14= SDSS high-z; 15=AMI; 16= WISE; 20 = legacy identification from the \pszone\ 2013 release; 21 = MCXC; 22= SPT; 23=ACT; 24= \redmapper; 25= \pszone\ counterpart with redshift updated in \citep{planck2015-XXXVI}; 30= NED.

\item The redshift source is the most robust external validation listed in the VALIDATION field.

\item Definition of $M_\text{sz}$. The hydrostatic mass, $M_{500}$, assuming the best-fit $Y-M$ scaling relation of \cite{arn10}  as a prior that cuts a plane through the parameter contours (see Sect.~\ref{section:Msz}).  The errors are 67\% confidence statistical errors and based on the \Planck\ measurement uncertainties only.  Not included in the error estimates  are the statistical errors on the scaling relation, the intrinsic scatter in the relation, or systematic errors in data selection for the scaling relation fit.

\item WISE confirmation flag is assigned by visual inspection and defined to be one of $[-10,-2,-1,0,1,2,3]$, where -10 denotes no information and the other values are discussed in Section~\ref{sec:WISE}.  Bayesian evidence for AMI counterpart defined in paper.

\end{enumerate}

\subsection{Individual algorithm catalogues}

The table contains the following columns:

\begin{center}
\begin{tabular}{L{1in}llp{9cm}}
\hline\hline
Column Name & Data Type & Units & Description\\
\hline
INDEX \dotfill         & Integer(4) & \hspace{8pt}\ldots         & Index of detection (see note 1)\\
NAME    \dotfill       & String     & \hspace{8pt}\ldots         & Name of detection (see note 1)\\
GLON    \dotfill       & Real(8)    & degrees    & Galactic longitude ($0\deg \le l < 360\deg$)\\
GLAT    \dotfill       & Real(8)    & degrees    & Galactic latitude ($-90\deg \le b \le 90\deg$)\\
RA       \dotfill      & Real(8)    & degrees    & Right ascension (J2000)\\
DEC     \dotfill       & Real(8)    & degrees    & Declination (J2000)\\
POS\_ERR  \dotfill     & Real(4)    & arcmin     & Uncertainty in position (see note 2)\\
SNR      \dotfill      & Real(4)    & \hspace{8pt}\ldots         & \snr\ of detection (see note 3)\\
TS\_MIN   \dotfill     & Real(4)    & arcmin     & Minimum $\theta_\text{s}$ in second extension HDU (see note 4)\\
TS\_MAX  \dotfill      & Real(4)    & arcmin     & Maximum $\theta_\text{s}$ in second extension HDU (see note 4)\\
Y\_MIN     \dotfill    & Real(4)    & arcmin$^2$ & Minimum \yfiver\ in second extension HDU (see note 4)\\
Y\_MAX  \dotfill       & Real(4)    & arcmin$^2$ & Maximum \yfiver\ in second extension HDU (see note 4)\\
\hline
\end{tabular}
\end{center}

Notes:

\begin{enumerate}

\item The index and name are taken from the union catalogue. The matching entries in the individual catalogues have the same index and name to facilitate cross-referencing.

\item The value given here is the 95\% confidence interval of the distribution of radial displacement.

\item The SNR column contains the native signal-to-noise ratio determined by the detection pipeline.

\item These entries define the limits of the grid used to evaluate the 2D probability distribution of $\theta_\text{s}$ and \yfiver\ in the second extension HDU (see below).

\end{enumerate}

\subsubsection*{Second extension HDU}

The second extension HDU contains a three-dimensional image with the two-dimensional probability distribution in $\theta_\text{s}$ and \yfiver\ for each cluster. The probability distributions are evaluated on a $256 \times 256$ linear grid between the limits specified in the first extension HDU. The limits are determined independently for each detection. The dimensions of the 3D image are $ 256 \times 256 \times n$, where $n$ is the number of detections in the catalogue. The second dimension is $\theta_\text{s}$ and the first dimension is \yfiver.

\subsubsection*{Third extension HDU}

The third extension HDU contains a three-dimensional image with the $M_\text{sz}$ observable information per cluster as a function of assumed redshift.  The image dimensions are $100 \times 4 \times n$, where $n$ is the number of detections in the catalogue.  The first dimension is the assumed redshift.  The second dimension has size $4$: the first element is the assumed redshift value for the $M_\text{sz}$ fields.  The second element is the $M_\text{sz}$ lower 67\% confidence bound, the third element is the $M_\text{sz}$ estimate and the fourth element is the $M_\text{sz}$ upper 67\% confidence bound, all in units of $10^{14} M_\odot$.  These errors are based on the Planck measurement uncertainties only.  Not included in the error estimates  are the statistical errors on the scaling relation, the intrinsic scatter in the relation, or systematic errors in data selection for the scaling relation fit.

\subsection{Selection function file format}

The selection function information is stored in FITS files. The filenames of the catalogues are of the form PSZ2-selection\_Rx.xx.fits, where x.xx is the release number.

\subsubsection*{First extension HDU}

The first extension HDU contains the survey region, denoted by an $N_\text{side}=2048$ ring-ordered HEALPix map in GALACTIC coordinates.  Pixels in the survey region have the value 1.0 while areas outside of the survey region have value 0.0.

\subsubsection*{Second extension HDU}

The second extension HDU contains a three-dimensional image containing the survey completeness probability distribution for various thresholds.  The information is stored in an image of size $30 \times 32 \times 12$.  The first dimension is \yfive, the second dimension is $\theta_{500}$ and the third dimension is the signal-to-noise threshold.  The units are percent and lie in the range 0-100 and denote the detection probability of a cluster lying within the given \yfive-$\theta_{500}$ bin.

\subsubsection*{Third extension HDU}

The second extension HDU contains the \yfive\ grid values for the completeness data cube held in the second extension.  It has length 30 and spans the range $1.12480\times 10^{-4} -7.20325\times 10^{-2}$ arcmin$^2$ in logarithmic steps.

\subsubsection*{Fourth extension HDU}

The fourth extension HDU contains the $\theta_{500}$ grid values for the completeness data cube held in the second extension.  It has length 32 and spans the range 0.9416-35.31 arcmin in logarithmic steps.

\subsubsection*{Fifth extension HDU}

The fifth extension HDU contains the signal-to-noise threshold grid values for the completeness data cube held in the second extension.  It has length 12 and contains thresholds at intervals of 0.5 from 4.5 to 10.0.

\newpage

\section{Detail of missing PSZ1 detections}
\label{appendix:missing_dets}
\begin{table}[H]
    \caption{Detail of the 291 \pszone\ detections not present in the \psztwo\ catalogue.  The TYPE column lists the reason why the detection was dropped. TYPE 1 lost detections are low-\snr\ detections lost due to changes in the noise realisation.  The  \snr\_non\_blind field contains the non-blind \snr\ for the $Y$ signal in the full-mission maps at the location and size of the \pszone\ detection and is provided for all TYPE 1 lost detections (whereas the field \snr\ is for the \pszone).  TYPE 2 are lost behind the new point source mask.  TYPE 3 are cut due to IR contamination. TYPE 4 are cut by internal \pws\ consistency criteria.  Each of these types are discussed in Sect.~\ref{sec:missing_detections}.}
\begin{tiny}
\begin{center}
    \begin{tabular}{L{0.5in}ccrclccc}
  \hline\hline
\multicolumn{1}{c}{I\textsc{ndex}} & \multicolumn{1}{c}{\snr\ }      & \multicolumn{1}{c}{P\textsc{ipeline}} & \multicolumn{1}{c}{P\textsc{ipe\_det}} & \multicolumn{1}{c}{I\textsc{d}\_\textsc{ext}}                & \hspace{-0.2in}
R\textsc{edshift} &\snr\ (non-blind) & \multicolumn{1}{c}{T\textsc{ype}} \\
\hline
4\dotfill     & 6.04 & 3        & 101       &         \ldots               &     \ldots     &       \ldots          & 3      \\
8 \dotfill    & 4.92  & 1        & 100       &           \ldots             &    \ldots      & 0.34        & 1      \\
9 \dotfill    & 5.76  & 1        & 100       &           \ldots             &    \ldots      &         \ldots        & 3      \\
13 \dotfill   & 4.52  & 1        & 100       & ZwCl 1454.5+0656       & 0.429    & 3.73         & 1      \\
28  \dotfill  & 4.70  & 2        & 10        &         \ldots               & 0.46     & 3.54         & 1      \\
30  \dotfill  & 4.72  & 1        & 100       &       \ldots                 &     \ldots     & 1.83         & 1      \\
32  \dotfill  & 4.54  & 1        & 100       &       \ldots                 &  \ldots        & 1.65         & 1      \\
34  \dotfill  & 4.50  & 1        & 100       &       \ldots                 &    \ldots      & 1.42         & 1      \\
38  \dotfill  & 4.65 & 1        & 100       &         \ldots               &     \ldots     & 2.48         & 1      \\
40 \dotfill   & 4.89  & 2        & 10        &         \ldots               &    \ldots      & 2.90         & 1      \\
41  \dotfill  & 4.81 & 2        & 10        &           \ldots             &     \ldots     & 2.90         & 1      \\
43  \dotfill  & 4.76 & 2        & 10        &          \ldots              &     \ldots     & 2.69         & 1      \\
52  \dotfill  & 5.71 & 2        & 10        &         \ldots               & 0.39     &        \ldots         & 4      \\
58  \dotfill  & 4.56  & 1        & 100       &        \ldots                &   \ldots       & 2.06         & 1      \\
59 \dotfill  & 4.63 & 1        & 100       &          \ldots              &     \ldots     & 3.27         & 1      \\
60 \dotfill   & 4.84 & 2        & 10        & RXC J1917.5-1315       & 0.177    & 3.11         & 1      \\
61 \dotfill   & 5.06  & 2        & 111       &        \ldots                & 0.650893 & 2.91         & 1      \\
62  \dotfill  & 4.95  & 2        & 11        & ACO S 1010             & 0.28     & 4.02         & 1      \\
66  \dotfill  & 5.20 & 2        & 10        &            \ldots            &      \ldots    & 3.19         & 1      \\
68  \dotfill  & 4.67  & 3        & 1         &            \ldots            &      \ldots    & 3.45         & 1      \\
77  \dotfill  & 4.58  & 1        & 100       & RXC J1453.1+2153       & 0.1186   & 2.96         & 1      \\
82  \dotfill  & 4.98  & 2        & 10        &        \ldots                &    \ldots      & 2.84         & 1      \\
83  \dotfill  & 4.96 & 2        & 10        &           \ldots             &      \ldots    & 3.53         & 1      \\
84  \dotfill  & 4.84  & 2        & 10        &         \ldots               &      \ldots    & 3.80         & 1      \\
86  \dotfill  & 5.23 & 1        & 100       &           \ldots             &      \ldots    & 2.82         & 1      \\
89  \dotfill  & 4.69  & 3        & 101       & WHL J248.764+15.4836   & 0.4725   & 3.27        & 1      \\
90  \dotfill  & 4.68  & 1        & 100       &            \ldots            &    \ldots      & 2.11         & 1      \\
97  \dotfill  & 4.82  & 1        & 100       & WHL J252.649+16.8253   & 0.3612   & 2.56         & 1      \\
98  \dotfill  & 4.70 & 3        & 1         &              \ldots          &      \ldots    & 3.54         & 1      \\
104  \dotfill & 4.54  & 2        & 10        &           \ldots             &   \ldots       & 1.94          & 1      \\
111 \dotfill  & 4.57  & 2        & 10        &          \ldots              &     \ldots     & 1.97         & 1      \\
112 \dotfill  & 4.61  & 1        & 100       &         \ldots               &   \ldots       & 1.59         & 1      \\
121 \dotfill  & 4.70 & 2        & 110       &           \ldots             &    \ldots      & 1.22         & 1      \\
126 \dotfill  & 4.67  & 2        & 10        &           \ldots             &    \ldots      & 1.91         & 1      \\
128 \dotfill  & 4.54 & 3        & 1         & RXC J1623.5+2634       & 0.4274   & 3.79         & 1      \\
131  \dotfill & 5.23  & 2        & 11        & AMF J320.551-6.81740   & 0.5344   & 3.59         & 1      \\
136 \dotfill  & 4.74 & 1        & 100       &         \ldots               &      \ldots    &      \ldots           & 3      \\
142 \dotfill  & 4.52  & 3        & 101       &        \ldots                &    \ldots      & 1.26         & 1      \\
157 \dotfill & 4.64 & 1        & 100       &          \ldots              &      \ldots    & 2.27         & 1      \\
158  \dotfill & 5.29  & 2        & 10        &        \ldots                &     \ldots     & 2.08         & 1      \\
162  \dotfill & 4.60 & 1        & 100       &         \ldots               &     \ldots     & 2.34         & 1      \\
165  \dotfill & 5.11  & 3        & 101       &         \ldots               &     \ldots     &       \ldots          & 3      \\
170  \dotfill & 5.33  & 2        & 11        &            \ldots            &     \ldots     & 4.32          & 1      \\
175  \dotfill & 4.82  & 1        & 100       &            \ldots            & 0.1944   & 2.84         & 1      \\
176  \dotfill & 5.72 & 2        & 10        &              \ldots          &      \ldots    &       \ldots          & 3      \\
184  \dotfill & 4.76 & 3        & 1         &                \ldots        &       \ldots   & 2.67         & 1      \\
193  \dotfill & 4.55   & 1        & 100       &           \ldots             &     \ldots     & 2.83         & 1      \\
199 \dotfill  & 4.55 & 3        & 1         &             \ldots           &      \ldots    & 2.12         & 1      \\
203 \dotfill  & 4.65 & 2        & 10        &           \ldots             &    \ldots      & 1.98        & 1      \\
211  \dotfill & 4.58 & 3        & 1         &             \ldots           &     \ldots     & 3.31         & 1      \\
212  \dotfill & 4.54 & 3        & 1         &             \ldots           &     \ldots     & 3.72         & 1      \\
213  \dotfill & 4.70 & 2        & 10        &             \ldots           &    \ldots      & 2.93         & 1      \\
223  \dotfill & 4.51 & 2        & 10        &             \ldots           & 0.3341   & 3.16         & 1      \\
233 \dotfill  & 4.79 & 2        & 10        & ZwCl 2151.0+1325       & 0.205    & 3.97          & 1      \\
237  \dotfill & 4.78 & 1        & 100       & ACO  2429              &     \ldots     & 3.36         & 1      \\
251 \dotfill  & 5.17 & 1        & 100       &           \ldots             &     \ldots     & 2.59         & 1      \\
257  \dotfill & 4.90  & 2        & 11        &           \ldots             &     \ldots     & 3.68         & 1      \\
260  \dotfill & 4.78 & 2        & 110       & WHL J242.728+51.2267   & 0.4096   & 3.60         & 1      \\
262 \dotfill & 4.52 & 1        & 100       &               \ldots         &      \ldots    & 3.63         & 1      \\
267 \dotfill  & 6.37  & 2        & 111       & ACTJ2327.4-0204        & 0.705    &       \ldots          & 2      \\
\hline
\end{tabular}
\end{center}
\end{tiny}
\label{table_lost_dets1}
\end{table}
\begin{table}[H]
    \caption{Continuation of Table~\ref{table_lost_dets1}.}
\begin{tiny}
\begin{center}
    \begin{tabular}{L{0.5in}n{2}{2}crcl@{\hskip 0.4in}n{2}{2}c}
  \hline\hline
\multicolumn{1}{c}{I\textsc{ndex}} & \multicolumn{1}{c}{\snr\ }      & \multicolumn{1}{c}{P\textsc{ipeline}} & \multicolumn{1}{c}{P\textsc{ipe\_det}} & \multicolumn{1}{c}{I\textsc{d}\_\textsc{ext}}                & \hspace{-0.2in}
R\textsc{edshift} & \multicolumn{1}{c}{\hspace{-0.3in}\snr\ (non-blind)} & \multicolumn{1}{c}{T\textsc{ype}} \\
\hline
271 \dotfill & 4.71  & 2        & 10        &           \ldots             &    \ldots      & 3.84092         & 1      \\
272 \dotfill  & 4.57  & 3        & 1         & ZwCl 1746.2+5429       & 0.31     & 3.9982          & 1      \\
276 \dotfill  & 4.65 & 2        & 11        &         \ldots               &     \ldots     & 2.88368         & 1      \\
278 \dotfill & 4.55 & 2        & 10        &          \ldots              & 0.306807 & 2.32043         & 1      \\
300 \dotfill & 5.07  & 2        & 10        &         \ldots               & 0.1132   & 4.09432         & 1      \\
305 \dotfill  & 4.82 & 2        & 10        &         \ldots               &     \ldots     & 2.12439         & 1      \\
306 \dotfill  & 4.90 & 1        & 100       &        \ldots                &    \ldots      & 3.50522         & 1      \\
309 \dotfill & 5.37 & 2        & 110       & ZwCl 1602.3+5917       & 0.2544   & 4.0135          & 1      \\
311 \dotfill & 4.64 & 2        & 110       &           \ldots             &    \ldots      &       \multicolumn{1}{l}{\hspace{1pt}\ldots}          & 3      \\
314 \dotfill  & 5.18  & 1        & 100       &          \ldots              &    \ldots      & 1.77555         & 1      \\
317 \dotfill & 5.46 & 1        & 100       &            \ldots            &      \ldots    & 1.0888          & 1      \\
321 \dotfill & 4.58   & 1        & 100       & ZwCl 1604.4+6113       & 0.3447   & 4.06313         & 1      \\
327 \dotfill & 5.78 & 3        & 101       & RXC J2318.4+1843       & 0.0389   &                 & 4      \\
331 \dotfill  & 4.79  & 3        & 101       &             \ldots           &      \ldots    & 2.53009         & 1      \\
333 \dotfill  & 4.71  & 3        & 1         & WHL J286.905+64.5511   & 0.3561   & 3.89137         & 1      \\
336 \dotfill  & 6.55 & 2        & 111       &            \ldots            &       \ldots   &         \multicolumn{1}{l}{\hspace{1pt}\ldots}        & 3      \\
349 \dotfill & 4.87 & 2        & 11        &            \ldots            &     \ldots     &         \multicolumn{1}{l}{\hspace{1pt}\ldots}        & 2      \\
361 \dotfill & 4.71  & 2        & 10        &           \ldots             &     \ldots     & 1.27901         & 1      \\
365 \dotfill  & 4.82  & 3        & 101       & RXC J1834.1+7057       & 0.0824   & 3.43465         & 1      \\
367 \dotfill & 4.52 & 2        & 10        & ZwCl 1748.0+7125       &     \ldots     & 2.72016         & 1      \\
370 \dotfill  & 5.08  & 2        & 110       &         \ldots               &       \ldots   &         \multicolumn{1}{l}{\hspace{1pt}\ldots}        & 3      \\
371 \dotfill  & 4.55  & 2        & 110       &         \ldots               &    \ldots      &         \multicolumn{1}{l}{\hspace{1pt}\ldots}        & 3      \\
372 \dotfill  & 4.63  & 2        & 11        &          \ldots              &      \ldots    & 4.02674         & 1      \\
373 \dotfill  & 4.50 & 2        & 10        &            \ldots            &      \ldots    &         \multicolumn{1}{l}{\hspace{1pt}\ldots}        & 2      \\
375 \dotfill & 4.78  & 3        & 101       &          \ldots              &     \ldots     & 3.02917         & 1      \\
376 \dotfill & 4.54 & 2        & 10        & AMF J359.521+15.1625   & 0.1785   & 2.56247         & 1      \\
382 \dotfill  & 4.73 & 2        & 110       &          \ldots              &     \ldots     &        \multicolumn{1}{l}{\hspace{1pt}\ldots}         & 3      \\
387 \dotfill  & 4.66  & 2        & 111       &         \ldots               &     \ldots     &        \multicolumn{1}{l}{\hspace{1pt}\ldots}         & 2      \\
396 \dotfill  & 4.55 & 2        & 10        &             \ldots           & 0.25     & 1.78124         & 1      \\
397 \dotfill & 6.89 & 2        & 111       &            \ldots            &      \ldots    &          \multicolumn{1}{l}{\hspace{1pt}\ldots}       & 3      \\
398 \dotfill  & 4.55  & 2        & 10        &           \ldots             &     \ldots     &          \multicolumn{1}{l}{\hspace{1pt}\ldots}       & 3      \\
400 \dotfill  & 4.58  & 2        & 111       &          \ldots              & 0.533998 & 2.74459         & 1      \\
405 \dotfill & 4.81  & 1        & 100       & WHL J358.170+38.9803   & 0.27     & 2.79469         & 1      \\
412 \dotfill & 4.76 & 3        & 1         &               \ldots         &     \ldots     & 3.08399         & 1      \\
426 \dotfill & 5.85  & 2        & 111       &           \ldots             &    \ldots      &         \multicolumn{1}{l}{\hspace{1pt}\ldots}        & 2      \\
430 \dotfill & 4.54  & 3        & 1         &               \ldots         &      \ldots    & 3.59742         & 1      \\
436 \dotfill  & 4.96 & 3        & 101       &           \ldots             &    \ldots      & 3.37978         & 1      \\
437 \dotfill  & 4.51153   & 1        & 100       &        \ldots                &    \ldots      & 3.4165          & 1      \\
444 \dotfill & 4.721222  & 2        & 110       &         \ldots               &    \ldots      & 3.54217         & 1      \\
445 \dotfill  & 4.657247  & 2        & 10        & Abell 98S              & 0.104    & 2.86664         & 1      \\
446 \dotfill  & 5.346884  & 2        & 11        &         \ldots               &      \ldots    &         \multicolumn{1}{l}{\hspace{1pt}\ldots}        & 2      \\
456 \dotfill  & 4.8859854 & 1        & 100       &       \ldots                 &     \ldots     & 3.07575         & 1      \\
458 \dotfill  & 4.857769  & 3        & 101       &        \ldots                &      \ldots    & 2.81502         & 1      \\
462 \dotfill  & 5.401291  & 2        & 110       &             \ldots           &      \ldots    &         \multicolumn{1}{l}{\hspace{1pt}\ldots}        & 2      \\
466 \dotfill  & 5.9113874 & 2        & 11        &          \ldots              &      \ldots    &         \multicolumn{1}{l}{\hspace{1pt}\ldots}        & 3      \\
468 \dotfill  & 4.9521785 & 2        & 10        &          \ldots              &   \ldots       & 3.00375         & 1      \\
469 \dotfill  & 4.898088  & 2        & 11        &          \ldots              & 0.423    & 2.90179         & 1      \\
476 \dotfill  & 4.909587  & 2        & 10        &           \ldots             &      \ldots    & 2.66482         & 1      \\
478 \dotfill  & 5.421913  & 3        & 1         &             \ldots           &     \ldots     &          \multicolumn{1}{l}{\hspace{1pt}\ldots}       & 3      \\
479 \dotfill  & 4.814969  & 2        & 11        &             \ldots           &    \ldots      & 4.24698         & 1      \\
483 \dotfill  & 4.5689936 & 2        & 10        &            \ldots            &     \ldots     & 3.39778         & 1      \\
488 \dotfill  & 4.5772305 & 1        & 100       & RXC J0115.2+0019       & 0.045    & 3.60322         & 1      \\
489 \dotfill  & 4.57693   & 3        & 1         &               \ldots         &      \ldots    & 2.75025         & 1      \\
490 \dotfill  & 5.2208633 & 2        & 10        &          \ldots              &   \ldots       & 3.17854         & 1      \\
491 \dotfill  & 4.5107737 & 1        & 100       & RXC J0152.9+3732       & 0.2993   & 3.61788         & 1      \\
504 \dotfill  & 4.9024134 & 1        & 100       &          \ldots              &      \ldots    & 2.99012         & 1      \\
505 \dotfill  & 4.531147  & 2        & 10        &            \ldots            & 0.172448 & 2.87619         & 1      \\
517 \dotfill  & 4.6261425 & 2        & 10        &           \ldots             &      \ldots    & 2.33451         & 1      \\
522 \dotfill & 6.621042 & 3   &   110 & \ldots & & \multicolumn{1}{l}{\hspace{1pt}\ldots}& 3 \\
524 \dotfill  & 4.955422  & 3        & 1         & RXC J0209.5+1946       & 0.0657   & 3.5731          & 1      \\
527 \dotfill  & 4.52441   & 1        & 100       &         \ldots               & 0.38508  & 1.53553         & 1      \\
529 \dotfill  & 8.404867  & 2        & 111       &         \ldots               &        \ldots  &       \multicolumn{1}{l}{\hspace{1pt}\ldots}          & 2      \\
534 \dotfill  & 4.7526336 & 3        & 1         &           \ldots             &       \ldots   &        \multicolumn{1}{l}{\hspace{1pt}\ldots}         & 3      \\
538 \dotfill  & 4.6985216 & 1        & 100       &         \ldots               &     \ldots     & 3.61937         & 1      \\
539 \dotfill  & 5.5858    & 2        & 111       & ACO   307              &       \ldots   & 3.51864         & 1      \\
544 \dotfill  & 4.7949095 & 2        & 10        &         \ldots               &      \ldots    & 3.56745         & 1      \\
555 \dotfill  & 4.5136585 & 3        & 1         &         \ldots               &       \ldots   & 2.81332         & 1      \\
556 \dotfill  & 5.27531   & 2        & 111       &          \ldots              & 0.532786 & 3.53787         & 1      \\
557 \dotfill  & 4.7586565 & 1        & 100       &        \ldots                &     \ldots     & 1.96562         & 1      \\
559 \dotfill  & 4.674867  & 1        & 100       &          \ldots              &      \ldots    & 2.48062         & 1      \\
562 \dotfill  & 4.6401553 & 2        & 10        & RXC J0157.4-0550       & 0.1289   & 4.35153         & 1      \\
564 \dotfill  & 6.557654  & 1        & 100       &           \ldots             &     \ldots     &       \multicolumn{1}{l}{\hspace{1pt}\ldots}          & 3      \\
565 \dotfill  & 5.2208285 & 2        & 111       & RXC J0137.4-1259       & 0.2143   & 5.1613          & 1      \\
576 \dotfill  & 4.5137296 & 1        & 100       &            \ldots            &      \ldots    & 2.81468         & 1      \\
586 \dotfill  & 5.319431  & 2        & 10        &              \ldots          &        \ldots  & 3.03603         & 1      \\
590 \dotfill  & 4.8859153 & 2        & 11        &             \ldots           &      \ldots    & 4.82751         & 1      \\
\hline
\end{tabular}
\end{center}
\end{tiny}
\end{table}
\begin{table}[H]
    \caption{Continuation of Table~\ref{table_lost_dets1}.}
\begin{tiny}
\begin{center}
   \begin{tabular}{L{0.5in}n{2}{2}crcl@{\hskip 0.4in}n{2}{2}c}
  \hline\hline
\multicolumn{1}{c}{I\textsc{ndex}} & \multicolumn{1}{c}{\snr\ }      & \multicolumn{1}{c}{P\textsc{ipeline}} & \multicolumn{1}{c}{P\textsc{ipe\_det}} & \multicolumn{1}{c}{I\textsc{d}\_\textsc{ext}}                & \hspace{-0.2in}
R\textsc{edshift} & \multicolumn{1}{c}{\hspace{-0.3in}\snr\ (non-blind)} & \multicolumn{1}{c}{T\textsc{ype}} \\
\hline
592 \dotfill  & 4.6193595 & 2        & 10        & RXC J0822.1+4705       & 0.1303   & 3.47281         & 1      \\
604 \dotfill  & 4.6777244 & 1        & 100       & RXC J0248.0-0332       & 0.1883   & 3.19088         & 1      \\
605 \dotfill  & 5.0145736 & 3        & 101       &             \ldots           &    \ldots      & 2.67592         & 1      \\
607 \dotfill  & 4.940518  & 1        & 100       & RXC J0956.0+4107       & 0.587    & 2.95738         & 1      \\
611 \dotfill  & 4.816886  & 2        & 111       &           \ldots            &     \ldots     &      \multicolumn{1}{l}{\hspace{1pt}\ldots}           & 3      \\
612 \dotfill  & 5.870266  & 2        & 11        &           \ldots             &     \ldots     &        \multicolumn{1}{l}{\hspace{1pt}\ldots}         & 2      \\
616 \dotfill  & 5.085939  & 2        & 10        &           \ldots             &     \ldots     & 2.20452         & 1      \\
618 \dotfill  & 5.4275565 & 2        & 11        &          \ldots              &    \ldots      & 3.43311         & 1      \\
621 \dotfill  & 4.78395   & 2        & 11        & RXC J0326.8+0043       & 0.45     & 2.89631         & 1      \\
622 \dotfill  & 5.426397  & 2        & 10        &          \ldots              & 0.494731 & 1.51139         & 1      \\
626 \dotfill  & 5.9038296 & 2        & 111       &       \ldots                 &    \ldots      &          \multicolumn{1}{l}{\hspace{1pt}\ldots}       & 3      \\
629 \dotfill  & 4.509565  & 1        & 100       &         \ldots               &    \ldots      & 2.06723         & 1      \\
639 \dotfill  & 5.103071  & 3        & 1         &           \ldots             &      \ldots    & 4.27439         & 1      \\
642 \dotfill  & 4.6462555 & 1        & 100       & WHL J164.029+34.0043   & 0.3805   & 3.50245         & 1      \\
645 \dotfill  & 4.509731  & 1        & 100       &        \ldots                &    \ldots      & 2.59932         & 1      \\
650 \dotfill  & 4.556457  & 1        & 100       &        \ldots                & 0.47     & 1.09522         & 1      \\
652 \dotfill  & 4.6217184 & 1        & 100       &      \ldots                  &    \ldots      & 2.59036         & 1      \\
653 \dotfill  & 5.187331  & 3        & 1         &          \ldots              &       \ldots   &        \multicolumn{1}{l}{\hspace{1pt}\ldots}         & 3      \\
658 \dotfill  & 4.568522  & 1        & 100       &        \ldots                &    \ldots      & 1.78067         & 1      \\
659 \dotfill  & 4.8158317 & 1        & 100       &         \ldots               &    \ldots      & 3.23821         & 1      \\
669 \dotfill  & 4.849164  & 2        & 10        & RXC J1110.7+2842       & 0.0314   & 2.96398         & 1      \\
670 \dotfill  & 4.8620453 & 1        & 100       & WHL J56.0261-13.5512   & 0.5757   & 3.72685         & 1      \\
671 \dotfill  & 4.696646  & 3        & 1         &            \ldots            &      \ldots    &          \multicolumn{1}{l}{\hspace{1pt}\ldots}       & 2      \\
672 \dotfill  & 4.9377847 & 2        & 110       & WHL J161.821+27.9906   & 0.4333   & 3.28498         & 1      \\
678 \dotfill  & 4.5387063 & 1        & 100       &        \ldots                & 0.37572  & 2.00976         & 1      \\
679 \dotfill  & 4.551286  & 1        & 100       &          \ldots              &    \ldots      & 3.31052         & 1      \\
683 \dotfill  & 5.3817525 & 2        & 10        &         \ldots               &     \ldots     &        \multicolumn{1}{l}{\hspace{1pt}\ldots}         & 2      \\
684 \dotfill  & 6.6154532 & 3        & 101       &        \ldots                &    \ldots      &       \multicolumn{1}{l}{\hspace{1pt}\ldots}          & 2      \\
697 \dotfill  & 4.6185327 & 3        & 1         & ACO   457              &      \ldots    & 3.80583         & 1      \\
698 \dotfill  & 4.7205544 & 2        & 10        &        \ldots                &    \ldots      &         \multicolumn{1}{l}{\hspace{1pt}\ldots}        & 2      \\
699 \dotfill  & 4.77476   & 1        & 100       &          \ldots              &      \ldots    & 2.73017         & 1      \\
704 \dotfill  & 4.563896  & 1        & 100       & WHL J131.956+13.5279   & 0.3487   & 3.79252         & 1      \\
705 \dotfill  & 4.6742454 & 1        & 100       &          \ldots              &    \ldots      & 0.271007        & 1      \\
712 \dotfill  & 4.653271  & 1        & 100       &           \ldots             &       \ldots   & 1.52425         & 1      \\
719 \dotfill  & 4.8215613 & 2        & 10        & WHL J158.665+20.5346   & 0.4674   & 2.94774         & 1      \\
721 \dotfill  & 4.5309896 & 1        & 100       & ACO S  270             &   \ldots       & 2.00543         & 1      \\
722 \dotfill  & 4.7083845 & 3        & 1         &            \ldots            &    \ldots      & 2.76103         & 1      \\
725 \dotfill  & 4.562224  & 2        & 10        &            \ldots            &     \ldots     & 3.60607         & 1      \\
728 \dotfill  & 5.289766  & 1        & 100       & RXC J0906.4+1020       & 0.1328   & 3.6668          & 1      \\
729 \dotfill  & 4.620416  & 1        & 100       & WHL J140.630+11.6581   & 0.2609   & 2.87712         & 1      \\
735 \dotfill  & 4.5158153 & 3        & 1         &           \ldots             &       \ldots   & 3.3707          & 1      \\
736 \dotfill  & 5.0232625 & 2        & 10        &         \ldots               &      \ldots    & 3.94745         & 1      \\
737 \dotfill  & 4.517171  & 2        & 10        &            \ldots            &      \ldots    & 3.18925         & 1      \\
743 \dotfill  & 4.5700955 & 2        & 10        &           \ldots             &    \ldots      & 3.44528         & 1      \\
748 \dotfill  & 4.5529675 & 3        & 1         &           \ldots             &      \ldots    & 3.84224         & 1      \\
749 \dotfill  & 4.9690437 & 3        & 101       & ACO  3218              &     \ldots     & 4.02593         & 1      \\
750 \dotfill  & 5.1315746 & 2        & 10        &          \ldots              & 0.31     & 3.37397         & 1      \\
751 \dotfill  & 4.8075237 & 1        & 100       & CXOMP J091126.6+055012 & 0.7682   & 3.81998         & 1      \\
753 \dotfill  & 4.5188074 & 3        & 1         &             \ldots           &     \ldots     & 3.42246         & 1      \\
755 \dotfill  & 7.7706156 & 3        & 1         &            \ldots            &    \ldots      &        \multicolumn{1}{l}{\hspace{1pt}\ldots}         & 2      \\
760 \dotfill  & 4.6238937 & 2        & 110       & WHL J134.086+1.78038   & 0.7243   & 2.89771         & 1      \\
762 \dotfill  & 4.9262795 & 3        & 1         &           \ldots             &    \ldots      &          \multicolumn{1}{l}{\hspace{1pt}\ldots}       & 4      \\
766 \dotfill  & 7.831258  & 3        & 1         &            \ldots            &     \ldots     &           \multicolumn{1}{l}{\hspace{1pt}\ldots}      & 4      \\
770 \dotfill  & 4.63975   & 3        & 1         & RXC J1047.5+1513       & 0.2108   & 3.50397         & 1      \\
771 \dotfill  & 4.5217223 & 2        & 10        & WHL J124.638-6.42296   & 0.5123   & 3.19257         & 1      \\
775 \dotfill  & 6.9654646 & 3        & 1         &           \ldots             &    \ldots      &          \multicolumn{1}{l}{\hspace{1pt}\ldots}       & 2      \\
781 \dotfill  & 4.7938952 & 3        & 1         &           \ldots             &     \ldots     &          \multicolumn{1}{l}{\hspace{1pt}\ldots}       & 3      \\
782 \dotfill  & 4.6223207 & 3        & 1         & ZwCl 0919.7-0016       & 0.3538   & 2.639           & 1      \\
788 \dotfill  & 4.64205   & 2        & 10        & ACO S  403             &     \ldots     & 3.71697         & 1      \\
789 \dotfill  & 5.2240515 & 2        & 110       &        \ldots                &   \ldots       &         \multicolumn{1}{l}{\hspace{1pt}\ldots}        & 3      \\
792 \dotfill  & 4.5170665 & 2        & 10        &          \ldots              &     \ldots     & 1.3851          & 1      \\
794 \dotfill  & 5.8301225 & 1        & 100       &         \ldots               &    \ldots      &          \multicolumn{1}{l}{\hspace{1pt}\ldots}       & 3      \\
795 \dotfill  & 4.598188  & 2        & 111       & WHL J170.480+15.8014   & 0.5593   & 2.95346         & 1      \\
798 \dotfill  & 5.0543885 & 2        & 10        &         \ldots               &      \ldots    & 3.29023         & 1      \\
809 \dotfill  & 4.588916  & 1        & 100       &         \ldots               &      \ldots    & 2.84826         & 1     \\
811 \dotfill  & 4.6410813 & 3        & 1         &          \ldots        &    \ldots      &         \multicolumn{1}{l}{\hspace{1pt}\ldots}        & 3      \\
813 \dotfill  & 4.6434965 & 1        & 100       &       \ldots           &    \ldots      & 3.05506         & 1      \\
814 \dotfill  & 4.583767  & 3        & 1         &            \ldots      &      \ldots    & 4.49628         & 1      \\
820 \dotfill  & 4.5714545 & 1        & 100       & RXC J1013.7-0006 & 0.0927   & 2.4853          & 1      \\
827 \dotfill  & 8.142719  & 3        & 1         &           \ldots       &      \ldots    &          \multicolumn{1}{l}{\hspace{1pt}\ldots}       & 2      \\
830 \dotfill  & 4.5559297 & 2        & 10        &        \ldots          &     \ldots     & 2.32998         & 1      \\
832 \dotfill  & 4.5498447 & 2        & 10        & ACO S  539       &    \ldots      & 3.03818         & 1      \\
833 \dotfill  & 4.6228485 & 2        & 110       &       \ldots           &    \ldots      & 3.38266         & 1      \\
836 \dotfill  & 4.7564516 & 2        & 110       & RXC J0345.7-4112 & 0.0603   & 3.88946         & 1      \\
843 \dotfill  & 4.513602  & 3        & 1         &         \ldots         &    \ldots      & 2.78046         & 1      \\
\hline
\end{tabular}
\end{center}
\end{tiny}
\end{table}
\begin{table}[H]
    \caption{Continuation of Table~\ref{table_lost_dets1}.}
\begin{tiny}
\begin{center}
   \begin{tabular}{L{0.5in}n{2}{2}crcl@{\hskip 0.4in}n{2}{2}c}
  \hline\hline
\multicolumn{1}{c}{I\textsc{ndex}} & \multicolumn{1}{c}{\snr\ }      & \multicolumn{1}{c}{P\textsc{ipeline}} & \multicolumn{1}{c}{P\textsc{ipe\_det}} & \multicolumn{1}{c}{I\textsc{d}\_\textsc{ext}}                & \hspace{-0.2in}
R\textsc{edshift} & \multicolumn{1}{c}{\hspace{-0.3in}\snr\ (non-blind)} & \multicolumn{1}{c}{T\textsc{ype}} \\
\hline
844 \dotfill  & 4.7834334 & 3        & 1         &      \ldots            &       \ldots   & 3.91417         & 1      \\
845 \dotfill  & 7.1448226 & 3        & 1         &      \ldots            &  \ldots        &      \multicolumn{1}{l}{\hspace{1pt}\ldots}            & 4      \\
859 \dotfill  & 4.886414  & 3        & 1         &        \ldots          &    \ldots      & 1.86549         & 1      \\
860 \dotfill  & 4.638687  & 1        & 100       &      \ldots            &  \ldots        & 2.41334         & 1      \\
864 \dotfill  & 5.0417113 & 2        & 10        &      \ldots            &   \ldots       & 3.54132         & 1      \\
866 \dotfill  & 5.168761  & 3        & 1         &         \ldots         &      \ldots    &         \multicolumn{1}{l}{\hspace{1pt}\ldots}         & 3      \\
874 \dotfill  & 4.548713  & 1        & 100       &      \ldots            &   \ldots       & 3.45449         & 1      \\
884 \dotfill  & 4.6741834 & 1        & 100       &     \ldots             &  \ldots        & 0.99106         & 1      \\
885 \dotfill  & 4.5309443 & 3        & 1         &        \ldots          &    \ldots      & 3.0147          & 1      \\
886 \dotfill  & 4.715562  & 2        & 111       &      \ldots            &  \ldots        & 2.5028          & 1      \\
900 \dotfill  & 4.6036177 & 3        & 1         &       \ldots           &    \ldots      & 2.9232          & 1      \\
908 \dotfill  & 4.6843767 & 2        & 10        &     \ldots             & 0.45     & 3.14342         & 1      \\
909 \dotfill  & 4.6769667 & 1        & 100       &    \ldots              &   \ldots       & 1.94061         & 1      \\
913 \dotfill  & 4.872177  & 2        & 10        &       \ldots           &     \ldots     & 3.50097         & 1      \\
917 \dotfill  & 4.7799654 & 1        & 100       &     \ldots             &   \ldots       & 1.45693         & 1      \\
921 \dotfill  & 4.720672  & 2        & 10        &       \ldots           & 0.26     & 4.22762         & 1      \\
923 \dotfill  & 4.713845  & 1        & 100       &       \ldots           &     \ldots     & 1.56599         & 1      \\
925 \dotfill  & 5.002551  & 1        & 100       &       \ldots           &     \ldots     & 2.16267         & 1      \\
927 \dotfill  & 4.715874  & 1        & 100       &      \ldots            &    \ldots      & 1.83619         & 1      \\
928 \dotfill  & 4.6580725 & 3        & 1         &        \ldots          &     \ldots     & 2.65203         & 1      \\
933 \dotfill  & 4.5972934 & 2        & 10        &      \ldots            &  \ldots        & 3.05862         & 1      \\
949 \dotfill  & 4.864598  & 3        & 1         &         \ldots         &    \ldots      & 1.88233         & 1      \\
950 \dotfill  & 4.7590675 & 1        & 100       &     \ldots             &   \ldots       & 2.10619         & 1      \\
953 \dotfill  & 4.865524  & 3        & 1         &          \ldots        &     \ldots     & 3.98468         & 1      \\
964 \dotfill  & 4.946659  & 2        & 10        &        \ldots          & 0.14     & 2.8276          & 1      \\
965 \dotfill  & 4.5068207 & 1        & 100       &      \ldots            &     \ldots     & 1.70981         & 1      \\
966 \dotfill  & 4.5067983 & 3        & 1         &         \ldots         &      \ldots    & 2.1943          & 1      \\
968 \dotfill  & 4.535754  & 2        & 10        &          \ldots        &    \ldots      & 2.67392         & 1      \\
973 \dotfill  & 4.5617847 & 1        & 100       &        \ldots          &    \ldots      & 1.52403         & 1      \\
980 \dotfill  & 12.780052 & 2        & 111       & RXC J1217.6+0339 & 0.0766   &                 & 2      \\
992 \dotfill  & 5.522259  & 2        & 11        &         \ldots         &     \ldots     & 0.68348         & 1      \\
1010\dotfill  & 5.906472  & 2        & 11        &        \ldots          &    \ldots      &       \multicolumn{1}{l}{\hspace{1pt}\ldots}           & 3      \\
1016\dotfill  & 4.763515  & 2        & 110       &       \ldots           &    \ldots      & 2.73909         & 1      \\
1018\dotfill  & 4.6082454 & 3        & 1         &         \ldots         &     \ldots     & 3.96679         & 1      \\
1019\dotfill  & 4.981976  & 2        & 11        &         \ldots         &      \ldots    & 3.89276         & 1      \\
1031\dotfill  & 4.8077016 & 2        & 110       &      \ldots            &    \ldots      &     \multicolumn{1}{l}{\hspace{1pt}\ldots}             & 3      \\
1039\dotfill  & 4.50475   & 2        & 10        &         \ldots         &      \ldots    & 3.30443         & 1      \\
1048\dotfill  & 4.8961234 & 2        & 10        & ACO S 137        & 0.02764  & 2.49855         & 1      \\
1049\dotfill  & 4.948578  & 2        & 10        & ACO  1603        & 0.1314   & 2.85532         & 1      \\
1052\dotfill  & 5.7445407 & 3        & 1         &         \ldots         &      \ldots    &      \multicolumn{1}{l}{\hspace{1pt}\ldots}            & 3      \\
1055\dotfill  & 4.6209354 & 3        & 1         & RXC J0052.7-8015 & 0.1141   & 4.03375         & 1      \\
1059\dotfill  & 4.684562  & 1        & 100       &       \ldots           &    \ldots      & 2.7398          & 1      \\
1060\dotfill  & 6.810468  & 2        & 111       &        \ldots          &    \ldots      &        \multicolumn{1}{l}{\hspace{1pt}\ldots}          & 3      \\
1069\dotfill  & 4.8399434 & 2        & 10        &        \ldots          & 0.12941  & 2.2881          & 1      \\
1080\dotfill  & 4.605055  & 1        & 100       &        \ldots          &     \ldots     & 2.39711         & 1      \\
1081\dotfill  & 4.6       & 2        & 10        &            \ldots      &       \ldots   &              \multicolumn{1}{l}{\hspace{1pt}\ldots}    & 3      \\
1091\dotfill  & 5.2132697 & 2        & 11        &        \ldots          &    \ldots      &        \multicolumn{1}{l}{\hspace{1pt}\ldots}          & 3      \\
1092\dotfill  & 4.6357284 & 2        & 10        &        \ldots          &    \ldots      & 3.08333         & 1      \\
1094\dotfill  & 4.676179  & 1        & 100       &        \ldots          &    \ldots      & 2.56732         & 1      \\
1103\dotfill  & 6.148161  & 1        & 100       &        \ldots          &    \ldots      &        \multicolumn{1}{l}{\hspace{1pt}\ldots}          & 2      \\
1107\dotfill  & 5.0184035 & 2        & 10        &       \ldots           &    \ldots      & 2.38865         & 1      \\
1111\dotfill  & 4.831168  & 2        & 10        &        \ldots          &      \ldots    & 2.73575         & 1      \\
1119\dotfill  & 5.4801097 & 2        & 10        &       \ldots           &     \ldots     & 3.52593         & 1      \\
1132\dotfill  & 4.5146875 & 1        & 100       &         \ldots         &     \ldots     & 1.42893         & 1      \\
1133\dotfill  & 4.7255597 & 1        & 100       &        \ldots          & 0.25     & 2.80855         & 1      \\
1135\dotfill  & 4.781288  & 3        & 1         &         \ldots         &       \ldots   & 3.83267         & 1      \\
1144\dotfill  & 4.9033356 & 3        & 1         &       \ldots           &  \ldots        & 4.21885         & 1      \\
1152\dotfill  & 4.63114   & 3        & 1         &        \ldots          &      \ldots    & 3.60522         & 1      \\
1155\dotfill  & 4.7780237 & 2        & 111       & SPT-CLJ2148-6116 & 0.571    & 3.58897         & 1      \\
1159\dotfill  & 4.7621408 & 1        & 100       &      \ldots           &   \ldots       & 2.15125         & 1      \\
1162\dotfill  & 4.628621  & 1        & 100       &       \ldots           &   \ldots       & 2.73569         & 1      \\
1170\dotfill  & 4.6537085 & 2        & 10        &        \ldots          &    \ldots      & 3.63003         & 1      \\
1173\dotfill  & 4.550007  & 1        & 100       &         \ldots         &     \ldots     & 2.2385          & 1      \\
1174\dotfill  & 6.613522  & 2        & 111       &         \ldots         &     \ldots     &      \multicolumn{1}{l}{\hspace{1pt}\ldots}            & 2      \\
1177\dotfill  & 5.580667  & 2        & 10        &          \ldots        &      \ldots    &        \multicolumn{1}{l}{\hspace{1pt}\ldots}          & 3      \\
1178\dotfill  & 4.5030346 & 1        & 100       &        \ldots          &    \ldots      & 2.90078         & 1      \\
1180\dotfill  & 5.8775973 & 2        & 111       &         \ldots         &    \ldots      &     \multicolumn{1}{l}{\hspace{1pt}\ldots}             & 3      \\
1188\dotfill  & 4.7791066 & 2        & 10        &          \ldots        &     \ldots     & 3.03411         & 1      \\
1194\dotfill  & 4.7459564 & 2        & 10        &         \ldots         & 0.21     & 3.2567          & 1      \\
1196\dotfill  & 4.568303  & 1        & 100       &         \ldots         &     \ldots     & 2.75983         & 1      \\
1197\dotfill  & 4.775479  & 2        & 11        &       \ldots           &     \ldots     & 2.34573         & 1      \\
1198\dotfill  & 4.8042645 & 1        & 100       &      \ldots            &    \ldots      & 2.90578         & 1      \\
1199\dotfill  & 4.5996885 & 2        & 110       &        \ldots          &    \ldots      & 2.31421         & 1      \\
1203\dotfill  & 5.0193796 & 1        & 100       & ACO S 808        & 0.049131 & 3.40921         & 1      \\
1204\dotfill  & 4.696981  & 1        & 100       &         \ldots         & 0.5      & 3.20218         & 1      \\
1212\dotfill  & 4.7399464 & 1        & 100       &        \ldots          &   \ldots       & 2.84324         & 1      \\
1215\dotfill  & 4.6261168 & 1        & 100       &        \ldots          &    \ldots      & 3.42905         & 1      \\
1217\dotfill  & 4.5237203 & 2        & 10        &         \ldots         &     \ldots     & 2.48797         & 1      \\
1219\dotfill  & 5.1635747 & 3        & 1         &          \ldots        &     \ldots     &       \multicolumn{1}{l}{\hspace{1pt}\ldots}           & 2      \\
1221\dotfill  & 5.468506  & 2        & 11        &         \ldots         &    \ldots      &        \multicolumn{1}{l}{\hspace{1pt}\ldots}          & 2     \\
\hline
\end{tabular}
\end{center}
\end{tiny}
\end{table}

\end{document}